\documentclass[preprint,nofootinbib,
preprintnumbers,
amsmath,amssymb]{revtex4}

\usepackage{graphicx,epsfig}
\usepackage{simplewick}
\usepackage{slashed}
\usepackage{color}
\usepackage{setspace}
\usepackage{hyperref}

\def\simg{{\ \lower-1.2pt\vbox{\hbox{\rlap{$>$}\lower6pt\vbox{\hbox{$\sim$}}}}\ }}
\def\siml{{\ \lower-1.2pt\vbox{\hbox{\rlap{$<$}\lower6pt\vbox{\hbox{$\sim$}}}}\ }} 

\newcommand{\eq}[1]{Eq.~\eqref{#1}}
\newcommand{\eqs}[2]{Eqs.~\eqref{#1} and \eqref{#2}}

\newcommand{\Sec}[1]{Sec.~\ref{#1}}

   \graphicspath{{./figs/}}
\usepackage{color}

\usepackage{relsize}
\newcommand{\babar}{{\mbox{\slshape B\kern-0.1em{\smaller A}\kern-0.1em
            B\kern-0.1em{\smaller A\kern-0.2em R}}}
\def\MSbar{\relax\ifmmode\overline                        %%%%%%%%%
            {\rm MS}\else{$\overline{\rm MS}${ }}\fi}     %%%%%%%%%
           }                                              %%%%%%%%%
\def\MSbar{\relax\ifmmode\overline                        %%%%%%%%%
            {\rm MS}\else{$\overline{\rm MS}${ }}\fi}     %%%%%%%%%
\def\1{\hbox{{1}\kern-.25em\hbox{l}}}
 \date{\today}
\def\be{\begin{equation}}
\def\ee{\end{equation}}
\def\bea{\begin{eqnarray}}
\def\eea{\end{eqnarray}}
\def\bear{\begin{array}}
\def\eear{\end{array}}

\def\al{\alpha}
\def\nn{\nonumber}

\newcommand{\MS}{\overline{\rm MS}}

\newcommand{\m}{{\overline m}}

\def\lQ{\Lambda_{\rm QCD}}

\begin{document}
\title{Hyperasymptotic approximation to the top, bottom and charm pole mass}

\author{Cesar Ayala}
\affiliation{Department of Physics, Universidad T{\'e}cnica Federico
Santa Mar{\'\i}a (UTFSM),  
Casilla 110-V, 
Valpara{\'\i}so, Chile}
\author{Xabier Lobregat}
\author{Antonio Pineda}
\affiliation{Grup de F\'\i sica Te\`orica, Dept. F\'\i sica and IFAE-BIST, Universitat Aut\`onoma de Barcelona,\\ 
E-08193 Bellaterra (Barcelona), Spain}

\date{\today}

\begin{abstract}
We construct hyperasymptotic expansions for the heavy quark pole mass regulated using the principal value (PV) prescription. We apply such hyperasymptotic expansions to the $B/D$ meson masses, and $\bar \Lambda $ computed in the lattice. The issue of the uncertainty of the (top) pole mass is critically reexamined. The present theoretical uncertainty in the relation between $\m_t$, the $\MS$ top mass, and $m_{t, \rm PV}$, the top pole mass regulated using the PV prescription, is numerically assessed to be 
$
\delta m_{t,\rm PV}= 28\;{\rm MeV}$ for $\m_t =163$ GeV.
\end{abstract}

\maketitle
\tableofcontents

\vfill
\newpage

\section{Introduction}

We construct hyperasymptotic expansions for the heavy quark pole mass regulated using the principal value (PV) prescription along the lines of \cite{HyperI}. We generalize the discussion of that reference by including possible ultraviolet renormalons. We then apply such expansions to various observables. The name hyperasymptotic we borrow from \cite{BerryandHowls}. For a treatise of hyperasymptotic expansions in the context of ordinary differential equations see \cite{Boyd99}.

In \cite{HyperI} we studied observables characterized by having a large scale $Q \gg \lQ$, and for which the operator product expansion (OPE) is believed to be a good approximation. We computed them within an hyperasymptotic expansion. More specifically, the perturbative part of the OPE was summed up using the PV prescription: $S_{\rm PV}$. The difference between $S_{\rm PV}$ and the full non-perturbative (NP) result is assumed to exactly scale as the intrinsic NP terms of the OPE. In general terms:
\be
{\rm Observable}(\frac{Q}{\lQ})
=
S_{\rm PV}(\al_X(Q))+K_X^{\rm (PV)}\al_X^{\gamma}(Q)\frac{\Lambda_X^{d}}{Q^d}\left(1+{\cal O}(\al_X(Q))\right)
+{\cal O}(\frac{\Lambda_X^{d'}}{Q^{d'}})
\,,
\label{Obtruncated}
\ee 
where the last term refers to genuine higher order terms in the OPE ($d' > d>0$).
Then, since $S_{\rm PV}$ can not be computed exactly, we obtain it approximately along an hyperasymptotic expansion (a combination of (truncated) perturbative sums and of NP corrections). This is possible if enough terms of the perturbative expansion are known, and if the divergent structure of the leading renormalons of the observable is also known. This allows us to have a clear (parametric) control on the error of the computation. Two alternative methods were considered in \cite{HyperI} depending on how the truncation of the leading perturbative sum 
\be
S_T(\al)=\sum_{n=0}^Np^{(X)}_n\al_X^{n+1}(\mu)
\ee
is made:
\begin{enumerate}
\item[{1)}]
$N$ and $\mu \sim Q$ large but finite:
\be
\label{eq:NP}
N=N_P \equiv |d|\frac{ 2\pi}{\beta_0\alpha_X(\mu)}\big(1-c\,\alpha_X(\mu)\big)
\,,
\ee
\item[{2)}]
$N \rightarrow \infty$ and $\mu \rightarrow \infty$ in a correlated way. We considered two options:
\be
\label{eq:muinfty}
{\rm A)} \quad N+1=N_S(\alpha)\equiv |d|\frac{ 2\pi}{\beta_0\alpha_X(\mu)}
 \,; \qquad 
{\rm B)} \quad N=N_A(\alpha) \equiv |d|\frac{ 2\pi}{\beta_0\alpha_X(\mu)}\big(1-c'\alpha_X(Q)\big)
\ee
\end{enumerate}
where $c'>0$ but $c$ is arbitrary otherwise. Note that in case 1), $c$ can partially simulate changes on the scale or scheme 
of $\alpha_X$. $d$ is the dimension associated to a given renormalon. Note that in this paper $d$ can be positive (infrared renormalons) or negative (ultraviolet renormalons), unlike in \cite{HyperI}, where only positive $d$'s were considered. Note also that genuine NP corrections are only associated to positive $d$'s.

We will not study the modifications the inclusion of ultraviolet renormalons produce in case 2). In this paper we are mainly concerned in the scenario where the leading renormalon is of infrared nature and subleading renormalons can be ultraviolet and/or infrared. This is the case of the pole mass. In such scenario the precision we can obtain in case 2) is limited by the approximate knowledge of the leading infrared renormalon and we cannot add further to the discussion given in \cite{HyperI}. Different is the case 1), which we discuss in the next section. 

The structure of the paper is as follows. In Sec. \ref{Sec:Scheme} we review the general case. Compared with \cite{HyperI} we include the possible effect of ultraviolet renormalons.  In Sec. \ref{Sec:mOS} we study the pole mass of a heavy quark in the large $\beta_0$ approximation. We use it as toy-model to test our methods. 
We then move to real QCD. In Sec. \ref{Sec:Lambda} we study the $B/D$ meson mass and lattice evaluations of $\bar \Lambda$.  Finally, in Sec. \ref{Sec:Top} we do a dedicated study of the top mass. 

In general we will avoid to make explicit the scheme ($X$) and scale ($\mu$) dependence unless necessary. 

\section{General formulas}
\label{Sec:Scheme}

\subsection{$N$ large and $\mu \sim Q \gg \lQ$. \eq{eq:NP}. Case 1)}

This case was already discussed at length in \cite{HyperI}. We now give the general expression after the inclusion of ultraviolet renormalons (for a more detailed discussion see \cite{Ayala:2019lak}). It can be written in the following way
\be
\label{SPVtotal}
S_{\rm PV}(Q)=S_P+\sum_{\{|d|\}}S_{|d|}+\sum_{\{d>0\}}\Omega_d+\sum_{\{d<0\}}\Omega_d
\,,
\ee
where
\be
S_P\equiv \sum_{n=0}^{N_P(|d_{min}|)}p_n\al^{n+1}(\mu) \equiv S_{|d|=0}
\,,
\ee
and ($|d|>0$)
\be
S_{|d|}\equiv \sum_{n=N_P(|d|)+1}^{N_P(|d'|)}(p_n-p_n^{(as)})\al^{n+1}(\mu)
\,,
\ee
where the asymptotic behavior associated to renormalons with dimensions $\leq |d|$ is included in $p_n^{(as)}$, and $d'$ is the dimension of the closest renormalon to the origin in the Borel plane fulfilling that $|d'| > |d|$. 
$\Omega_d$ is a modification of the definition of terminant given in \cite{Dingle} that is more suitable to our case. Whereas in \cite{Dingle} ($p_N \alpha^N \times$ the) terminant refers to the completation of the superasymptotic approximation to give the complete result, here $\Omega_d$ is the completion of the part of the perturbative series associated to the singularity located at $u\equiv \frac{\beta_0 t}{4\pi}=\frac{d}{2}$ in the Borel plane using the PV prescription. For the case of infrared renormalons ($d>0$) the general analytic expression of $\Omega_d$ can be found in \cite{HyperI}. For a generic ultraviolet renormalon ($d<0$) that produces the asymptotic behavior 
\be
p_n^{(as)}=Z_{O_d}^X\frac{\mu^d}{Q^d}
\frac{\Gamma(n+b'+1)}{\Gamma(b'+1)}\left(\frac{\beta_0}{2\pi d}\right)^n
\bigg\{1+c_1\frac{b'}{n+b'}+c_2\frac{b'^2}{(n+b')(n+b'-1)}+\dots\bigg\}
\,,
\ee
$\Omega_{d<0}$ reads 
\be
\Omega_{d<0}=\Delta \Omega_{UV}(db)+c_1\Delta\Omega_{UV}(db-1)+\cdots
\,,
\ee
where\footnote{An sketch of how these computations are done is given in Appendix \ref{Sec:Integral}.} 
(we define $\eta_c \equiv -b'+\frac{2\pi |d|c}{\beta_0}-1$ where $b'=db-\gamma$)
\bea
\Delta\Omega_{UV}(db)&=&Z_{O_d}^X\frac{\mu^d}{Q^d}(-1)^{N_P+1}\frac{1}{\Gamma(b'+1)}\left(\frac{\beta_0}{2\pi |d|}\right)^{N_P+1}\alpha^{N_P+2}\int_0^{\infty}dx\,\frac{e^{-x}x^{N_P+1+b'}}{1+\frac{x\beta_0\alpha}{2\pi |d|}}
\\
\nn
&=&Z_{O_d}^X\frac{\mu^d}{Q^d}(-1)^{N_P+1}\frac{\pi}{\Gamma(b'+1)}\left(\frac{\beta_0}{ |d|}\right)^{-b'-1/2}\alpha(\mu)^{1/2-b'}e^{\frac{-2\pi |d|}{\beta_0\alpha(\mu)}}\bigg\{
1
\\
\nn
&&
+\frac{\alpha(\mu)}{\pi}\frac{\beta_0}{12|d|}\left[-1+3\eta_c^2\right]
\\
&&
\nn
+\frac{\alpha^2(\mu)}{\pi^2}\frac{\beta_0^2}{1152|d|^2}\bigg[13-48\eta_c
-60\eta_c^2+48\eta_c^3+36\eta_c^4\bigg]
+\mathcal{O}(\alpha^3)
\bigg\}
\,.
\eea
Joining all terms together we have
\be
\Omega_{d<0}=\sqrt{\alpha(\mu)}K_X^{(P)}\frac{Q^{|d|}}{\mu^{|d|}}e^{\frac{-2\pi|d|}{\beta_0\alpha(\mu)}}\left(\frac{\beta_0\alpha(\mu)}{4\pi}\right)^{-b'}\bigg\{1+\bar{K}_{X,1}^{(P)}\alpha(\mu)+\bar{K}_{X,2}^{(P)}\alpha^2(\mu)+\mathcal{O}\left(\alpha^3(\mu)\right)\bigg\}
\,,
\ee
where
\bea
K_X^{(P)}&\equiv& Z^X_{\mathcal{O}_d}(-1)^{N_p+1}\left(\frac{\beta_0}{\pi^2|d|}\right)^{-1/2}\frac{1}{\Gamma(b'+1)}\left(\frac{2}{|d|}\right)^{-b'}
\,,
\\
\bar{K}_{X,1}^{(P)}&\equiv&\left(\frac{2}{\pi}\right)^{1/2}\left(c_1\frac{\beta_0 b'}{2 \sqrt{2 \pi } |d|}+\frac{\beta_0}{12|d|\sqrt{2\pi}}(-1+3\eta_c^2)\right)
\,,
\\
\nn
\bar{K}_{X,2}^{(P)}&\equiv&\left(\frac{2}{\pi}\right)^{1/2}\bigg(
c_2\frac{b'^2\beta_0^2}{4\sqrt{2}|d|^2\pi^{3/2}}
+c_1\frac{b'\beta_0^2(-1+3(\eta_c+1)^2)}{24\sqrt{2}|d|^2\pi^{3/2}}
\\
&&
+\frac{\beta_0^2}{1152|d|^22^{1/2}\pi^{3/2}}\bigg[13-48\eta_c
-60\eta_c^2+48\eta_c^3+36\eta_c^4\bigg]\bigg)
\,.\eea

Note that in this case $\mu$ is in the denominator. If we set the anomalous dimension to zero ($b'=db$), $\Omega_{d<0} \sim \sqrt{\al(\mu)}\frac{\lQ^{|d|}Q^{|d|}}{\mu^{2|d|}}$ (unlike for infrared renormalons where $\Omega_{d>0} \sim \sqrt{\al(\mu)}
\frac{\lQ^{d}}{Q^{d}}$). If one takes $\mu$ very large this term will be quite small. In practice, if we take $\mu \sim Q$, we may need this term. 

$S_{\rm PV}$ will be computed truncating the hyperasymptotic expansion in a systematic way. This means truncating \eq{SPVtotal} as follows (note that we always define $D$ to be positive): 
\bea
\label{SPVDN}
S^{(D,N)}_{\rm PV}(Q)&=&\sum_{\{|d|\}}S_{|d|<D}+\sum_{\{|d|\leq D\}}\Omega_d
\\
\nn
&&
+
\sum_{n=N_P(D)+1}^{N_P(D)+N}(p_n-p_n^{(as)})\al^{n+1}(\mu)
\,.
\eea
For each value of the couple $(D,N)$, we can state the parametric accuracy of $S^{(D,N)}_{\rm PV}(Q)$. For instance for $S^{(0,N_P)}$ the error would be (up to a numerical  and a $\sqrt{\al_X}$ factor)
\be
\label{error0NP}
\delta S^{(0,N_P)} \sim {\cal O}\left(e^{-|d_{min}|\frac{2\pi}{\beta_0\alpha_X(Q)}}\right)
\,.
\ee
This is what is commonly named the superasymptotic approximation. 
For $S^{(|d_{min}|,0)}$ the parametric form of the error reads (up to a numerical  and a possible $\al^{3/2}_X$ factor):
\be
\label{errordmin}
\delta S^{(|d_{min}|,0)} \sim {\cal O}\left(e^{-|d_{min}|\frac{2\pi}{\beta_0\alpha_X(Q)}\left(1+\ln(|d/d_{min}|\right)}\right)
\,,
\ee
where $d$ is the location of the next renormalon closest to the origin. This corresponds to the first term in the hyperasymptotic approximation. The expression for the error in the general case $S^{(D,N)}_{\rm PV}(Q)$ reads ($N \not=N_P$ but large)
\be
\label{errorgeneral}
\delta S^{(D,N)} \sim {\cal O}\left(e^{-D\frac{2\pi}{\beta_0\alpha_X(Q)}\left(1+\ln(|d/D|\right)}\al_X^N\right)
\,,
\ee
where $d$ is the location of the next renormalon closest to the origin after $D$.

\subsection{$m_{\rm PV}$, general formulas}
\label{Sec:mPV}
For the case of the heavy quark mass, which we discuss at length in this paper, we have ($\m = m_{\MS}(m_{\MS})$)
\be
\label{mPV}
m_{\rm PV}(\m)=m_P+\m \Omega_m+\sum_{n=N_P+1}^{2N_P}(r_n-r_n^{(as)})\al^{n+1}(\mu)+
\m\Omega_2+\m \Omega_{-2}+{\cal O}\left(e^{-2\frac{2\pi}{\beta_0\al}\left(1+\ln(3/2)\right)}\right)
\,,
\ee
where 
\be
m_P\equiv \m+\sum_{n=0}^{N_P} r_n\al^{n+1}(\mu)\;;
\ee 
the coefficients $r_n$ for $n \leq 3$ were computed in \cite{Tarrach:1980up,Chetyrkin:1999ys,Melnikov:2000qh,Marquard:2015qpa};
\begin{equation}
\label{Omegam}
\Omega_m=\sqrt{\alpha_X(\mu)}K_X^{(P)}\frac{\mu}{\m}e^{-\frac{2\pi}{\beta_0 \alpha_X(\mu)}}
\left(\frac{\beta_0\alpha_X(\mu)}{4\pi}\right)^{-b}
\bigg(1+\bar K_{X,1}^{(P)}\alpha_X(\mu)+\bar K_{X,2}^{(P)}\alpha_X^2(\mu)+\mathcal{O}\left(\alpha_X^3(\mu)\right)\bigg)
\,,\end{equation}
where now $K_X^{(P)}$ and $K_{X,i}^{(P)}$ read
\bea
K_X^{(P)}&=&-\frac{Z^X_m2^{1-b}\pi}{\Gamma(1+b)}\beta_0^{-1/2}\bigg[-\eta_c+\frac{1}{3}\bigg]
\,,
\\
\bar K_{X,1}^{(P)}&=&\frac{\beta_0/(\pi)}{-\eta_c+\frac{1}{3}}\bigg[-b_1 b \left(\frac{1}{2}\eta_c+\frac{1}{3}\right)
-\frac{1}{12}\eta_c^3+\frac{1}{24}\eta_c-\frac{1}{1080}\bigg]
\,,
\\
\bar K_{X,2}^{(P)}&=&\frac{\beta_0^2/\pi^2}{-\eta_c+\frac{1}{3}}
\bigg[-w_2 (b -1) b \left(\frac{1}{4}\eta_c+\frac{5}{12}\right)
+b_1b\left(-\frac{1}{24}\eta_c^3-\frac{1}{8}\eta_c^2
-\frac{5}{48}\eta_c-\frac{23}{1080}\right)
\nn
\\
&&
-\frac{1}{160}\eta_c^5
-\frac{1}{96}\eta_c^4+\frac{1}{144}\eta_c^3
+\frac{1}{96}\eta_c^2-\frac{1}{640}\eta_c-\frac{25}{24192}\bigg]
\,,
\eea
where we have applied the general expression obtained in \cite{HyperI} to this case. In particular ($b$ and $s_n$ are defined in \cite{HyperI}),
\be
\eta_c=-b+\frac{2\pi c}{\beta_0}-1 \;, \quad b_1=s_1 \;, \qquad w_2=\left(\frac{s_1^2}{2}-s_2\right)\frac{b}{b-1}
\,.
\ee
Finally, 
\be
\label{rnas}
r_n^{\rm (as)}(\mu)=Z^X_{m} \mu \,\left({\beta_0 \over
2\pi}\right
)^n \,\sum_{k=0}^\infty c_k{\Gamma(n+1+b-k) \over
\Gamma(1+b-k)}
\,.
\ee
The coefficients $c_k$ are pure functions of the $\beta$-function coefficients, as first shown in \cite{Beneke:1994rs}. They can be found in \cite{Beneke:1998ui,Pineda:2001zq,Ayala:2014yxa}. At low orders they read ($c_0=1$) 
\be
c_1=s_1 \,,\quad 
c_2=\frac{1}{2}\frac{b}{b-1}(s_1^2-2s_2)
\,,
\quad
c_3=\frac{1}{6}\frac{b^2}{(b-2)(b-1)}(s_1^3-6s_1s_2+6s_3)
\,.
\ee
Note that
\be
m_{\rm OS}^{(N)}=\m+\sum_{n=0}^{N} r_n\al^{n+1}(\mu)
\,.
\ee
Therefore, $m_P$ is nothing but the pole mass truncated to order $N=N_P$. 

Our knowledge of the other terminants, $\Omega_2$ and $\Omega_{-2}$, is  limited. We do not know the renormalization group structure of $\Omega_{-2}$, except in the large $\beta_0$. On the other hand, the renormalization group structure of $\Omega_{2}$ is exactly known (provided the coefficients of the beta function are known to all orders). The reason is that it is linked to the kinetic operator of the HQET Lagrangian, the Wilson coefficient of which is protected by reparameterization invariance \cite{Luke:1992cs}. Therefore, it has no anomalous dimension and the Wilson coefficient is 1 in dimensional regularization to all orders in perturbation theory. Still, in the large $\beta_0$ approximation, the coefficient $Z^X_{2}$ is equal to zero. If it is different from zero beyond the large $\beta_0$ approximations has been a matter of debate \cite{Neubert:1996zy}. We will retake this discussion in the following sections.  

We also give the formulas that apply to \eq{eq:muinfty}, i.e. to case 2): the limit $(N,\mu) \rightarrow \infty$. A general discussion can be found in \cite{HyperI}. It was argued that the limit 2A) was likely to be logarithmic divergent (see also \cite{Sumino:2005cq}), and no formulas could be found that are valid beyond the large $\beta_0$ approximation. Therefore, we will not study this case further. For the limit 2B) formulas with NP exponential accuracy were found in \cite{HyperI} generalizing results from \cite{VanAcoleyen:2003gc}. These formulas were valid beyond the large $\beta_0$ approximation. For the specific case of the pole mass they read
\be
\label{mPVHyperAco}
m_{\rm PV}= m_A+K_X^{(A)}\Lambda_X+{\cal O}(\al\Lambda_X)
\,,
\ee
where 
\be
m_A=\m+\lim_{\mu \rightarrow \infty; 2B)}\sum_{n=0}^{N_A}r_n\al^{n+1}(\mu)
\,,
\ee
and 
\be
K^{(A)}_X=\frac{2\pi }{\beta_0}Z_m^X\bigg(\frac{\beta_0}{4\pi}\bigg)^{b}
\int_{-c',\rm PV}^{\infty}dx\,e^{-\frac{2\pi }{\beta_0}x}\frac{1}{(-x)^{1+b}}
\,.
\ee
As discussed in \cite{HyperI}, there is more than one way to take the $\mu \rightarrow \infty$; 2B) limit. One is to take Eq. (67) of \cite{HyperI} for $N_A$ instead of limit B) of 
\eq{eq:muinfty}. Both methods are general but require the knowledge of $r_n$ and the beta function coefficients to all orders. This potentially limits their applicability in practice. 
Another option is to interpret the $\mu \rightarrow \infty$ limit as a change of scheme (where $\mu_0 \sim \m$):
\be
\label{1loop}
\al_{X'}(\mu)=\frac{\al_X(\mu_0)}{1+\frac{\beta_0}{2\pi}\al_X(\mu_0)\ln(\frac{\mu}{\mu_0})} \quad {\rm and} \quad N_A(\alpha) \equiv \frac{ 2\pi}{\beta_0\alpha_{X'}(\mu)}\big(1-c'\alpha_X(\mu_0)\big)
\,.
\ee 
This method still requires the knowledge of $r_n$ to all orders. On the other hand, there is no need to know the $\beta$-function to all orders. 

Irrespectively of which of the above methods we use to take the $\mu \rightarrow \infty$ limit we have
\be
\label{eq:limit}
\lim_{\mu \rightarrow \infty; 2B)}\sum_{n=0}^{N_A}r_n\al^{n+1}(\mu)=
\int_0^{\frac{4\pi}{\beta_0\chi}}dt e^{-t/\alpha_X(\mu_0)} B[m_{\rm PV}-\m](t)
\,,
\ee
where $2/\chi=1-c'\al(\mu_0)$. The right-hand side of \eq{eq:limit} can not be computed exactly. An approximated determination can be obtained by approximating the Borel transform to ($u\equiv \frac{\beta_0 t}{4\pi}$)
\be
\label{eq:Borelapprox}
B[m_{\rm PV}-\m](t)=\sum_{n=0}^{N_{max}}\frac{(r_n-r_n^{(as)})}{n!}t^n+\frac{Z_m\mu}{(1-2u)^{1+b}}\left(1+c_1(1-2u)++c_2(1-2u)^2+\cdots\right)
\,,
\ee
where $N_{max}$ is the number of perturbative coefficients that are known.

\section{$m_{\rm PV}(\m)$ in the large $\beta_0$ approximation}
\label{Sec:mOS}
Here the discussion runs parallel to the discussion for the static potential in the large $\beta_0$ approximation made in Section III of \cite{HyperI}. Nevertheless, we do not have the same analytic control as for the static potential. Note also that now we have ultraviolet renormalons. Moreover, the pole mass has the extra complication that it is ultraviolet divergent and needs renormalization. This makes the Borel transform  more complicated and we do not have the exact $\mu$ factorization one has in the static potential. We take the Borel transform from \cite{Beneke:1994sw,Ball:1995ni,Neubert:1994vb}:
\be
\label{eq:BorelmPV}
B[m_{\rm PV}-\m](u)
=
\m \frac{C_F}{4\pi}\bigg[\bigg(\frac{\m^2}{\mu^2}\bigg)^{-u}e^{-c_{\MS}u}6(1-u)\frac{\Gamma(u)\Gamma(1-2u)}{\Gamma(3-u)}-\frac{3}{u}+R(u)\bigg]
\,,
\ee
where $c_{\MS}=-5/3$, $u=\frac{\beta_0}{4\pi}t$ and
\be
R(u)=\sum_{n=1}^{\infty}\frac{1}{(n!)^2}\frac{d^n}{dz^n}G(z)\bigg|_{z=0}u^{n-1}=-\frac{5}{2}+\frac{35}{24}u+\mathcal{O}(u^2)
\,,
\ee
\be
G(u)=-\frac{1}{3}(3+2u)\frac{\Gamma(4+2u)}{\Gamma(1-u)\Gamma^2(2+u)\Gamma(3+u)}
\,.
\ee
This expression has been derived in the $\MS$ scheme. Whereas the scheme dependence of the first term in \eq{eq:BorelmPV} can be reabsorbed in changes of $\mu$ and $c_{\MS}$ (it would then be equivalent to a change of scale), controlling the scheme dependence of $R(u)$ is more complicated. We will not care much, as $R(u)$ has to do with the high energy behavior, and should only affect $m_P$, the finite sum. Therefore, when we change from the $\MS$ to the lattice scheme, we will just change $c_{\MS} \rightarrow c_{\rm latt}$ and leave $R(u)$ unchanged. Strictly speaking then, the object we compute in the lattice scheme is not the pole mass, still it will have the same infrared behavior. The fact that we will obtain the same result after subtracting $m_P$ from $m_{\rm PV}$ in both cases will be a nice confirmation that high-energy cancellation has effectively taken place and what is left is low energy\footnote{To make an analogy, the situation is similar to determinations of the infrared behavior of the energy of an static source in lattice perturbation theory. In \cite{Bauer:2011ws,Bali:2013pla,Bali:2013qla} two different discretizations were used for the static quark propagators. This affected the ultraviolet, but let the infrared behavior unchanged, as it was nicely seen in those simulations. See also the discussion in \cite{Hayashi:2019mlb}.}. The value of $c_{\rm latt}$ that we use is the same to the one used in \cite{HyperI}. To determine it we take the $n_f=0$ number for a Wilson action of $d_1=5.88359$ \cite{Hasenfratz:1980kn} and use $c_{\rm latt}=-2(\frac{5}{6}+\frac{2\pi d_1}{\beta_0})$. This is enough for our purposes, as we only use this scheme for checking the consistency between the results obtained with different schemes. Note that this yields two values of $c_{\rm latt}$ if we introduce the $n_f$ dependence of $\beta_0$: $c_{\rm latt}(n_f=0)=-8.38807$ and $c_{\rm latt}(n_f=3)=-9.88171$.

\subsection{$N$ large and $\mu \sim \m \gg \lQ$. \eq{eq:NP}. Case 1)}

\begin{center}
\begin{figure}[htb!]
\includegraphics[width=0.73\textwidth]{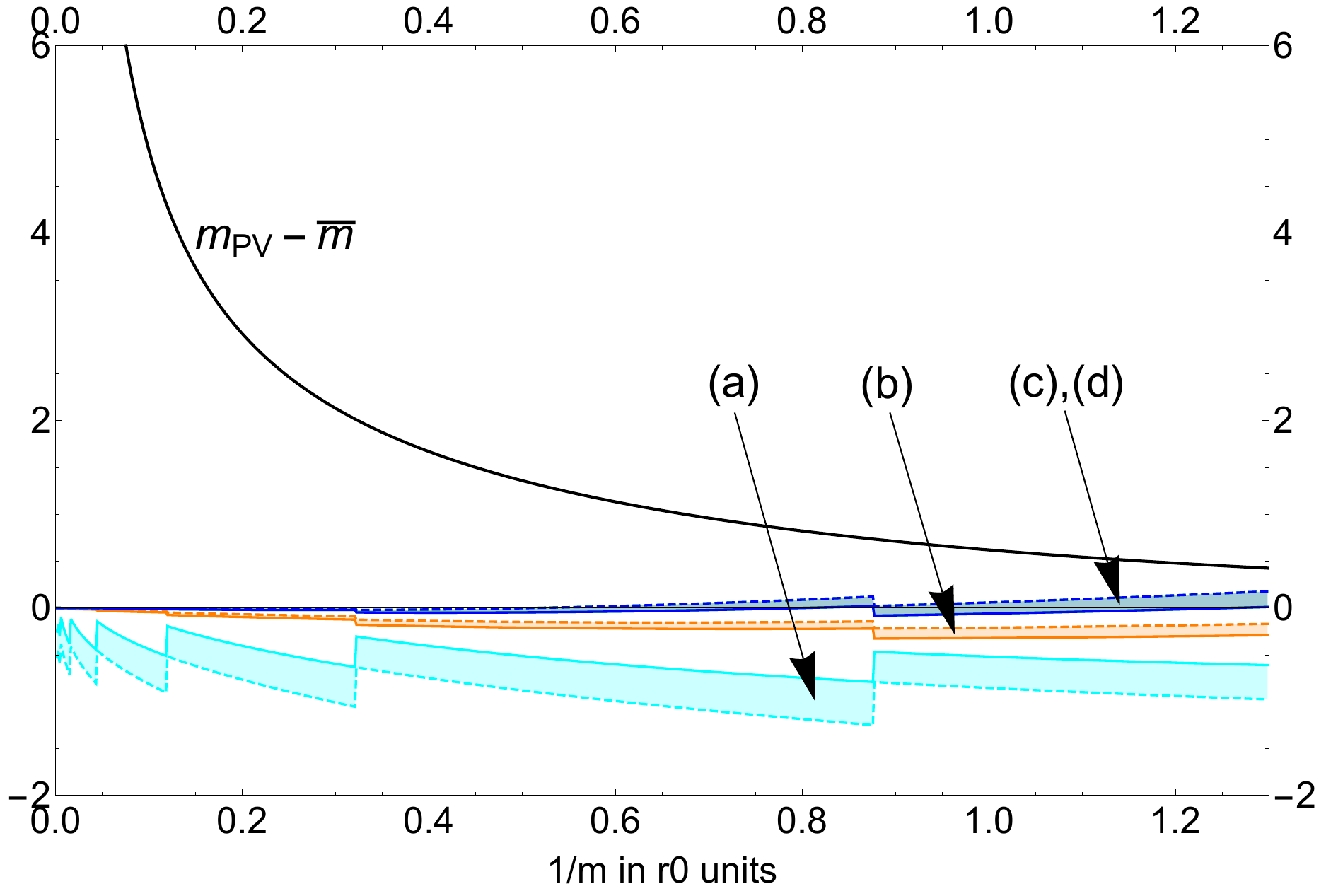}
\put(-360,100){\rotatebox{90}{\large $r_0^{-1}$}}
%\put(0,139){{\large (d)}}
%\put(0,125){{\large (c)}}
%\put(0,165){{\large (b)}}
\vspace{0.1in}
\includegraphics[width=0.8\textwidth]{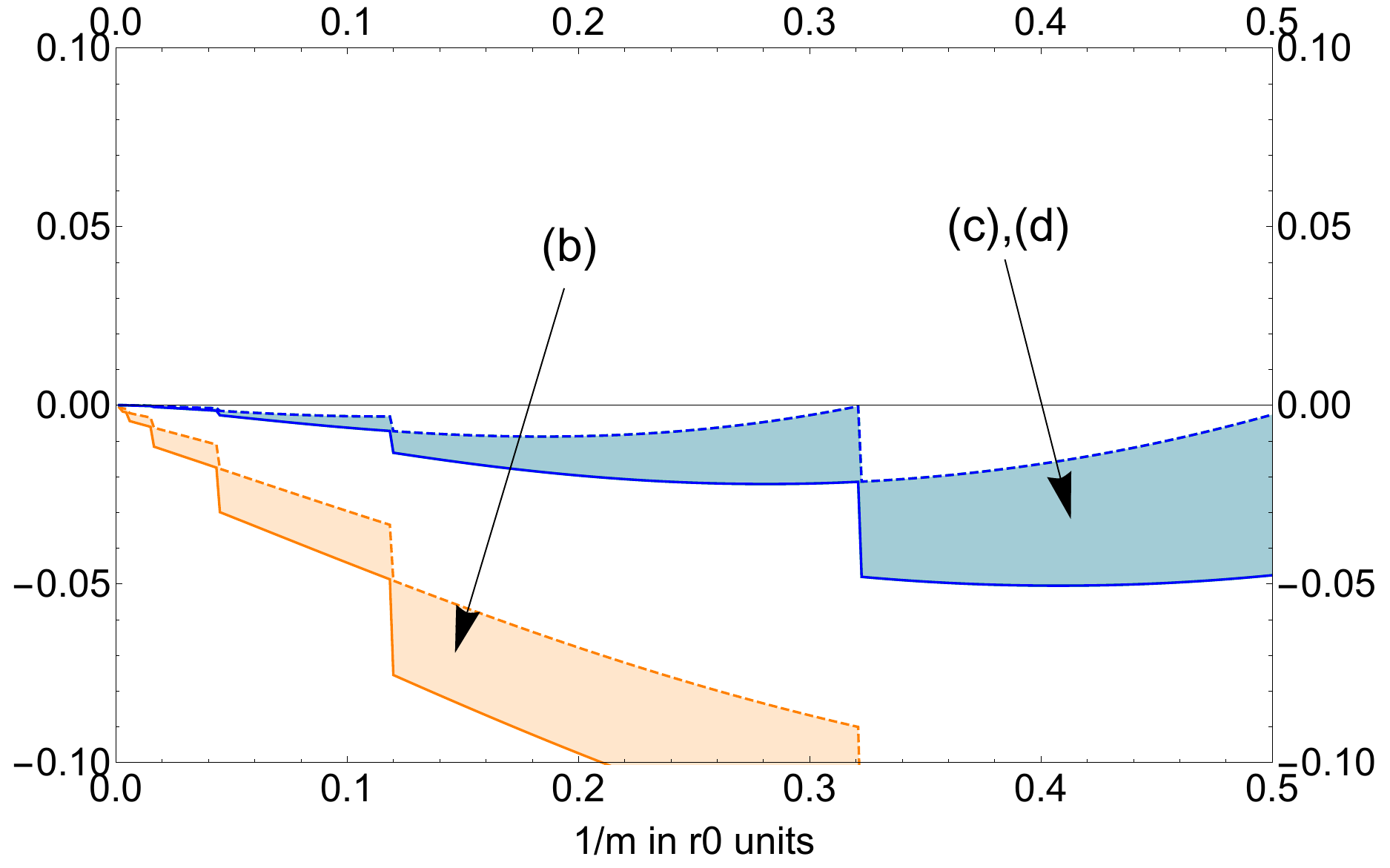}
\put(-378,100){\rotatebox{90}{\large $r_0^{-1}$}}
%\includegraphics[width=0.76\textwidth]{PlottingSmallestPositivecAndNegativeHyperZomed.png}
%\put(-140,189){{\large (b)}}
%\put(0,105){{\large (d)}}
%\put(0,75){{\large (c)}}
\caption{{\bf Upper panel}: We plot $m_{\rm PV}-\m$ (black line) and the differences: (a) $m_{\rm PV}-m_P$ (cyan), (b) $m_{\rm PV}-m_P-\m\Omega_m$ (orange),
 (c) $m_{\rm PV}-m_P-\m\Omega_m-\sum_{n=N_P+1}^{2N_P} (r_n-r_n^{(\rm as)}) \al^{n+1}$ (green),
and  (d) $m_{\rm PV}-m_P-\m\Omega_m-\sum_{n=N_P+1}^{2N_P} (r_n-r_n^{(\rm as)}) \al^{n+1}-\m\Omega_{-2}$ (blue)
in the large $\beta_0$ approximation using the lattice scheme with $n_f=0$ light flavours. For each difference, 
the bands are generated by the difference of the prediction produced by the smallest positive or negative possible values of $c$ that yields integer values for $N_P$. The (c) and (d) bands are one on top of the other. {\bf Lower panel}: As in the upper panel but in a smaller range. $r_0^{-1} \approx 400$ MeV. The value of $N_P$ depends on the scale $1/m$ we use. For instance for $c$ positive, $N_P=9$ for $1/m \in [0.003,0.0045]$, $N_P=8$ for $1/m \in [0.006,0.0015]$,
$N_P=7$ for $1/m \in [0.0165,0.0435]$, $N_P=6$ for $1/m \in [0.045,0.01185]$, $N_P=5$ for $1/m \in [0.12,0.321]$,
$N_P=4$ for $1/m \in [0.3225,0.876]$ and $N_P=3$ for $1/m \in [0.8775,1.299]$.
\label{Fig:MPVb0nf0latt}}
\end{figure}
\end{center}

\begin{center}
\begin{figure}
\includegraphics[width=0.73\textwidth]{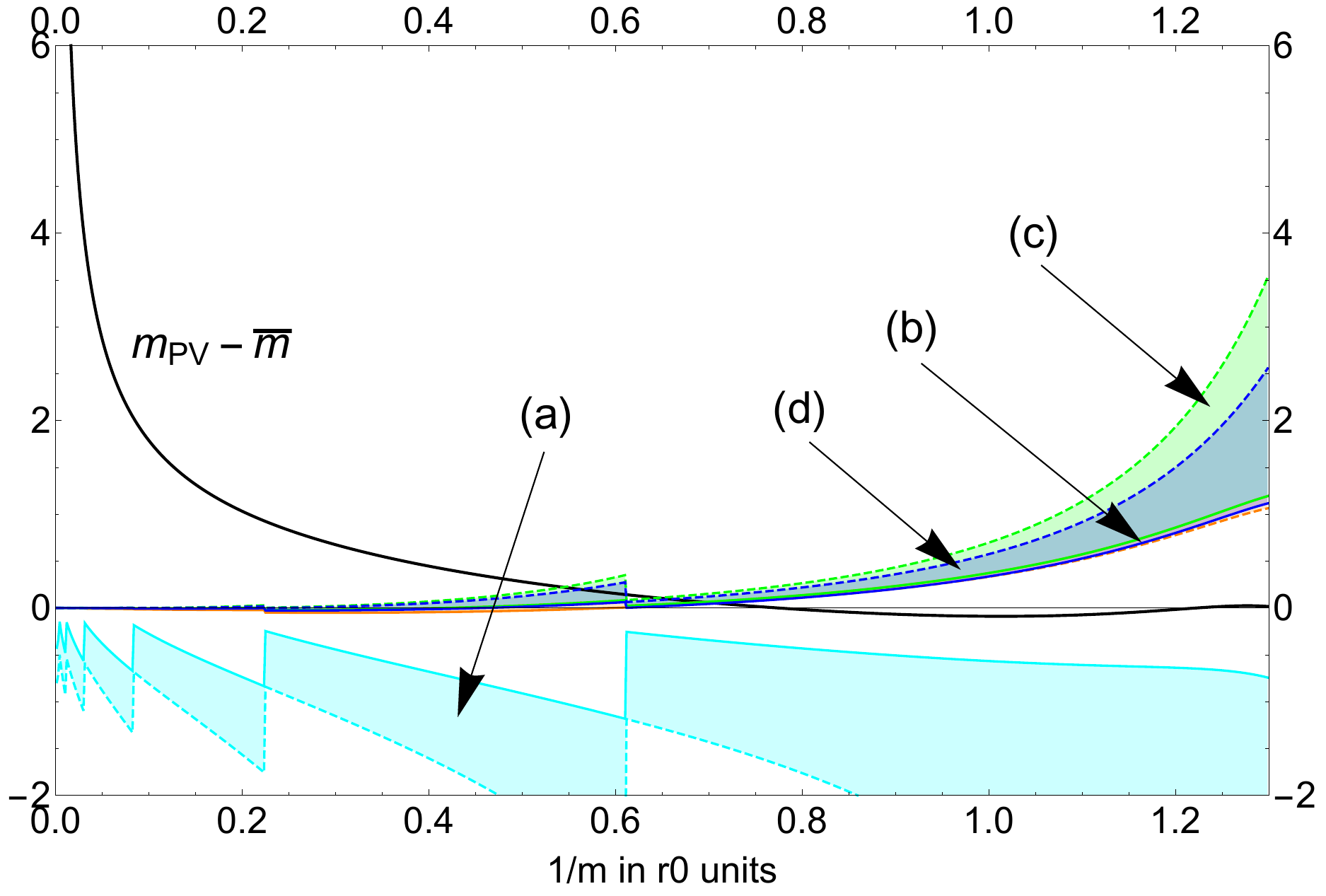}
\put(-360,100){\rotatebox{90}{\large $r_0^{-1}$}}
%\includegraphics[width=0.76\textwidth]{PlottingSmallestPositivecAndNegativeHyperMS.png}
%\put(0,220){{\large (a)}}
%\put(0,129){{\large (d)}}
%\put(0,105){{\large (c)}}
%\put(0,65){{\large (b)}}
\vspace{0.1in}
\includegraphics[width=0.8\textwidth]{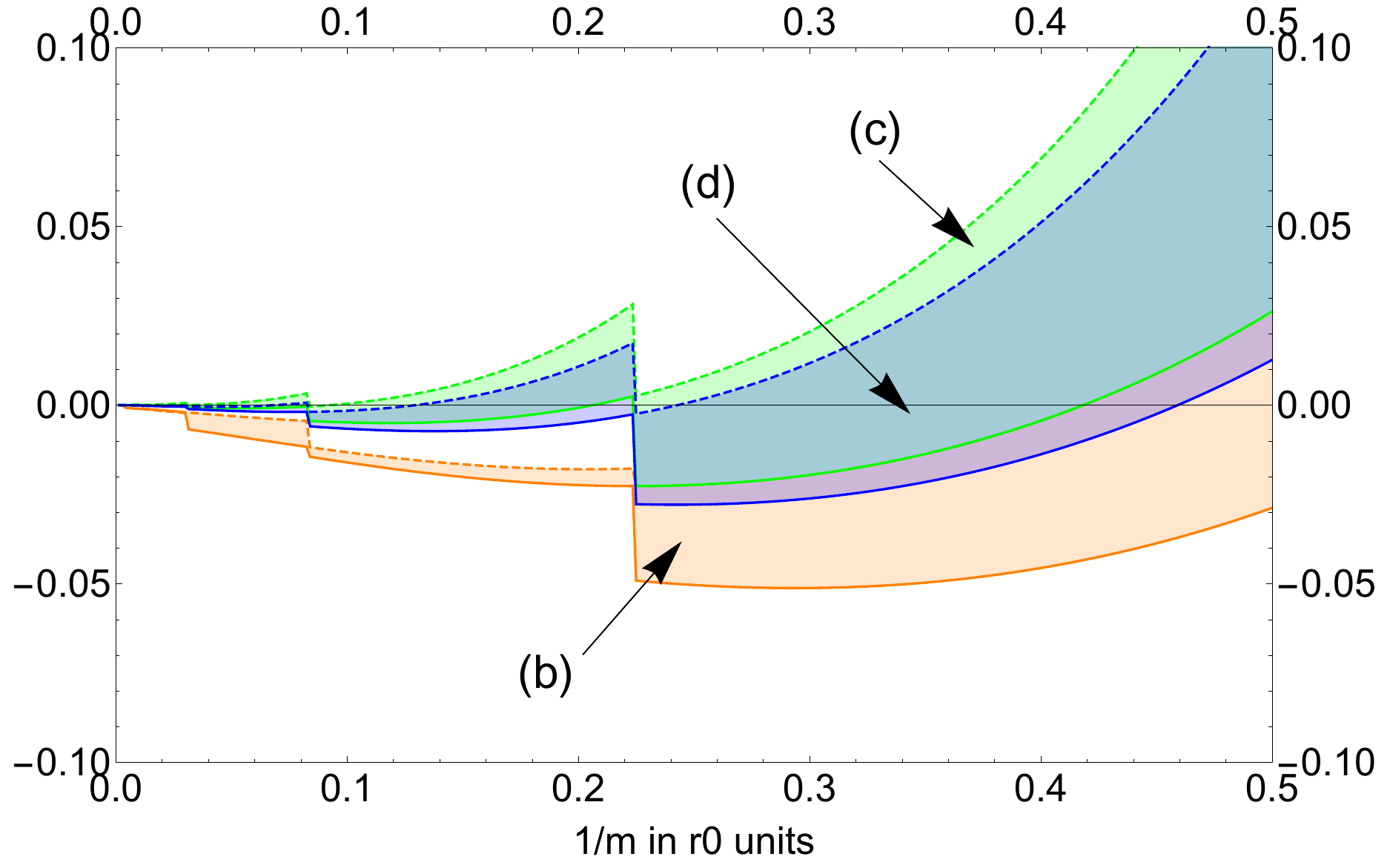}
\put(-378,100){\rotatebox{90}{\large $r_0^{-1}$}}
%\includegraphics[width=0.76\textwidth]{PlottingSmallestPositivecAndNegativeHyperZoomed.png}
%\put(0,145){{\large (b)}}
%\put(0,110){{\large (d)}}
%\put(0,55){{\large (c)}}
\caption{As in Fig. \ref{Fig:MPVb0nf0latt} but in the $\MS$ scheme. The values of $N_P$ for $c$ positive are, for instance, 
$N_P=6$ for $1/m =0.003$, $N_P=5$ for $1/m \in [0.0045,0.0105]$, $N_P=4$ for $1/m \in [0.012,0.03]$, $N_P=3$ for $1/m \in [0.0315,0.0825]$, $N_P=2$ for $1/m \in [0.084,0.2235]$, $N_P=1$ for $1/m \in [0.225,0.6105]$ and $N_P=0$ for $1/m \in [0.612,1.5]$. 
\label{Fig:MPVb0nf0MS}}
\end{figure}
\end{center}

\begin{center}
\begin{figure}
\includegraphics[width=0.73\textwidth]{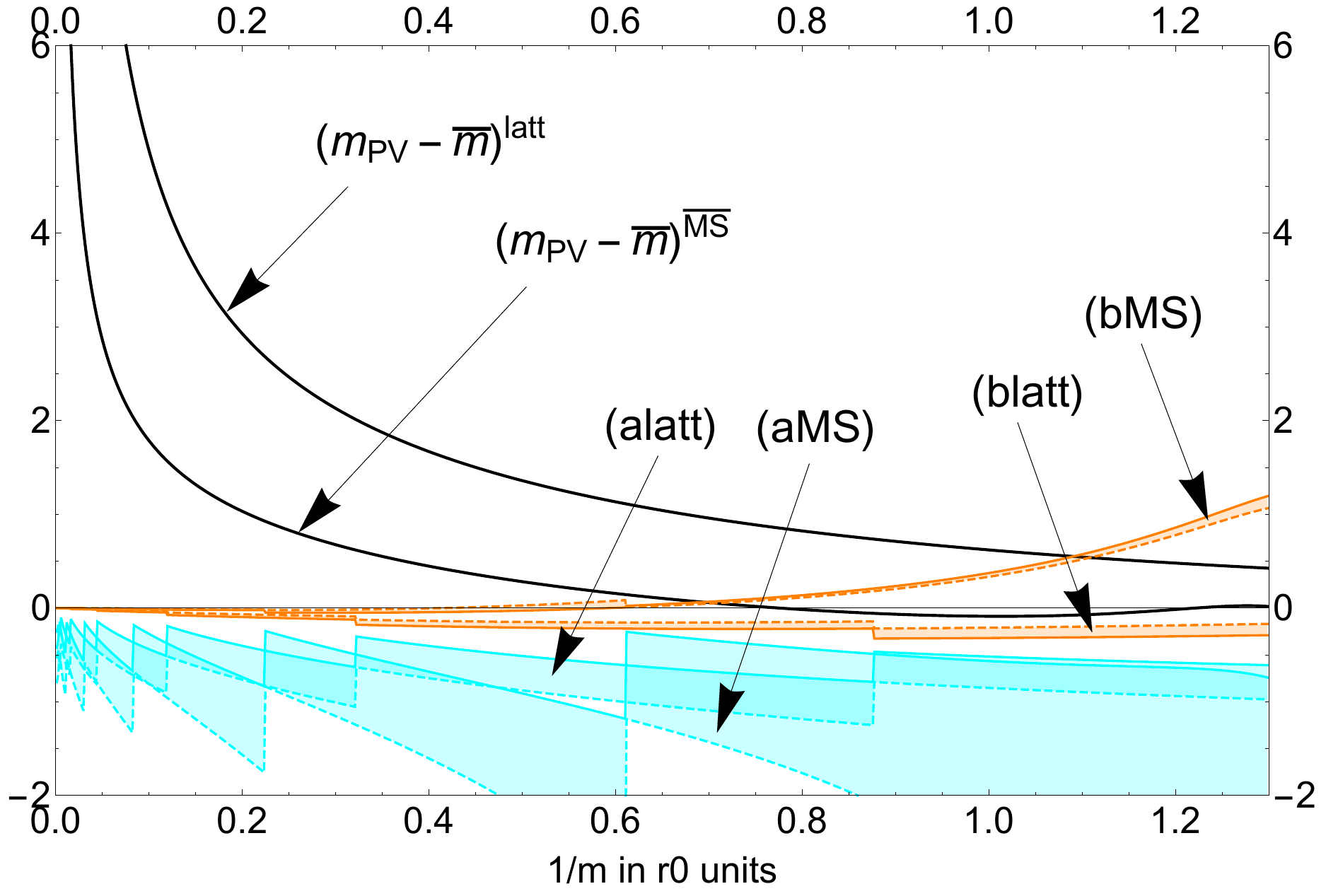}
\put(-360,100){\rotatebox{90}{\large $r_0^{-1}$}}
%\includegraphics[width=0.76\textwidth]{PlottingSmallestPositivecAndNegativeWithoutDeltaVCombined.png}
%\put(0,220){{\large (aMS)}}
%\put(0,190){{\large (alatt)}}
%\put(0,159){{\large (blatt)}}
%\put(0,105){{\large (bMS)}}
\vspace{0.1in}
\includegraphics[width=0.8\textwidth]{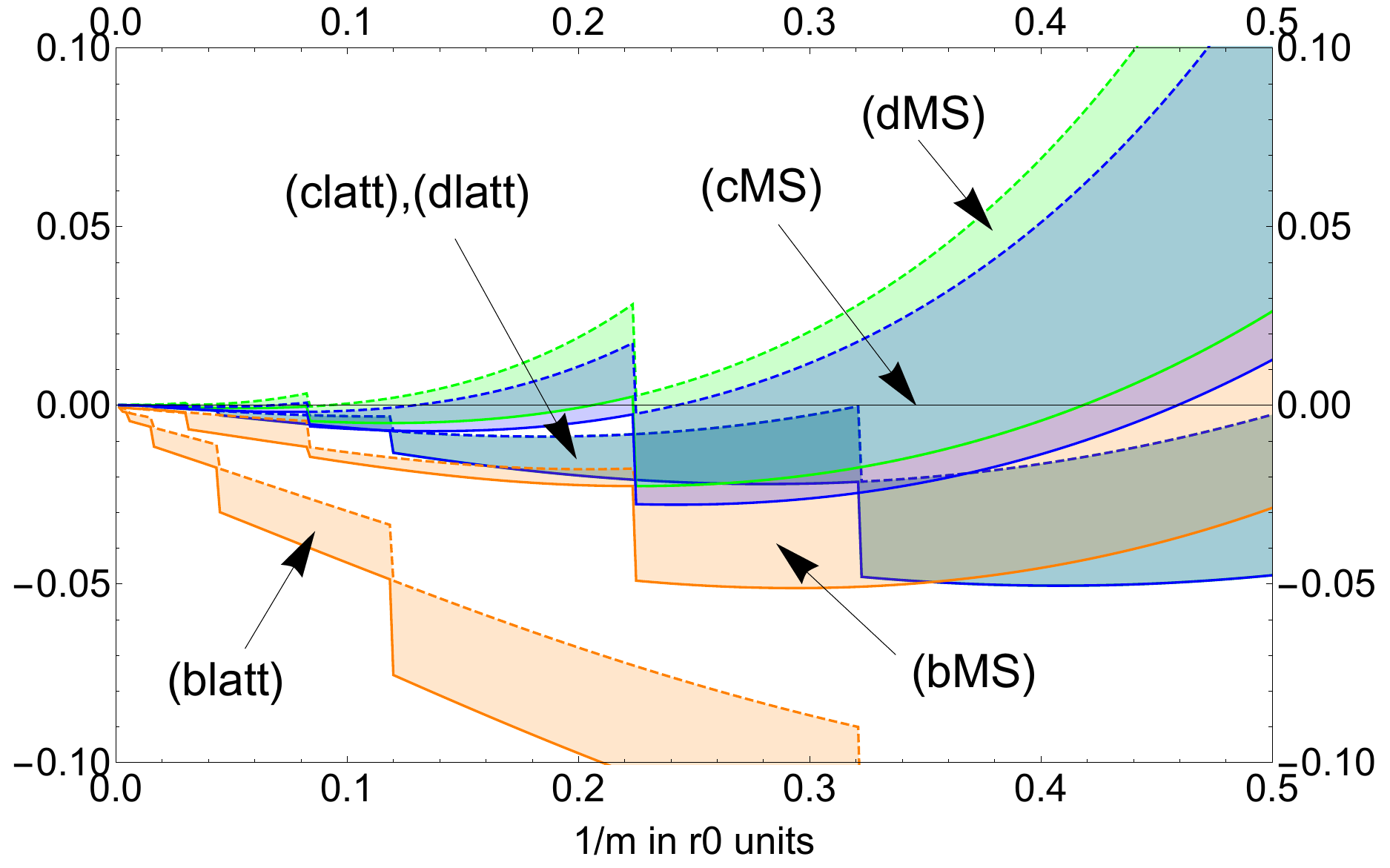}
\put(-378,100){\rotatebox{90}{\large $r_0^{-1}$}}
%\includegraphics[width=0.76\textwidth]{PlottingSmallestPositivecAndNegativeHyperZoomedCombined.png}
%\put(-160,200){{\large (blatt)}}
%\put(0,145){{\large (bMS)}}
%\put(0,70){{\large (clatt)}}
%\put(0,25){{\large (cMS)}}
%\put(0,103){{\large (dlatt)}}
%\put(0,112){{\large (dMS)}}
\caption{\label{Fig:MPVb0nf0lattvsMS} Comparison of lattice and $\MS$ scheme results for $n_f=0$ obtained in Figs. \ref{Fig:MPVb0nf0latt} and \ref{Fig:MPVb0nf0MS}. {\bf Upper panel}: We plot $m_{\rm PV}-\m$ and the differences: (a) $m_{\rm PV}-m_P$, and (b) $m_{\rm PV}-m_P-\m\Omega_m$ 
in the lattice and $\MS$ scheme with $n_f=0$ light flavours. {\bf Lower panel}: Lower panel Figs. \ref{Fig:MPVb0nf0latt} and \ref{Fig:MPVb0nf0MS} combined.}
\end{figure}
\end{center}

We now take \eq{mPV} in the large $\beta_0$ approximation and truncate it at different orders in the hyperasymptotic expansion. We then compare such truncations with the exact solution. 
We can study (even if in the large $\beta_0$ approximation) up to which values of $\m$ the OPE is a good approximation of $m_{\rm PV}$. Remarkably enough we can actually check more than one term of the OPE (hyperasymptotic) expansion. Note that in the large $\beta_0$ approximation $\Omega_2=0$, but not $\Omega_{-2}$, which in the large $\beta_0$ approximation reads ($\eta_c^{(\beta_0)}=\frac{2\pi c}{\beta_0}-1$)
\be
\label{Omegaminustwo}
\Omega_{-2}=\sqrt{\alpha(\mu)}K_X^{(P)}
\frac{\Lambda_X^{2} \m^{2}}{\mu^{4}}
\bigg\{1+\bar{K}_{X,1}^{(P)}\alpha(\mu)+\bar{K}_{X,2}^{(P)}\alpha^2(\mu)+\mathcal{O}\left(\alpha^3(\mu)\right)\bigg\}
\,,
\ee
\be
K_X^{(P)}\equiv Z^X_{-2}(-1)^{N_p+1}\left(\frac{\beta_0}{2\pi^2}\right)^{-1/2}\,,
\qquad
Z_{-2}^X=-\frac{C_F e^{c_X}}{\pi}
\,,
\ee
\bea
\bar{K}_{X,1}^{(P)}&\equiv&
\frac{\beta_0}{24\pi}(-1+3\eta_c^{(\beta_0)2})
\,,
\\
\bar{K}_{X,2}^{(P)}&\equiv&
\frac{\beta_0^2}{4608\pi^{2}}\bigg[13-48\eta_c^{(\beta_0)}
-60\eta_c^{(\beta_0)2}+48\eta_c^{(\beta_0)3}+36\eta_c^{(\beta_0)4}\bigg]
\,.
\eea
We also explore the scheme dependence by performing the computation in the lattice and the $\MS$ scheme. We will do these analyses for the cases with $n_f=0$ and $n_f=3$. The first in view of comparing with quenched lattice simulations, the second to simulate a more physical scenario, for which we can draw some conclusions that could be applied beyond the large-$\beta_0$ limit. In Figs. \ref{Fig:MPVb0nf0latt}, \ref{Fig:MPVb0nf0MS}, and \ref{Fig:MPVb0nf0lattvsMS} we plot $m_{\rm PV}-\m$, $m_{\rm PV}-m_P$, 
$m_{\rm PV}-m_P-\m\Omega_m$, 
$m_{\rm PV}-m_P-\m\Omega_m-\sum_{n=N_P+1}^{2N_P} (r_n-r_n^{(\rm as)}) \al^{n+1}$,
and  $m_{\rm PV}-m_P-\m\Omega_m-\sum_{n=N_P+1}^{2N_P} (r_n-r_n^{(\rm as)}) \al^{n+1}-\m \Omega_{-2}$ with $n_f=0$ light flavours. In the counting of \eq{SPVDN} this corresponds to (0,0), (0,$N_P$), (1,0), (1,$N_P$), (2,0) precision. We do such computation in the lattice (Fig. \ref{Fig:MPVb0nf0latt}) and the $\MS$ (Fig. \ref{Fig:MPVb0nf0MS}) scheme. In Fig. \ref{Fig:MPVb0nf0lattvsMS} we compare the results in the lattice and $\MS$ scheme. 
We observe a very nice convergent pattern in all cases down to surprisingly small scales. To visualize the dependence on $c$, we show the band generated by the smallest positive and negative possible values of $c$ that yield integer values for $N_P$. The size of the band generated by the different values of $c$ (the $c$ dependence) decreases after introducing  $\Omega_m$ to its associated sum. On the other hand $\Omega_{-2}$ (an ultraviolet renormalon) gives a very small contribution, in particular in the lattice scheme. This is consistent with interpreting the lattice scheme as the $\MS$ scheme using a much higher renormalization scale $\mu$ for the scale of the strong coupling.


Let us discuss the results in more detail. We first observe that the $\m$ dependence of $m_{\rm PV}$ is basically eliminated in $m_{\rm PV}-m_P$, as expected. This happens both in the lattice and $\MS$ scheme. The latter shows a stronger $c$ dependence. This is to be expected, as in the $\MS$, we truncate at smaller orders in $N$. This makes the truncation error bigger.  As we can see in the upper panel of Fig. \ref{Fig:MPVb0nf0lattvsMS}, both schemes yield consistent predictions for $m_{\rm PV}-m_P$. We can draw some interesting observations out of this analysis. For $m_{\rm PV}-m_P$ is better to choose a larger factorization scale, if we have enough coefficients of the perturbative expansion. This is particularly so at large distances: We can still get good results at very large distances in the lattice scheme. 

We now turn to $m_{\rm PV}-m_P-\m\Omega_m$. Adding the new correction brings much better agreement with expectations (which we remind is to get zero).   
After the introduction of $\m\Omega_m$, the $\MS$ scheme yields more accurate results than the lattice scheme. This can already be seen in the upper panel of Fig. \ref{Fig:MPVb0nf0lattvsMS}, and in greater detail in the lower panel of Fig. \ref{Fig:MPVb0nf0lattvsMS}. 

$m_{\rm PV}-m_P-\m\Omega_m$ shows some dependence on $\m$, which is more pronounced in the lattice than in the $\MS$ scheme. As in the large $\beta_0$ approximation the difference between both schemes is somewhat equivalent to a change of scale, these results point to that $\mu=\m$ in $\MS$ scheme is close to the natural scale and minimize higher order corrections. Note that the lattice scheme computation is equivalent to the $\MS$ scheme choosing $\mu_{latt}= \mu_{\MS}e^{\frac{-c_{latt}}{2}}e^{\frac{c_{\MS}}{2}}$. This gives around a factor 30 (!). 
Once $\sum_{n=N_P+1}^{2N_P} (r_n-r_n^{(\rm as)}) \al^{n+1}$ is incorporated in the prediction most of the difference between schemes disappears. The effect of introducing $\Omega_{-2}$ is very small, in particular in the lattice scheme. This is to be expected, since the lattice scheme corresponds to a larger renormalization scale $\mu$. In any case, the difference between schemes gets smaller and smaller as we go to higher orders in the hyperasymptotic expansion, in particular at short distances.
We also want to stress that this analysis opens the window to apply perturbation theory at rather large distances. Note that in the upper panel plots in Figs. \ref{Fig:MPVb0nf0latt}, \ref{Fig:MPVb0nf0MS}, and \ref{Fig:MPVb0nf0lattvsMS}, we have gone to very large distances. 

As some concluding remarks let us emphasize the following points. $m_{\rm PV}-m_P$ is more or less constant with relatively large uncertainties. This is to be expected, as the next correction in magnitude is $\m\Omega_m$ which is approximately constant (mildly modulated by $\sqrt{\al(\mu)}$). After introducing this term the error is much smaller and we can see more structure. In particular we are sensitive to $\sum_{n=N_P+1}^{2N_P} (r_n-r_n^{(\rm as)}) \al^{n+1}$. Here we find (at the level of precision we have now) a sizable difference between lattice and $\MS$. This can be expected: $\sum_{n=N_P+1}^{2N_P} (r_n-r_n^{(\rm as)}) \al^{n+1}$ is the object we expect to be more sensitive to the scale.  

Another interesting observation is that truncated sums behave better in the lattice scheme than in the $\MS$ scheme. Nevertheless, this could be misleading. The sums are truncated at the minimal term. Therefore, one needs more terms in the lattice scheme. If the number of terms is not an issue (which could be the case with dedicated numerical stochastic perturbation theory (NSPT) \cite{DiRenzo:1994sy,DiRenzo:2004hhl} computations in the lattice scheme) then the lattice scheme looks better. But as soon as $\Omega_m$ is introduced in the computation $\MS$ behaves better (at least in the large $\beta_0$ approximation).

Overall, we observe a very nice convergence pattern up to (surprisingly) rather large scales in the lattice and $\MS$ scheme. The agreement with the theoretical prediction (which is zero) is perfect at short distances. The estimated error is also expected to be small. It will be interesting to see if this also happens beyond the large $\beta_0$. 

\begin{center}
\begin{figure}
\includegraphics[width=0.8\textwidth]{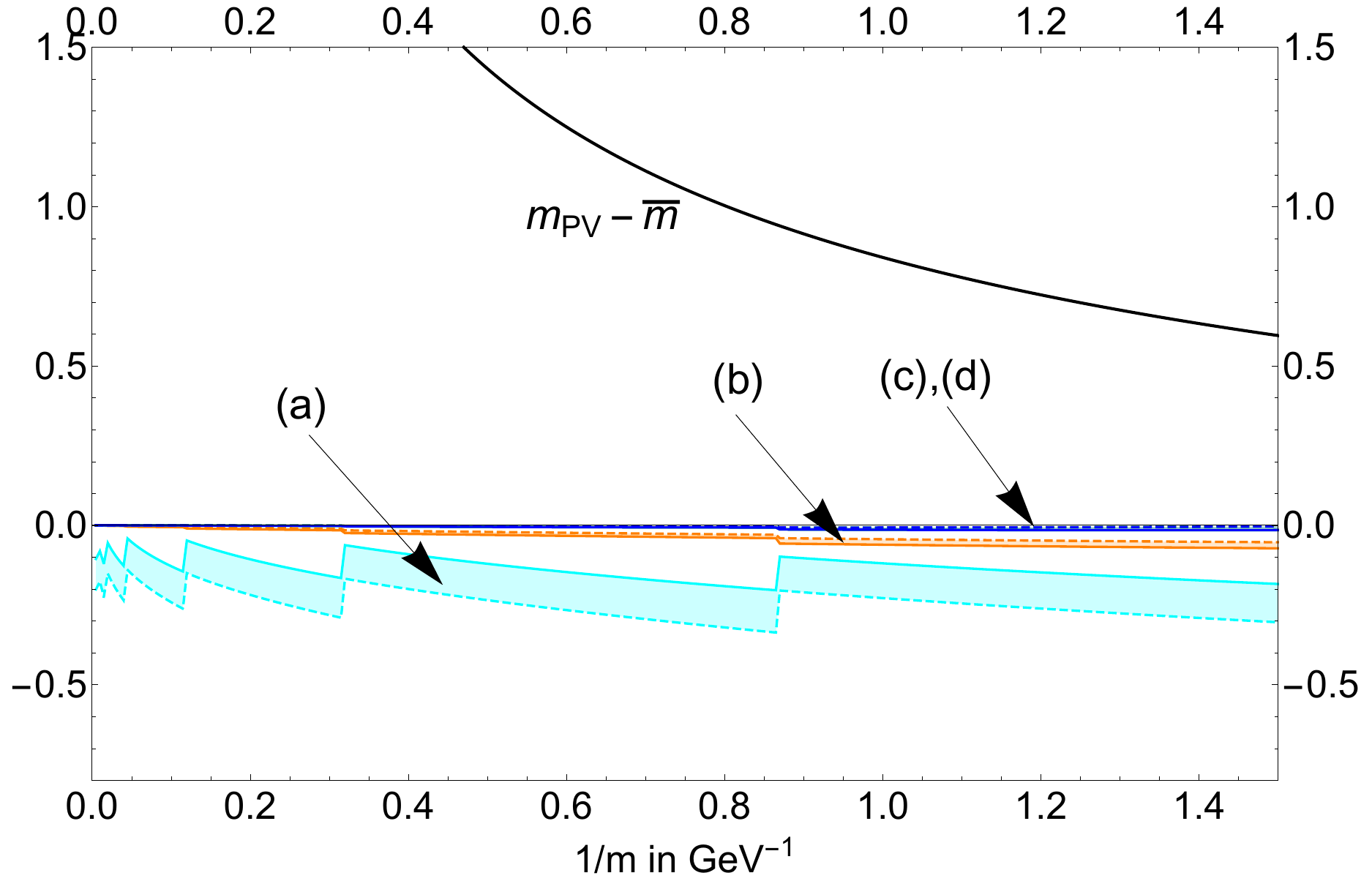}
\put(-390,100){\rotatebox{90}{\large GeV}}
%\includegraphics[width=0.75\textwidth]{PlottingSmallestPositivecAndNegativeHypernf3.png}
%\put(0,179){{\large (a)}}
%\put(0,155){{\large (c,d)}}
%\put(0,165){{\large (b)}}
\vspace{0.1in}
\includegraphics[width=0.825\textwidth]{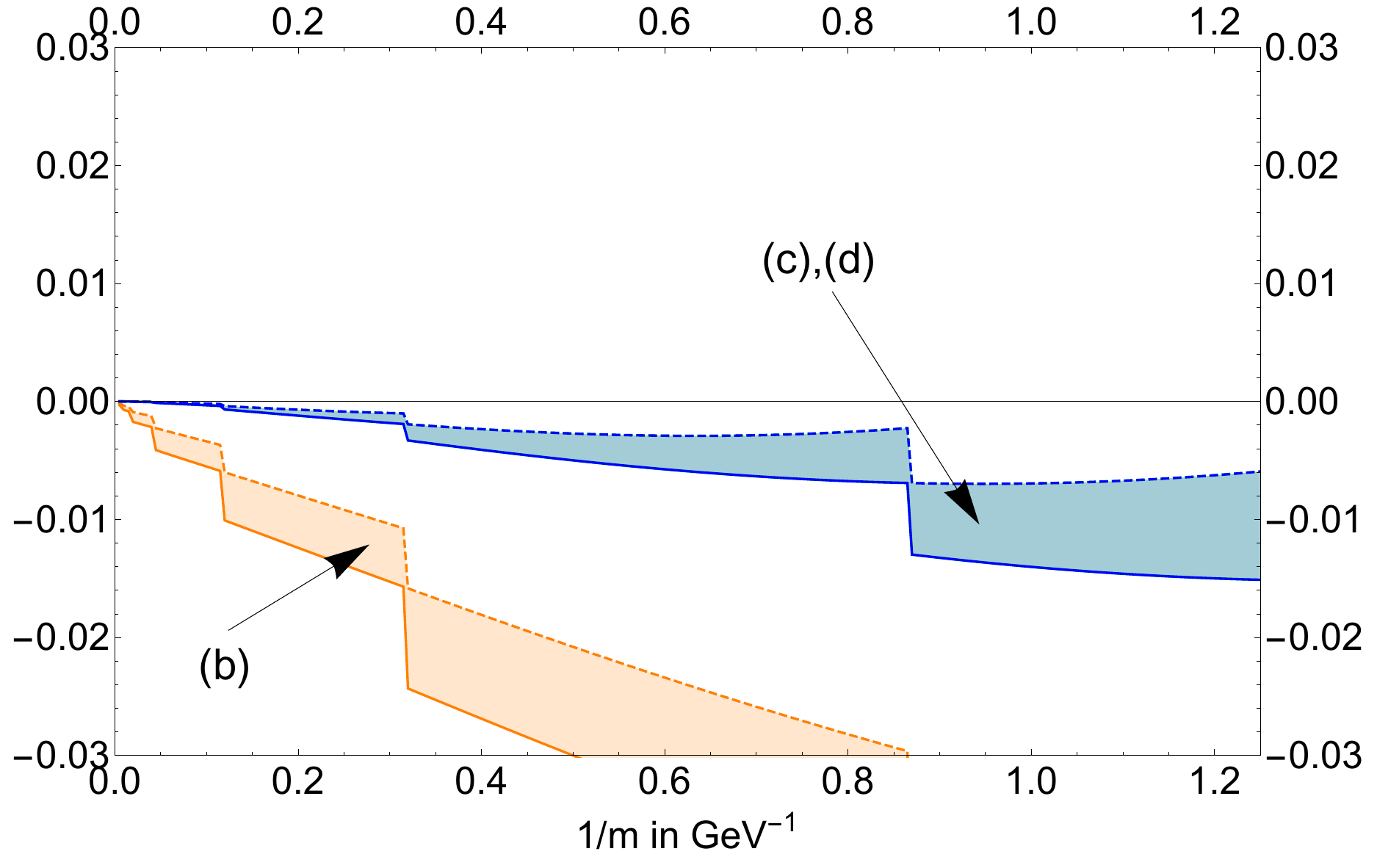}
\put(-405,100){\rotatebox{90}{\large GeV}}
%\includegraphics[width=0.75\textwidth]{PlottingSmallestPositivecAndNegativeHyperZoomednf3.png}
%\put(-220,189){{\large (b)}}
%\put(0,105){{\large (d)}}
%\put(0,75){{\large (c)}}
\caption{As in Fig. \ref{Fig:MPVb0nf0latt} but with $n_f=3$ light flavours.  The value of $N_P$ depends on the scale $1/m$ we use. For instance for $c$ positive, $N_P=1$ for $1/m = 0.005$, $N_P=10$ for $1/m \in [0.01,0.015]$, 
$N_P=9$ for $1/m \in [0.02,0.04]$, $N_P=8$ for $1/m \in [0.045,0.115]$,
$N_P=7$ for $1/m \in [0.12,0.315]$, $N_P=6$ for $1/m \in [0.32,0.865]$, and $N_P=5$ for $1/m \in [0.87,1.5]$.
\label{Fig:MPVb0nf3latt}}
\end{figure}
\end{center}

\begin{center}
\begin{figure}
\includegraphics[width=0.8\textwidth]{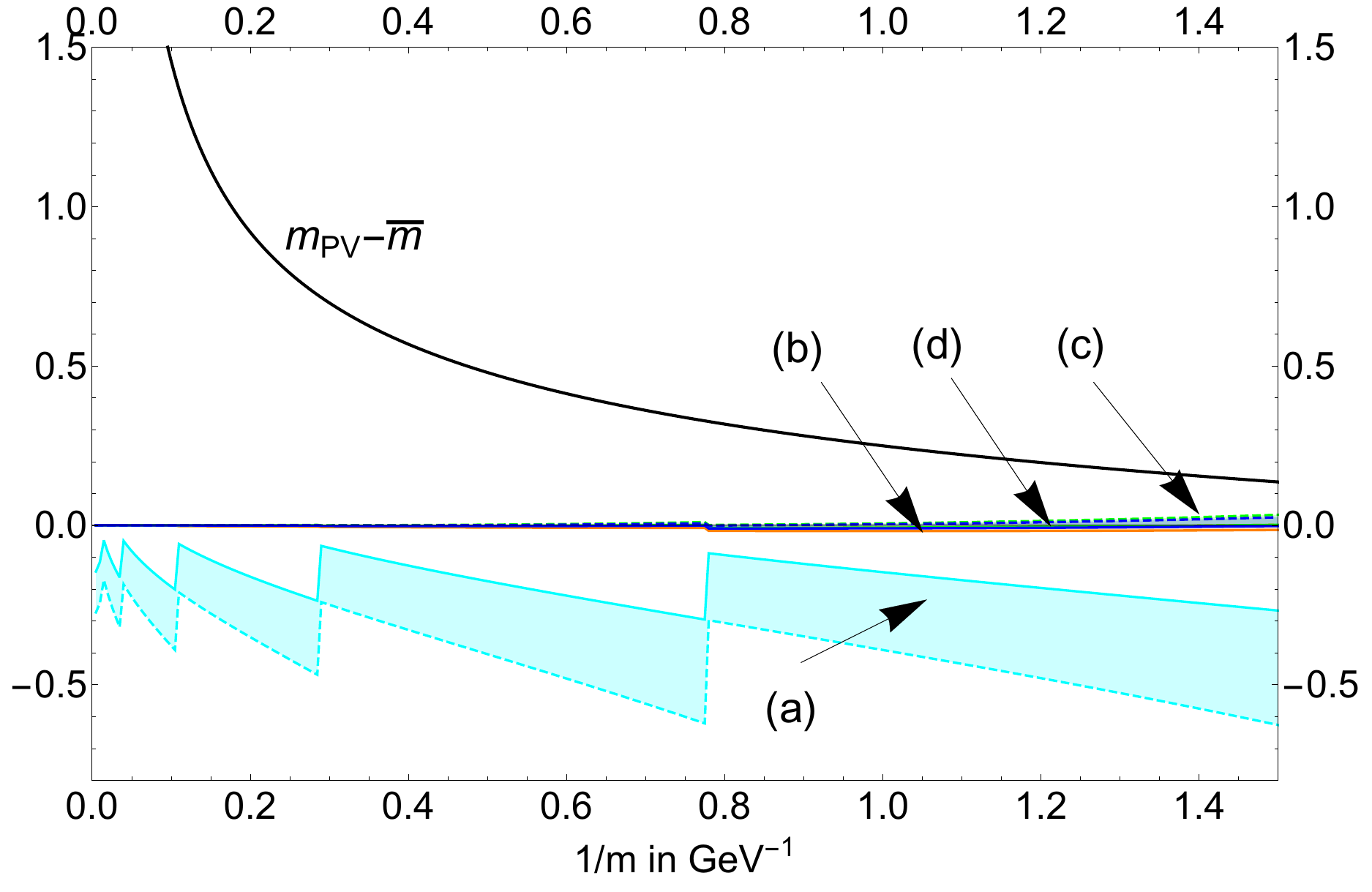}
\put(-390,100){\rotatebox{90}{\large GeV}}
%\includegraphics[width=0.75\textwidth]{PlottingSmallestPositivecAndNegativeHyperMSnf3.png}
%\put(0,199){{\large (a)}}
%\put(0,155){{\large (b,c,d)}}
\vspace{0.1in}
\includegraphics[width=0.83\textwidth]{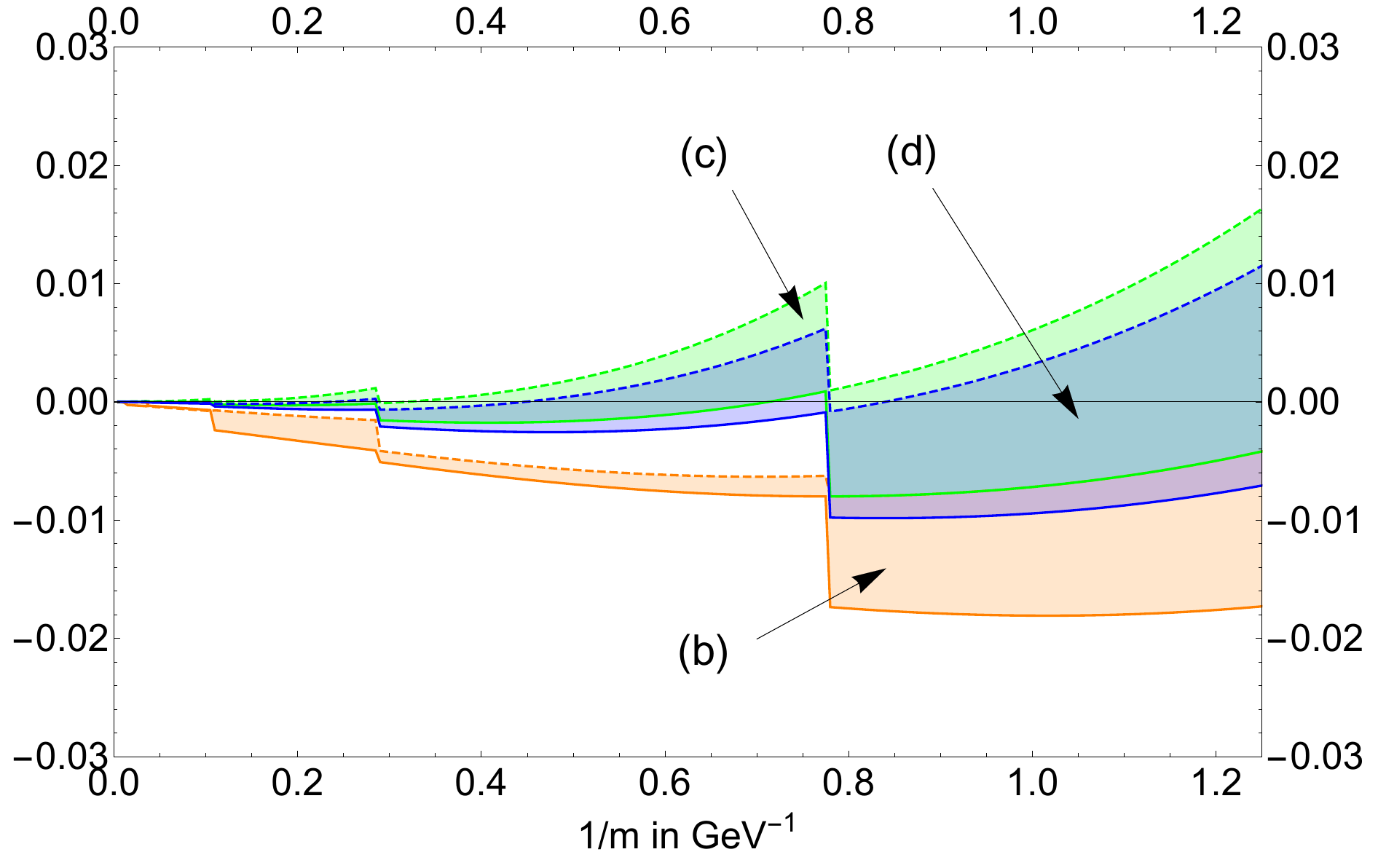}
\put(-405,100){\rotatebox{90}{\large GeV}}
%\includegraphics[width=0.75\textwidth]{PlottingSmallestPositivecAndNegativeHyperZoomedMSnf3.png}
%\put(0,199){{\large (b)}}
%\put(0,105){{\large (d)}}
%\put(0,75){{\large (c)}}
\caption{As in Fig. \ref{Fig:MPVb0nf0latt} but with $n_f=3$ light flavours and in the $\MS$ scheme. The value of $N_P$ depends on the scale $1/m$ we use. For instance for $c$ positive, 
$N_P=7$ for $1/m =0.005$, $N_P=6$ for $1/m=0.01$, $N_P=5$ for $1/m \in [0.015,0.035]$,
$N_P=4$ for $1/m \in [0.04,0.105]$, $N_P=3$ for $1/m \in [0.11,0.285]$, $N_P=2$ for $1/m \in [0.29,0.775]$, and  $N_P=1$ for $1/m \in [0.78,1.5]$.
 \label{Fig:MPVb0nf3MS}}
\end{figure}
\end{center}

\begin{center}
\begin{figure}
\includegraphics[width=0.81\textwidth]{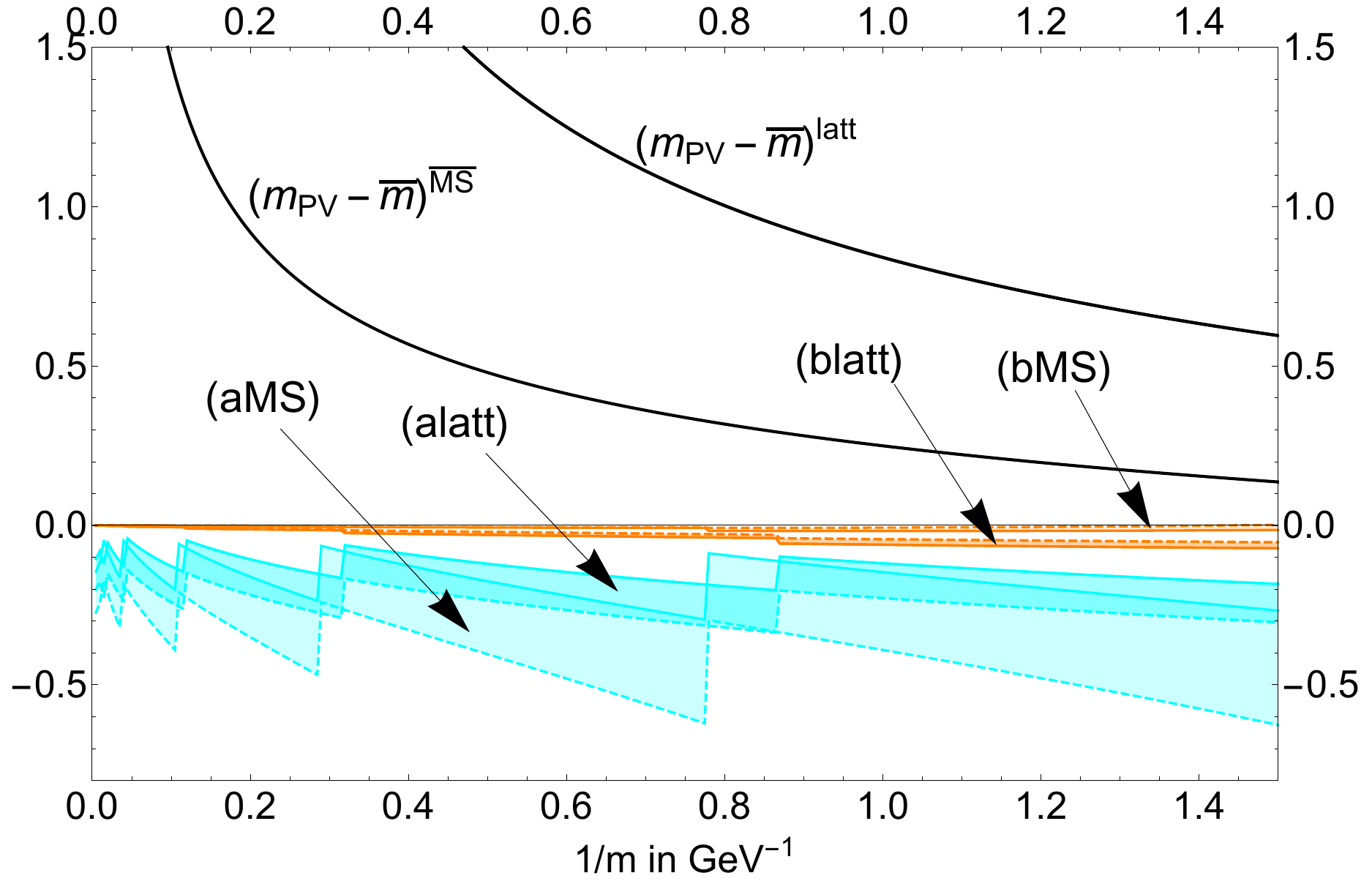}
\put(-390,100){\rotatebox{90}{\large GeV}}
%\includegraphics[width=0.75\textwidth]{PlottingSmallestPositivecAndNegativeExcludingDeltaVCombined.png}
%\put(0,205){{\large (aMS)}}
%\put(0,182){{\large (alatt)}}
%\put(0,163){{\large (blatt)}}
%\put(0,153){{\large (bMS)}}
\vspace{0.1in}
\includegraphics[width=0.84\textwidth]{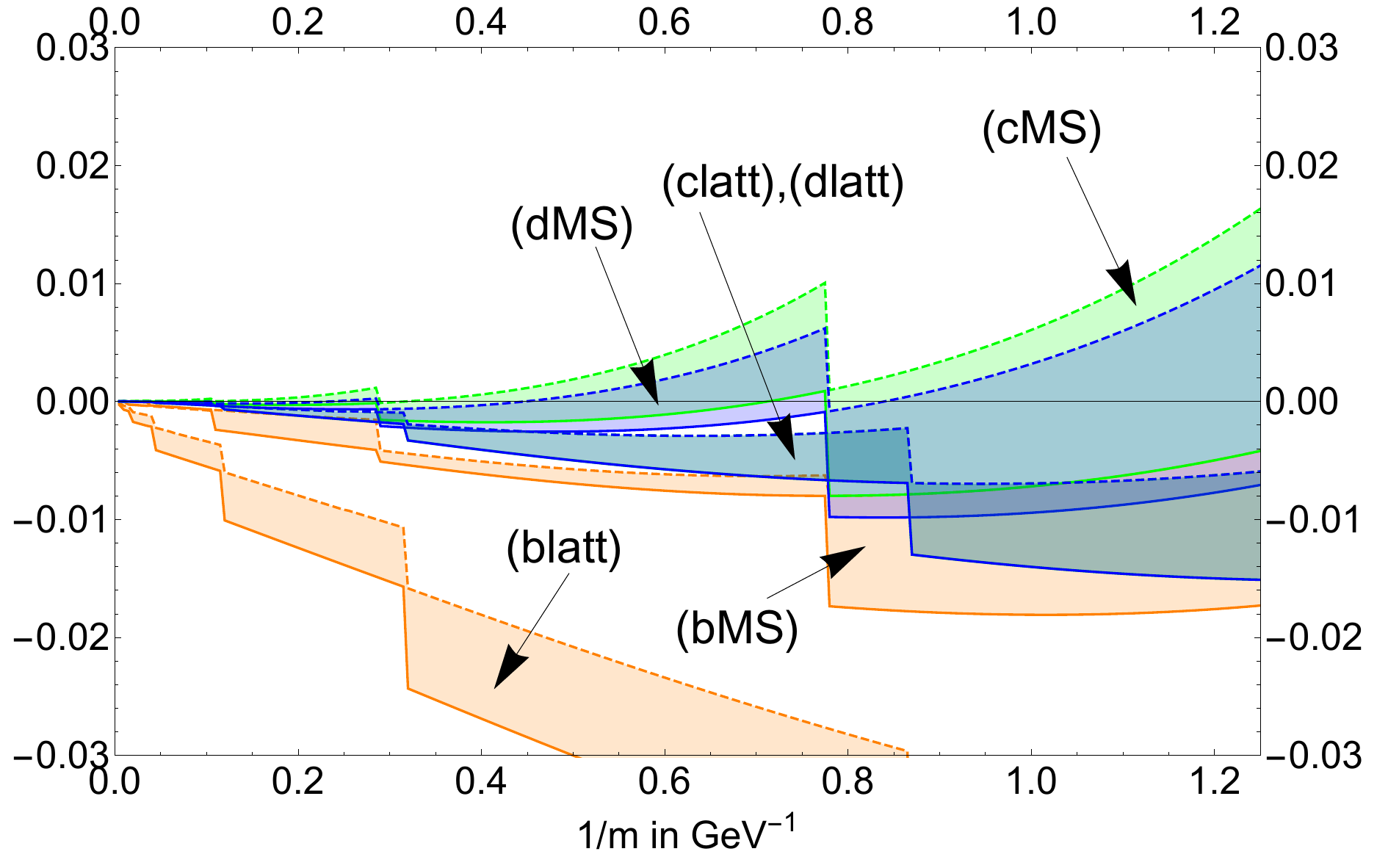}
\put(-405,100){\rotatebox{90}{\large GeV}}
%\includegraphics[width=0.75\textwidth]{PlottingSmallestPositivecAndNegativeHyperZoomedCombinednf3.png}
%\put(-230,200){{\large (blatt)}}
%\put(0,195){{\large (bMS)}}
%\put(0,85){{\large (clatt)}}
%\put(0,45){{\large (cMS)}}
%\put(0,103){{\large (dlatt)}}
%\put(0,112){{\large (dMS)}}
\caption{\label{Fig:MPVb0nf3lattvsMS} 
Comparison of lattice and $\MS$ scheme results for $n_f=3$. {\bf Upper panel}: We plot $m_{\rm PV}$ and the differences: (a) $m_{\rm PV}-m_P$, and (b) $m_{\rm PV}-m_P-\m\Omega_m$ 
in the lattice and $\MS$ scheme with $n_f=3$ light flavours. {\bf Lower panel}: Fig. \ref{Fig:MPVb0nf3latt} and Fig. \ref{Fig:MPVb0nf3MS} combined.}
\end{figure}
\end{center}

We now turn to the $n_f=3$ case. To easy the comparison with \cite{HyperI}, we use the same value: $\Lambda_{\MS}(n_f=3)=174$ MeV (which yields $\al(M_{\tau})\approx 0.3$). The general conclusions do not change if we fix $\Lambda_{\MS}$ (in the large $\beta_0$ approximation) using the world average value of $\al$. 
We note that $\lQ$ for the physical case ($n_f=3$) is smaller than for the $n_f=0$ case (if one sets the physical scale according to $r_0^{-1} \approx 400$ MeV). On top of that the running is less important. All this points to that the convergence should be even better than in the $n_f=0$ case (and it was quite good already there). We show our results in Figs. 
\ref{Fig:MPVb0nf3latt}, \ref{Fig:MPVb0nf3MS} and \ref{Fig:MPVb0nf3lattvsMS} (these are the analogous of Figs. \ref{Fig:MPVb0nf0latt}, \ref{Fig:MPVb0nf0MS} and \ref{Fig:MPVb0nf0lattvsMS} but with $n_f=3$). These plots confirm our expectations. Down to scales as low as 667 MeV we see no sign of breakdown of the OPE. This is so in both the lattice and the $\MS$ schemes. Note that the precision we get is extremely high as we go to small scales: Using truncation 
(c): $m_P+\m\Omega_m+\sum_{n=N_P+1}^{2N_P} (r_n-r_n^{(\rm as)}) \al^{n+1}$,
one gets $m_{\rm PV}$ in the $\MS$ scheme with a precision below 1 MeV at scales of the order of the mass of the bottom, and in the lattice scheme with a precision below 2 MeV. Using truncation (d): $m_P+\m\Omega_m+\sum_{n=N_P+1}^{2N_P} (r_n-r_n^{(\rm as)}) \al^{n+1}+\m\Omega_{-2}$, the precision does not significantly change, in particular in the lattice scheme. This reflects that ultraviolet renormalons play a minor role. The rest of the discussion follows parallel the one for $n_f=0$. 

In the above numerics, we have used the exact expression for $\Omega_m$ and $\Omega_{-2}$. In full QCD, we will not know the exact expression. 
Therefore, it makes sense to study how well the exact result is reproduced by its semiclassical expansion. We observed in \cite{HyperI} that $\Omega_m$ is very well saturated by the first terms of such expansion. Truncating the expansion produces differences much smaller than the typical precision of the different terms of the hyperasymptotic expansion. 
For $\Omega_{-2}$, we compare in Table \ref{Table:OmegaMnf0} and \ref{Table:OmegaMnf3} the exact result and the truncated semiclassical expansion for an illustrative set of values. 
We observe that the exact result is very well saturated by the first terms of the expansion computed in \eq{Omegaminustwo}. Truncating the expansion produces differences much smaller than the typical precision of the different terms of the hyperasymptotic expansion. 
As expected $n_f=3$ is better than $n_f=0$.  Note that in the large $\beta_0$ approximation we exactly have $\Lambda=\mu e^{-2\pi/(\beta_0\al(\mu))}$.

%%%%%%%%%%%%%%%%%%%%%%%%%%%%%%%%%%
%%%%%%%%%%%%%%%%%%%%%%%%%%%%%%%%%%
%%%%%%%%%%%%%%%%%%%%%%%%%%%%%%%%%%
\renewcommand{\arraystretch}{1.5}
\begin{table}[h]
\begin{tabular}{|c|c|c|c|c|c|}
\hline
\multicolumn{6}{|c|}{$\overline{\text{MS}}$-Scheme ($n_f=0$)} \\
\hline
 $\m$ in $r_0^{-1}$ & $c$ & $\m\Omega$Exact & $\left| \frac{\Omega\text{LO}}{\Omega\text{Exact}}-1\right|\times 10^2$ & $\left| \frac{\Omega \text{NLO}}{\Omega\text{Exact}}-1\right|\times 10^3$ & $\left| \frac{\Omega \text{NNLO}}{\Omega\text{Exact}}-1\right|\times 10^4$ \\ \hline
 0.6667 & 0.1786 & 0.2089 & 33.8725 & 147.64 & 3372.63 \\
 0.8333 & 0.5693 & 0.0572 & 8.1940 & 93.6387 & 922.993 \\
 1 & 0.8885 & 0.0362 & 14.1752 & 45.6019 & 16.0275 \\
 1.25 & 1.2791 & 0.0260 & 5.7282 & 13.6703 & 130.93 \\
 1.6667 & 0.0321 & 0.0199 & 12.0969 & 7.3723 & 12.2818 \\
 2.5 & 0.7419 & 0.0094 & 4.9357 & 7.7804 & 6.5465 \\
 5 & 0.2047 & 0.0042 & 2.9590 & 0.5254 & 4.8485 \\
 10 & 1.4182 & 0.0018 & 0.2254 & 2.2970 & 2.3334 \\
 100 & 0.1972 & 0.0001 & 1.2994 & 0.1190 & 0.3921 \\
\hline \hline
\multicolumn{6}{|c|}{Lattice-Scheme ($n_f=0$)} \\
\hline
 $\m$ in $r_0^{-1}$ & $c$ & $\m\Omega$Exact$\times 10^{9}$ & $\left| \frac{\Omega\text{LO}}{\Omega\text{Exact}}-1\right|\times 10^3$ & $\left| \frac{\Omega \text{NLO}}{\Omega\text{Exact}}-1\right|\times 10^4$ & $\left| \frac{\Omega \text{NNLO}}{\Omega\text{Exact}}-1\right|\times 10^{5}$ \\ \hline
0.6667 & 0.8101 & 0.33643 & 23.0068 & 12.0234 & 0.06620 \\
0.8333 & 1.2008 & 0.26320 & 13.94 & 5.2020 & 12.6513 \\
1 & 1.5200 & 0.21971 & 11.9588 & 13.8372 & 5.6692 \\
1.25 & 0.1599 & 0.17233 & 20.0735 & 0.3374 & 6.8322 \\
1.6667 & 0.6636 & 0.12061 & 15.0134 & 9.1010 & 3.3885 \\
2.5 & 1.3734 & 7.7980 & 1.2511 & 7.4891 & 5.2999 \\
5 & 0.8362 & 3.5950 & 14.9013 & 4.4610 & 0.4998 \\
10 & 0.2990 & 1.7262 & 4.2813 & 2.3906 & 2.5543 \\
100 & 0.8287 & 14.527 & 9.6569 & 1.8830 & 0.05590 \\
\hline
\end{tabular}
\caption{\label{Table:OmegaMnf0}
$\m\Omega_{-2}$ in the large $\beta_0$ approximation for $n_f=0$ in $r_0^{-1}$ units compared with \eq{Omegaminustwo} truncated at different powers of $\al$. Upper panel computed in the $\MS$ scheme. Lower panel in the lattice scheme. Lattice seems to be better but both schemes yield very good results.}
\end{table}

\renewcommand{\arraystretch}{1.5}
\begin{table}[h]
\begin{tabular}{|c|c|c|c|c|c|}
\hline
\multicolumn{6}{|c|}{$\overline{\text{MS}}$-Scheme ($n_f=3$)} \\
\hline
 $\m$ in GeV & $c$ & $\m\Omega$Exact & $\left| \frac{\Omega\text{LO}}{\Omega\text{Exact}}-1\right|\times 10^2$ & $\left| \frac{\Omega \text{NLO}}{\Omega\text{Exact}}-1\right|\times 10^3$ & $\left| \frac{\Omega \text{NNLO}}{\Omega\text{Exact}}-1\right|\times 10^4$ \\ \hline
 0.6667 & 0.4916 & 0.00375 & 3.6513 & 8.8280 & 16.5412 \\
 0.8333 & 0.8113 & 0.00274 & 5.0007 & 2.9016 & 7.864 \\
 1 & 1.0724 & 0.00223 & 1.6450 & 3.9632 & 11.1774 \\
 1.25 & 1.3921 & 0.00183 & 6.2621 & 3.6008 & 0.7661 \\
 1.6667 & 0.3717 & 0.00119 & 0.8631 & 2.7541 & 4.6880 \\
 2.5 & 0.9525 & 0.00072 & 2.2110 & 0.5893 & 3.1281 \\
 5 & 0.5130 & 0.00032 & 1.7608 & 1.5455 & 0.9363 \\
 10 & 0.0735 & 0.00015 & 2.8072 & 0.2396 & 0.5520 \\
 100 & 0.5069 & 0.00001 & 0.9405 & 0.4412 & 0.1384 \\
\hline \hline
\multicolumn{6}{|c|}{Lattice-Scheme ($n_f=3$)} \\
\hline
 $\m$ in GeV & $c$ & $\m\Omega$Exact$\times 10^{11}$ & $\left| \frac{\Omega\text{LO}}{\Omega\text{Exact}}-1\right|\times 10^3$ & $\left| \frac{\Omega \text{NLO}}{\Omega\text{Exact}}-1\right|\times 10^4$ & $\left| \frac{\Omega \text{NNLO}}{\Omega\text{Exact}}-1\right|\times 10^{5}$ \\ \hline
 0.6667 & 0.6457 & 0.13921 & 14.5539 & 5.0568 & 0.1286 \\
 0.8333 & 0.9653 & 0.10969 & 9.6413 & 1.9386 & 3.4992 \\
 1 & 1.2264 & 9.1458 & 6.7772 & 6.1070 & 1.8276 \\
 1.25 & 0.1137 & 7.2422 & 15.151 & 0.1208 & 2.0634 \\
 1.6667 & 0.5258 & 5.1755 & 9.9772 & 4.3766 & 1.2143 \\
 2.5 & 1.1065 & 3.3731 & 1.7019 & 3.6030 & 1.9541 \\
 5 & 0.6670 & 1.5918 & 10.8713 & 2.5255 & 0.06625 \\
 10 & 0.2275 & 77.282 & 4.1604 & 1.2278 & 1.1874 \\
 100 & 0.6609 & 674.45 & 7.7077 & 1.7587 & 4.5364 \\
\hline
\end{tabular}
\caption{\label{Table:OmegaMnf3}
$\m\Omega_{-2}$ in the large $\beta_0$ approximation for $n_f=3$ in GeV units compared with \eq{Omegaminustwo} truncated at different powers of $\al$. Upper panel computed in the $\MS$ scheme. Lower panel in the lattice scheme. Lattice seems to be better but both schemes yield very good results.}
\end{table}

\begin{center}
\begin{figure}
\includegraphics[width=0.95\textwidth]{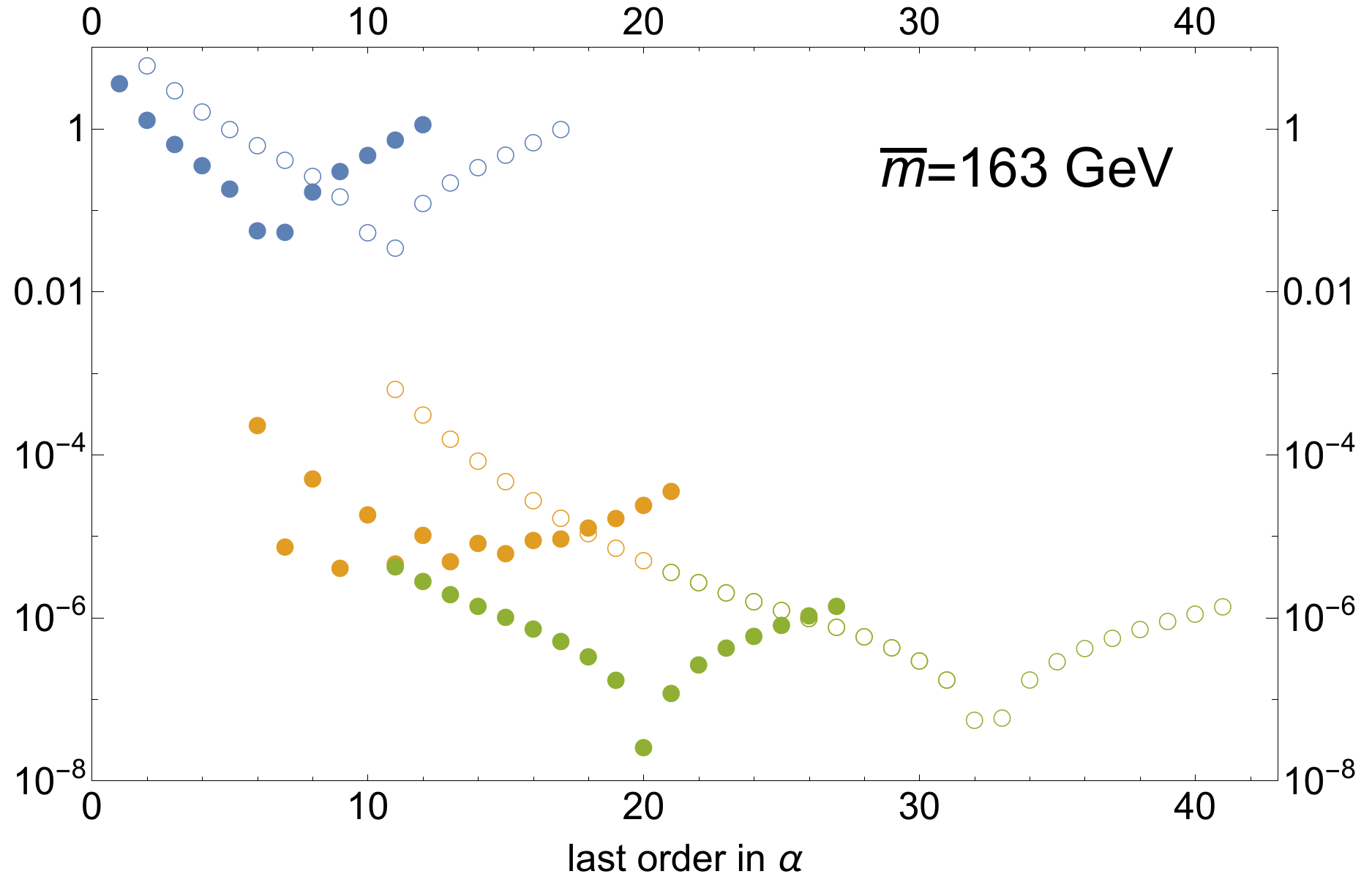}
\includegraphics[width=0.97\textwidth]{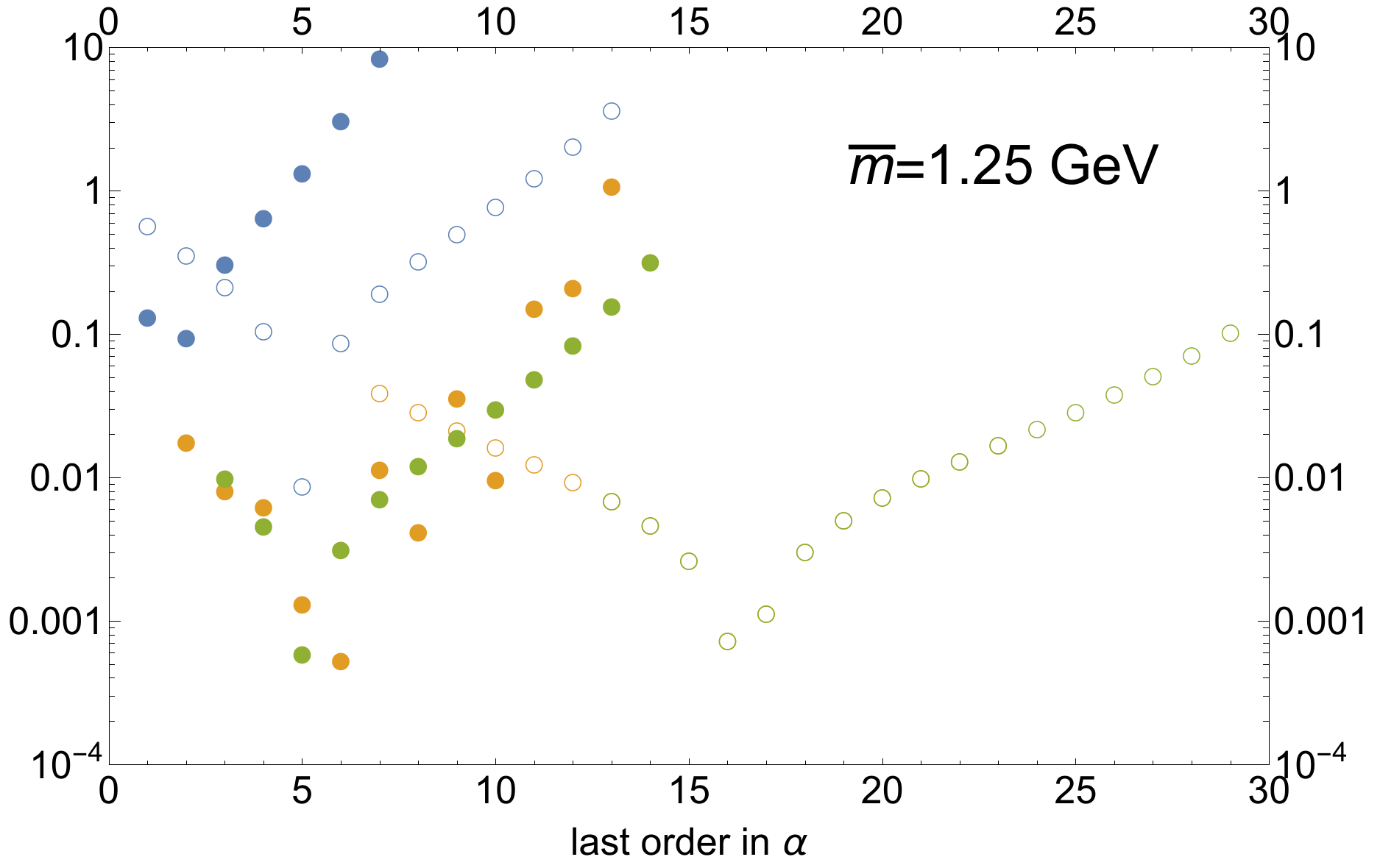}
\caption{\label{Fig:Boyd}
$|m_{\rm PV}-m_{\rm PV}^{\rm Hyperasymptotic}|$ for $\m=163$ GeV (upper panel) and $\m=1.25$ GeV (lower panel). Blue points are $|m_{\rm PV}-m_N|$. Orange points are $|m_{\rm PV}-m_P-\m\Omega_m-\sum_{n=N_P+1}^{N} (r_n-r_n^{(\rm as)}) \al^{n+1}|$ with $c=1.21/1.39$ 
% 1.20672/1.39206
and 
%1.36073/0.113682
$c=1.36/0.11$ (the smallest positive values that yield integer $N_P$) in the $\MS$ and lattice scheme respectively for $\m=163/1.25$ GeV. Green points are $|m_{\rm PV}-m_P-\m\Omega_m-\sum_{n=N_P+1}^{2N_P} (r_n-r_n^{(\rm as)}) \al^{n+1}-\m\Omega_{-2}-\sum_{n=2N_P+1}^{N} (r_n-r_n^{(\rm as)}) \al^{n+1}|$, where in the last sum the two first renormalons are subtracted. Change of color correspond to the inclusion of $\Omega_m$ and  $\Omega_{-2}$. Full points have been computed in the $\MS$ scheme and empty points in the lattice scheme. We work with $n_f=3$.}
\end{figure}
\end{center}

An alternative, very effective, presentation of the above results can be done by plotting the relative accuracy of the prediction at each order in $\al$, and at each order of the superasymptotic expansion. We note that we have one observable for each value of $\m$. Therefore, for illustration, we take two extreme cases. We use $m_{\rm PV}$ with $\m=1.25$ GeV and $\m=163$ GeV. For the theoretical prediction we take the smallest positive value of $c$ corresponding to lattice or $\MS$ scheme.\footnote{Taking different values of $c$ do not change the picture. The new points stand on top of the old ones where they overlap.} We use the exact expressions for  $\Omega_m$ and $\Omega_{-2}$. Nevertheless, the NNLO truncated expression for $\Omega_m$ is precise enough to yield the same result. For $\Omega_{-2}$ we could truncate earlier with no visible effect. We show the results in Fig. \ref{Fig:Boyd}. We stress that several terms of the hyperasymptotic expansion are included. First, we nicely see that, once reached the minimum, $N \sim N_P$, both schemes yield similar precision, but in the lattice scheme (bigger factorization scale $\mu$) more terms of the perturbative expansions are needed to reach the same precision. We can see a gap when $\Omega_m$ is included, with significant better precision in the $\MS$ scheme. One important lesson one may extrapolate from this exercise is that, if the number of perturbative coefficients is fixed, the smaller the renormalization scale $\mu$, the better. One can obtain much better precision for an equal number of perturbative coefficients. Another observation is that the minimal term determined numerically need not to coincide with the minimal term computed using $N=N_P$ (though it should not be much different). The difference reflects how much the exact coefficient is saturated by the asymptotic expression. The effect of $\Omega_{-2}$ is very small compared with the effect due to $\Omega_m$.  For the case of the top ($\m=163$ GeV) we can still see the sign alternating behavior of the perturbative series associated to the $d=-2$ renormalon in the $\MS$ scheme. In the lattice scheme the effect is so small that it cannot be seen and the precision is set by the next renormalon located at $d=3$. If one makes $\m$ small, $\m=1.25$ GeV, green and orange points mix in the $\MS$ scheme. This effect is more pronounced in the lattice scheme, where one can continuously move from the orange to the green points. The effect of the ultraviolet renormalon is very small and the precision is set by the $u=3/2$ renormalon. 

\subsection{$(N,\mu) \rightarrow \infty$. \eq{eq:muinfty}. Case 2)}
\label{Sec:MPVmuinfty}

We take \eq{mPVHyperAco} in the large $\beta_0$ limit by setting $b=0$. As before, we have 
 no analytic expressions to compare with (unlike the case of the static potential).
Therefore, we directly focus on taking the limit 2B) and numerically check its convergence and how it compares with method 1). 

\begin{center}
\begin{figure}[htb]
\includegraphics[width=0.76\textwidth]{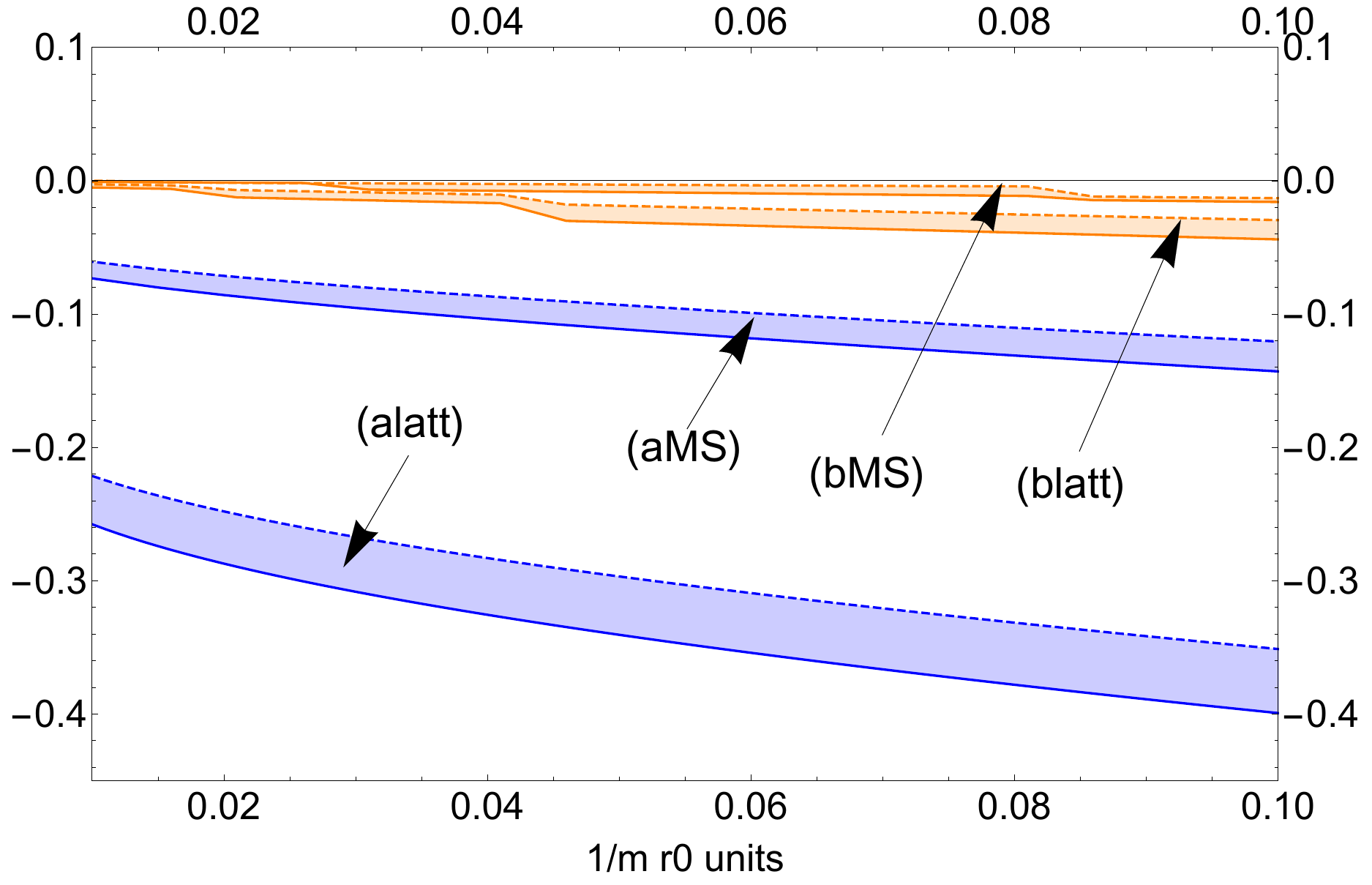}
\includegraphics[width=0.78\textwidth]{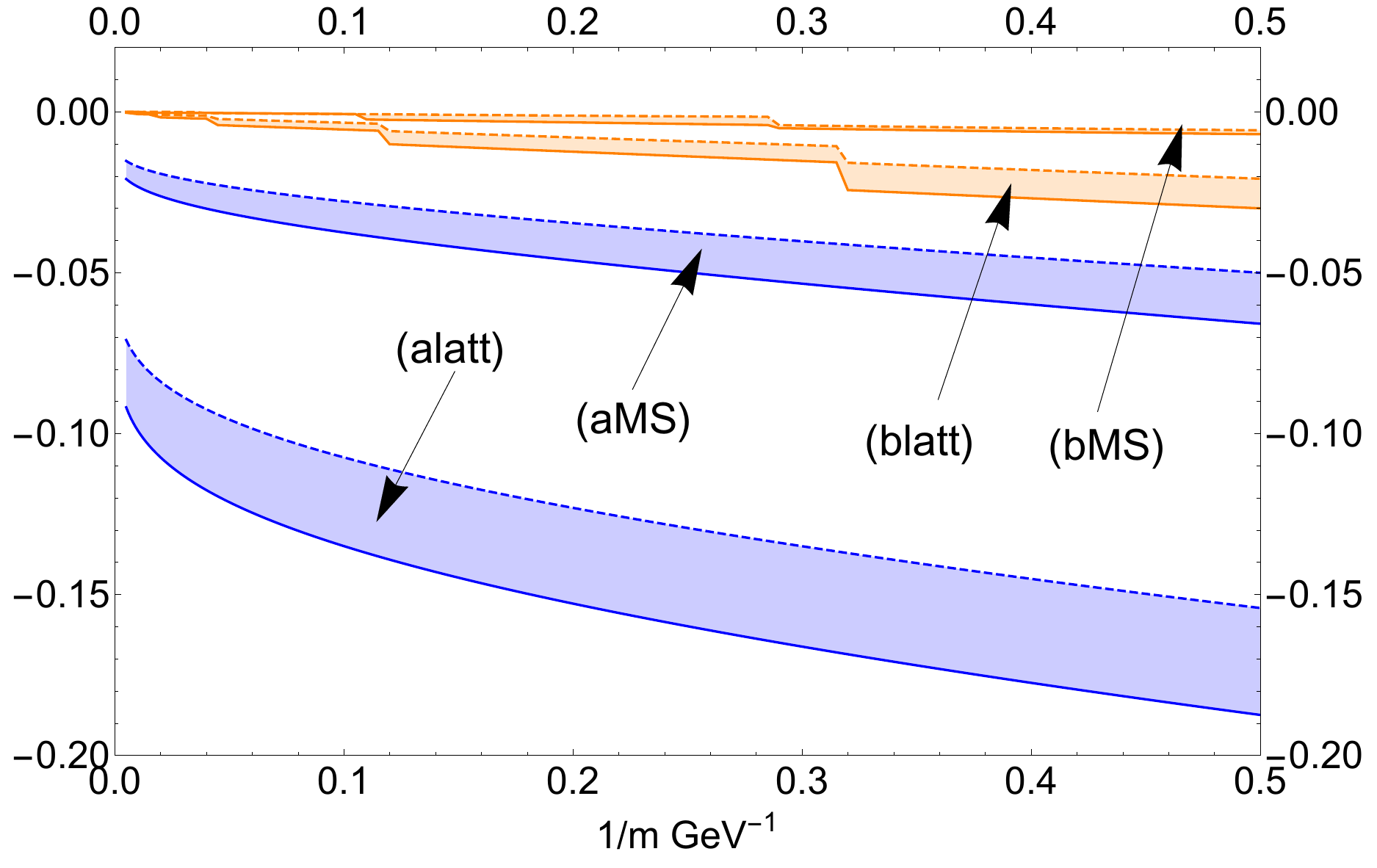}
%\put(0,195){{\large (alatt)}}
%\put(-200,25){{\large (blatt)}}
%\put(0,15){{\large (bMS)}}
%\put(0,63){{\large (aMS)}}
\caption{{\bf Upper panel}: We plot (a) $m_{\rm PV}-m_A-K_X^{(A)}\Lambda_X$ for $n_f=0$ in the lattice and $\MS$ scheme. For each case, we generate bands by computing $m_A$ with $c'=1$ and $c'=c'_{\rm min}=0.652$. 
%0.652150
We also compare with (b) $m_{PV}-m_P-\m\Omega_m$ obtained with method 1) with the bands generated for Fig. \ref{Fig:MPVb0nf0lattvsMS}. {\bf Lower panel}: As the upper panel with $n_f=3$, $c'_{\rm min}=0.534$ 
%0.533577
and taking the  the bands obtained with method 1) for Fig. 
\ref{Fig:MPVb0nf3lattvsMS} for (b) $m_{\rm PV}-m_P-\m\Omega_m$ .}
\label{Fig:BothmethodsMPVnf0}
\end{figure}
\end{center}

Method 2B) has the pleasant feature that the generated ${\cal O}(\lQ)$ correction complies with the OPE. It also yields results that do not depend on $N$ (and $\mu$) anymore. Still, it has some errors and does not reach the precision of method 1). There is a residual scheme dependence associated to uncomputed terms of ${\cal O}(\al \lQ)$. Part of it can be estimated by the residual dependence in $c'$. In order to estimate it, we compute $m_A$ for different values of $c'$. On the one hand $c'$ cannot be very large, as $c'\al(\m)$ should be relatively close to zero.  On the other hand we cannot make $c'\al(\m)$ to get arbitrary close to zero, as the ${\cal O}(\lQ)$ correction diverges logarithmically in $c'$. We also note that there is a value of $c'=c'_{\rm min}$ that makes that $K_X^{(A)}=0$ so that the 
${\cal O}(\lQ)$ correction vanishes. Therefore, we compute $m_A$ for different values of $c'$. For illustration we show some results in Fig. \ref{Fig:BothmethodsMPVnf0}. We draw lines for $m_{\rm PV}-m_A-K_X^{(A)}\Lambda_X$ at $c'=1$ and $c'=c_{\rm min}$ generating a band. We also explore the dependence on the scheme by comparing the results in the lattice and $\MS$ scheme.  We stress again that, in the large $\beta_0$ approximation, lattice and $\MS$ schemes basically correspond to a redefinition of $\mu$, but quite large indeed. On the other hand the final result is $\mu$ independent. Nevertheless, the way the $\mu \rightarrow \infty$ limit is taken is fixed by $N_A$, as defined in \eq{eq:muinfty}, which is dependent on $\mu$. This explains why different results are obtained. 

\begin{center}
\begin{figure}
\includegraphics[width=0.75\textwidth]{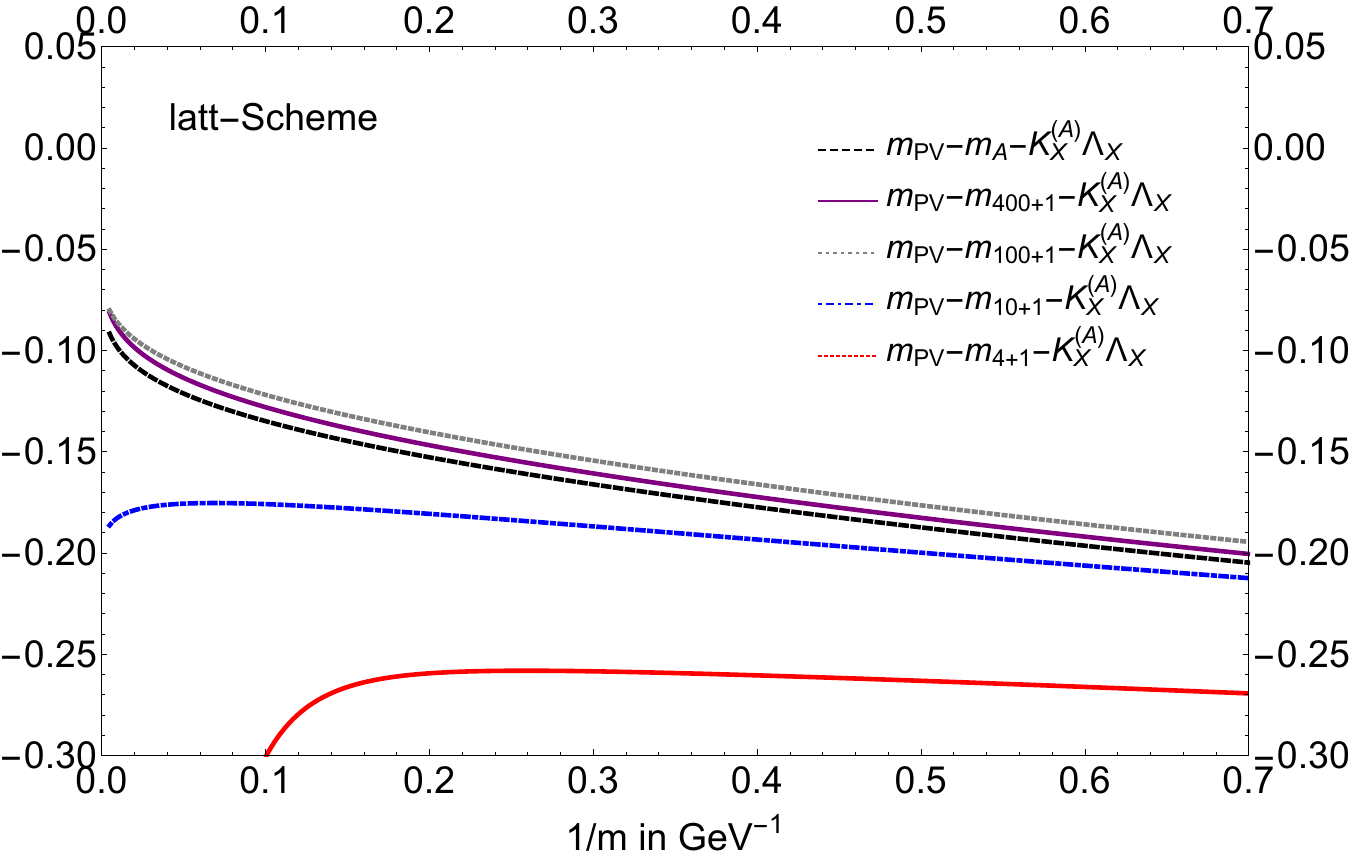}
\includegraphics[width=0.75\textwidth]{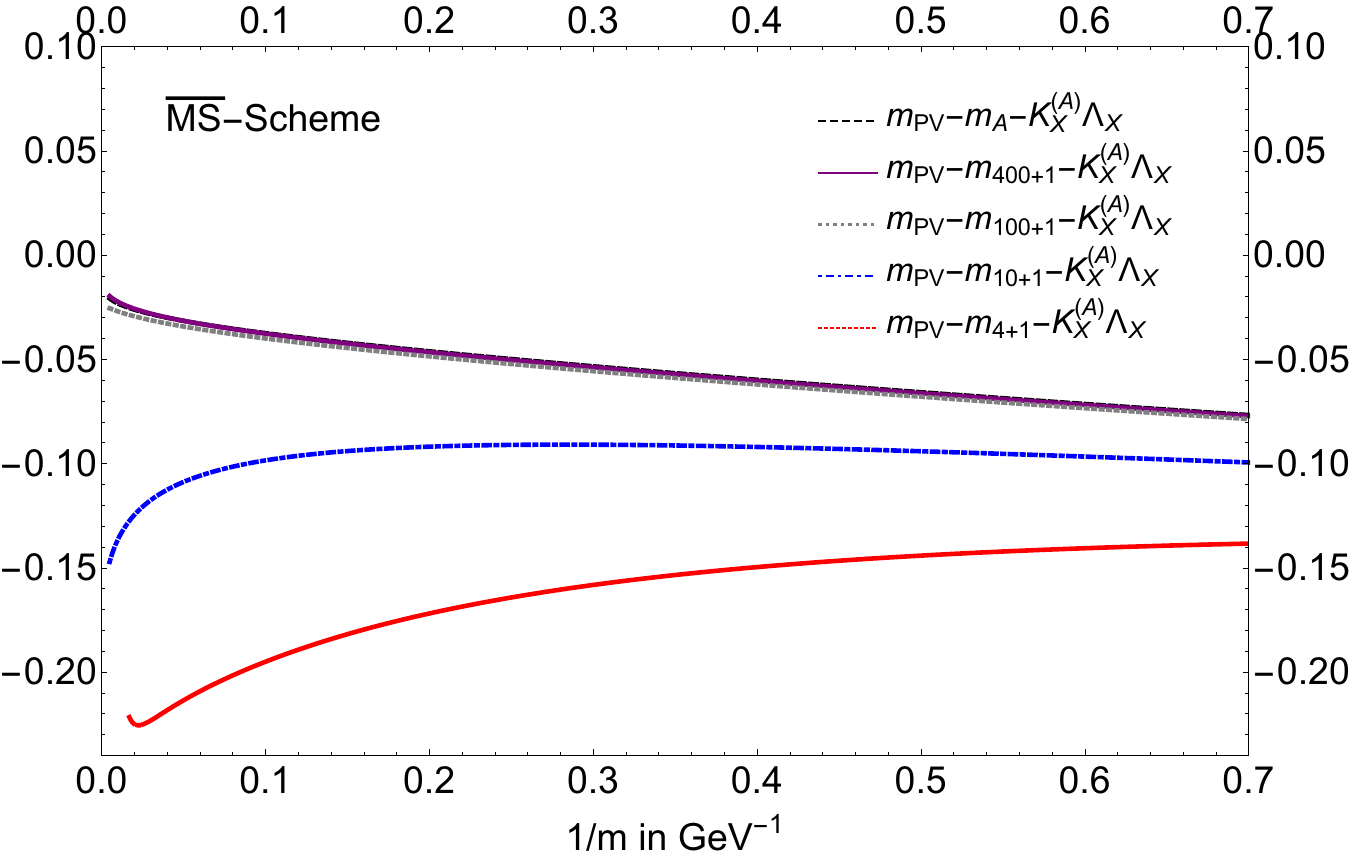}
\caption{{\bf Upper panel}: We plot $m_{\rm PV}-m_A-K_X^{(A)}\Lambda_X$ for $n_f=3$ in the lattice scheme with $c=1$ versus the truncated sums $m_{\rm PV}-\sum_{n=0}^{N_A}r_n\al^{n+1}(\mu)-K_X^{(A)}\Lambda_X$, where $\mu$ is fixed using $N_A$ defined in \eq{eq:muinfty}. {\bf Lower panel}: As in the upper panel but in the $\MS$ scheme.}
\label{Fig:Convergencec1nf3}
\end{figure}
\end{center}

In Fig. \ref{Fig:BothmethodsMPVnf0}, we also compare with results obtained using method 1), more specifically we compare with $m_{\rm PV}-m_P-\m\Omega_m$, as they both have analogous power accuracy (though method 1) is  parametrically more precise). For $\Omega_m$ we take the exact expression but using its approximated expression does not change the discussion, as the difference is very small. What we see is that the $\MS$ scheme yields more precise predictions than the lattice scheme, and that method 1) yields considerable better results than method 2B). 

Another issue specific of method 2B) is to determine how large we need to take $N$ (and consequently $\mu$) of the truncated sum such that it approximates well $m_A$. For illustrative purposes we show the convergence in Fig. \ref{Fig:Convergencec1nf3} for $n_f=3$ in the lattice and $\MS$ scheme. We find that we have to go to relatively large values of $\mu$ (and $N$) to get it precise. This can be a problem if one wants to go beyond the large 
$\beta_0$. This problem would be less severe if one can use the asymptotic expression for the coefficients beyond certain $n$. Nicely enough, we find that the use of asymptotic expression for the coefficients for $n>N^*$ ($\sim 3$ in the $\MS$ and $\sim$ 8 in the lattice scheme) is very efficient and basically yields the same results as the exact result.  Finally, we also remind that to approximate well $m_A$ by the truncated sum is more costly for small values of $c'$. 

\section{$\overline{\Lambda}_{\rm PV}$ from lattice and $B$ physics}
\label{Sec:Lambda}

We now abandon the large-$\beta_0$ approximation. Our aim is to determine $\bar \Lambda_{\rm PV}$.  We will determine it first in gluedynamics ($n_f=0$) in Sec. \ref{Sec:LambdaLatt}. To study the scheme dependence of the result it will be useful to estimate the higher order coefficients of the $\beta$ function in the Wilson action lattice scheme. We do so in the next section. 

\subsection{$\beta$-function coefficients in the Wilson action lattice scheme}
\begin{table}[htb]
\begin{tabular}{|c|c|c|c|}
\hline
  $\beta _3$ & $\beta _4$ & $\beta _5$ & $\beta _6$ \\
\hline
  $
 %-1.15766
-1.16(3)\times 10^6$ & $
%-1.34875
-1.35(10)\times 10^8$ & $
%-1.4401
-1.44(28)\times 10^{10}$ & $
%-1.4056
-1.41(60)\times 10^{12}$ \\
%\hline
% Ref.\cite{Ayala:2014yxa} & $-1.18287\times 10^6$ & $-1.44889\times 10^8$ & $-1.72058\times 10^{10}$ & $-2.11111\times 10^{12}$ \\
\hline
\end{tabular}
\caption{Estimates of the coefficients of the beta function for the bare coupling in the lattice scheme using renormalon dominance and $Z_m^{\MS}=0.62$ \cite{Bali:2013qla}. The error quoted in the table gives the difference with the values of the beta coefficients obtained if one uses instead $Z_m^{\MS}=0.6$ \cite{Ayala:2014yxa} (which yields more negative values), and it is only meant to illustrate the typical spread of values of the beta coefficients if one uses different values of $Z_m^{\MS}$.}
\label{Tab:alLatt}
\end{table}
\begin{center}
\begin{figure}[htb]
\includegraphics[width=0.95\textwidth]{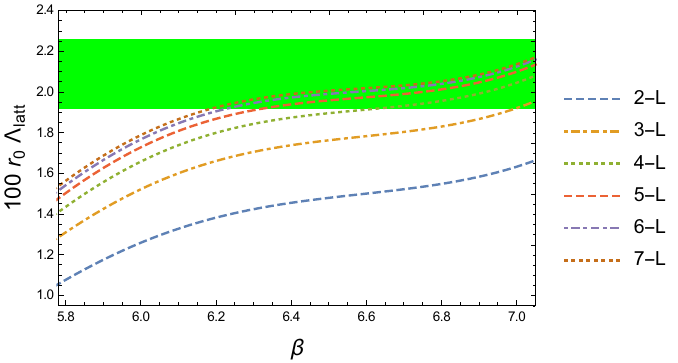}
\caption{Same caption as in Fig. 1 of \cite{Bali:2014sja} including more terms in the perturbative expansion using the $\beta$-function coefficients listed in Table \ref{Tab:alLatt}.}
\label{Fig:Lambda}
\end{figure}
\end{center}
 In \cite{Bali:2013pla,Bali:2013qla} it was shown that renormalon dominance allowed to give an accurate value for $\beta_3^{latt}$ assuming that $c_3$ (see \eq{deltamPVlatt}) is already saturated by the renormalon in the $\MS$ scheme.  We can estimate higher order terms of the $\beta$ function in the lattice scheme (using the Wilson action) by also assuming that for $n>3$ the coefficients $c_n$ in the $\MS$ scheme are saturated by the renormalon. We show such estimates in Table \ref{Tab:alLatt}. These coefficients of the $\beta$ function improve the agreement with the phenomenological parameterization of $\alpha_{latt}(1/a)$ obtained in \cite{Necco:2001xg} in the range $\beta \in (6,6.8)$ (see Fig. \ref{Fig:Lambda}). It is also worth mentioning that we observe a geometrical growth of the coefficients of the $\beta$ function. Elucubrative, this would indicate that the beta function in this scheme has a finite radius of convergence, and one can take the ansatz
\be
\beta^{\rm latt}(\al)=\nu\frac{d}{d\nu}\al\simeq
-2\al\left\{\sum_{n=0}^3\beta_n\left(\frac{\al}{4\pi}\right)^{n+1}-1.4\times 10^8\left(\frac{\al}{4\pi}\right)^{5}\frac{1}{1- 10^2 \frac{\al}{4\pi}}\right\}
\,,
\ee
which would have a pole at around $\beta =6/g^2 \simeq 3.8$. 

\subsection{$\bar \Lambda_{\rm PV}$ from lattice}
\label{Sec:LambdaLatt}
\begin{center}
\begin{figure}
\includegraphics[width=0.79\textwidth]{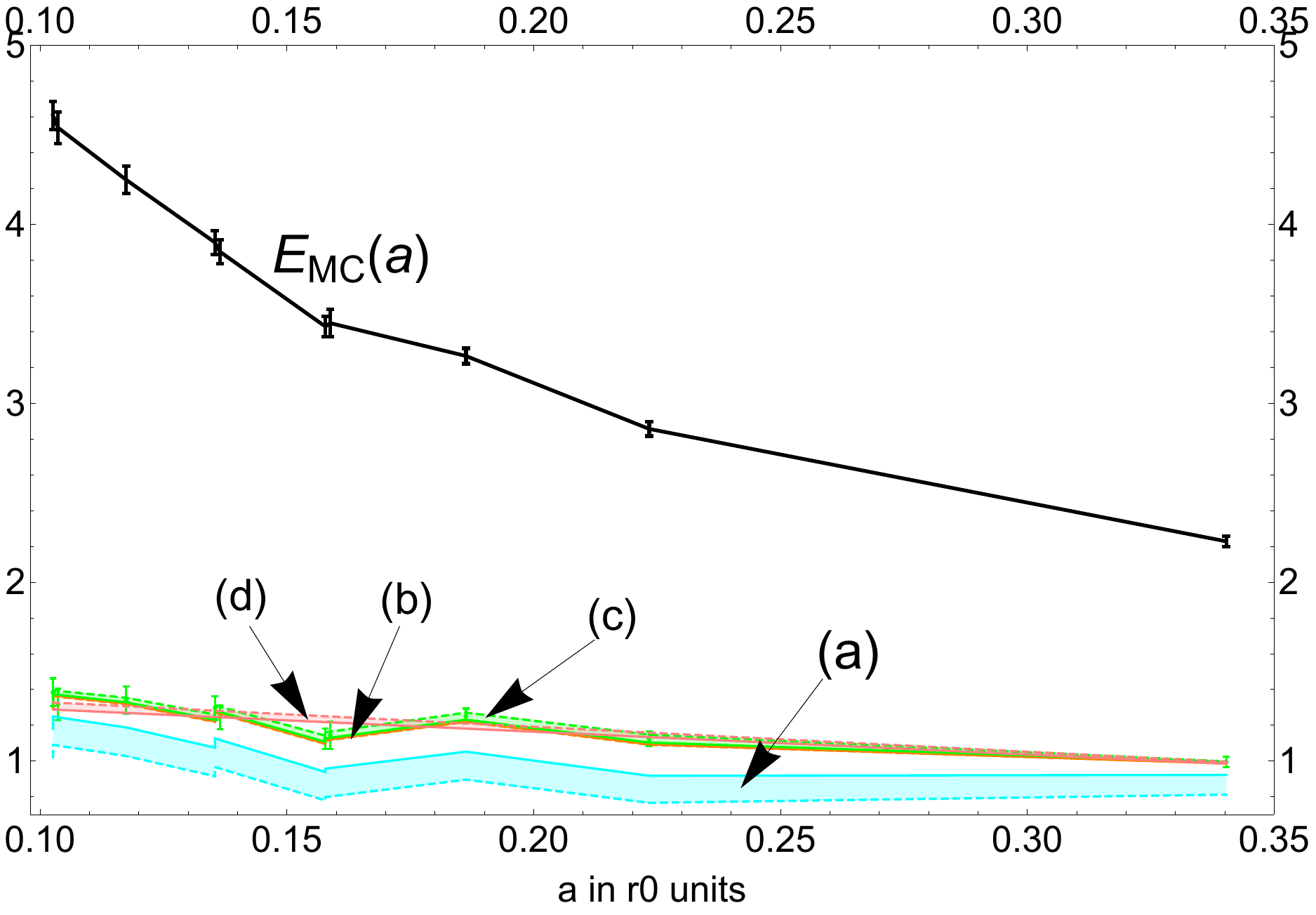}
\includegraphics[width=0.82\textwidth]{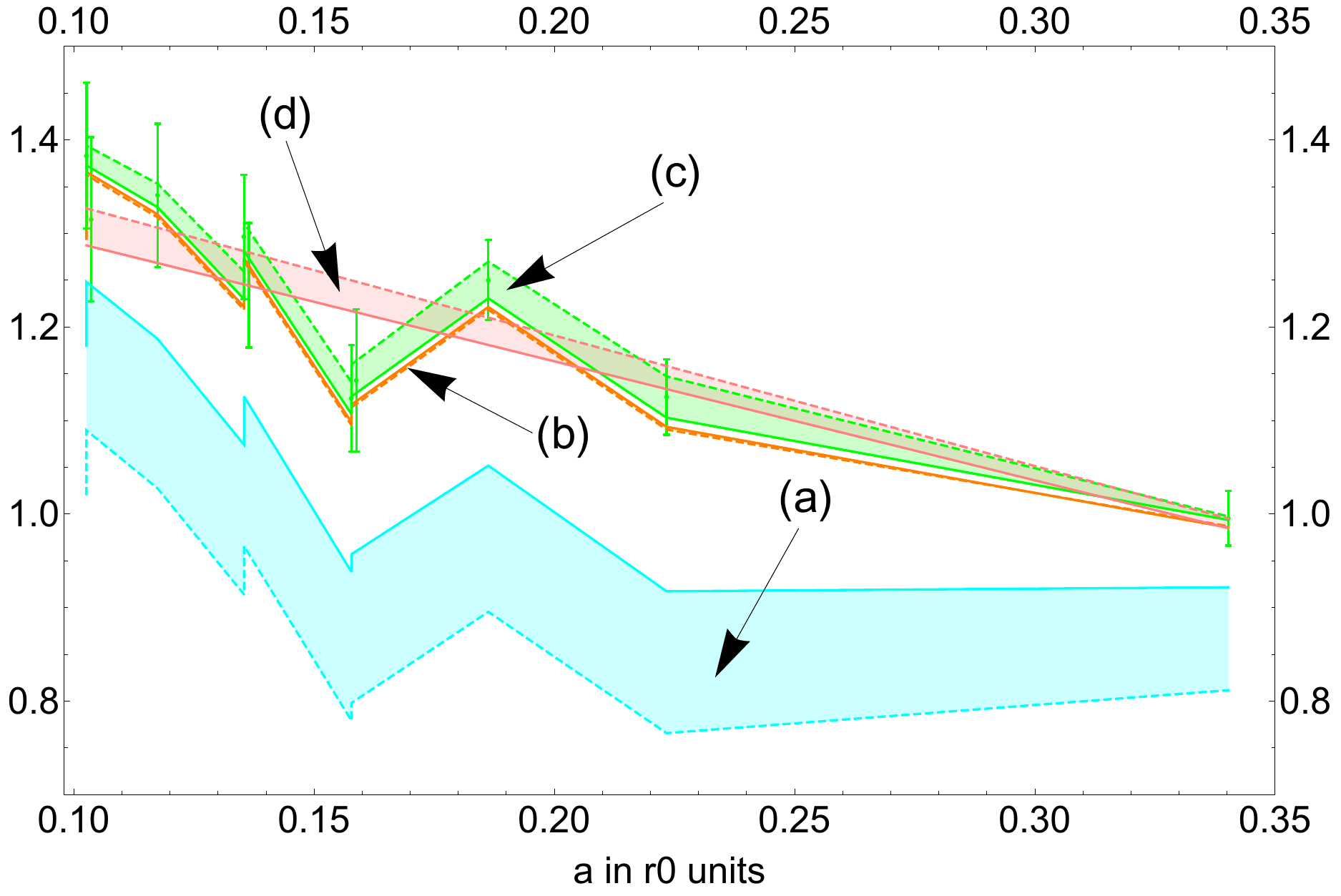}
\caption{{\bf Upper panel}: $E_{\rm MC}$ is the Montecarlo lattice data \cite{Duncan:1994uq,Allton:1994tt,Ewing:1995ih}. The continuous lines are drawn to guide the eye. The other lines correspond to \eq{Lambdabarlatt} truncated at different orders in the hyperasymptotic expansion. (a) $E_{\rm MC} (a)-\delta m_{P}(1/a)$, (b) $E_{\rm MC} (a)-\delta m_{P}(1/a)-\frac{1}{a}\Omega_m$, (c) $E_{\rm MC} (a)-\delta m_{P}(1/a)-\frac{1}{a}\Omega_m-\sum_{N_P+1}^{N'=2N_P}\frac{1}{a}[c_n-c_n^{\rm (as)}]\alpha^{n+1}$ (in this last case we include the error of the lattice points in the middle of the band), (d) is the fit of the right-hand-side of \eq{Lambdabarlatt} to $\bar \Lambda_{\rm PV}(n_f=0)-Ka$. For each difference and for the final fit, the bands are generated by the difference of the prediction produced by the smallest positive or negative possible values of $c$ that yield integer values for $N_P$. {\bf Lower panel}: As in the upper panel but in a smaller range. $r_0^{-1} \approx 400$ MeV.}
\label{Fig:barLambdaLatt}
\end{figure}
\end{center} 

 We determine $\bar \Lambda_{\rm PV}$ in gluedynamics ($n_f=0$) from the energy of a meson made of a static quark and a light valence quark:
\be
\label{EMC}
E_{\rm MC}(a)=\delta m^{\rm PV}_{\rm latt} +\bar \Lambda_{\rm PV} +{\cal O}(a \lQ^2)
\,.
\ee
$\delta m^{\rm PV}_{\rm latt}$ has the following asymptotic series in powers of $\al=\al_{\rm latt}(1/a)$:
\be
\label{deltamPVlatt}
\delta m^{\rm PV}_{\rm latt} \sim \sum_{n=0}^{\infty}\frac{1}{a}c_n\alpha^{n+1}
\,,
\ee
where $\frac{r_n^{\rm (as)}(\nu)}{\nu}= c_n^{\rm (as)}$, since $m_{\rm PV}$ and $\delta m^{\rm PV}_{\rm latt}$ have the same leading infared renormalon (located at $d=1$). 
The coefficients $c_n$ are known from $n=0\div19$ in the lattice scheme for a Wilson action \cite{Bauer:2011ws,Bali:2013pla,Bali:2013qla}. We then adapt Eq. (\ref{mPV}) to $\delta m^{\rm PV}_{\rm latt}$ to determine 
$\bar \Lambda_{\rm PV}$:
\be
\label{Lambdabarlatt}
\bar \Lambda_{\rm PV}(n_f=0)=E_{\rm MC} (a)-\delta m^{P}_{\rm latt}-\frac{1}{a}\Omega_m-
\sum_{N_P+1}^{N'=2N_P}\frac{1}{a}[c_n-c_n^{\rm (as)}]\alpha^{n+1}+{\cal O}(a\lQ^2)\,.
\ee
where $\delta m^{P}_{\rm latt}=\sum_{n=0}^{N_P}\frac{1}{a}c_n\alpha^{n+1}$. In the counting of \eq{SPVDN} this corresponds to (1,$N_P$) precision. The expression we use for $\Omega_m$ is \eq{Omegam} truncated to ${\cal O}(\al^3)$ here and in the rest of the paper. The error committed by this truncation is smaller than the error associated to $Z_m$. Therefore, we will neglect it in the following. 
The renormalon behavior associated to subleading renormalons of $E_{\rm MC} (a)$ is not well known, except that the next singularity in the Borel plane is expected to be at $|u|=1$ (d=2). Therefore, we stop the second perturbative expansion at $N'=2N_P$ such that the reminder should be of ${\cal O}(a\lQ^2)$. For the coefficients $c_n^{\rm (as)}$ we use $Z_m^{latt}(n_f=0)=17.9(1.0)$ \cite{Bali:2013qla}.  We also truncate the $1/n$ expansion in \eq{rnas} to ${\cal O}(1/n^3)$. This means using the estimates for $\beta_3$ and $\beta_4$ listed in Table \ref{Tab:alLatt}.
We take $E_{\rm MC}$(a) from
\cite{Duncan:1994uq,Allton:1994tt,Ewing:1995ih}. These points expand over the following energy range: $1/a \sim 2.93 \; r_0^{-1} \div 9.74 \;  r_0^{-1}$. We show our results in Fig. \ref{Fig:barLambdaLatt}. They follow the same logic than Figs. 1-6 in Sec. \ref{Sec:mOS}. We observe that the subtraction of the perturbative expansion accounts for most of the $1/a$ dependence. Still we have enough precision to be sensitive to ${\cal O}(a\lQ^2)$ effects. A strict fit setting the ${\cal O}(a\lQ^2)$ correction to zero gives a large $\chi^2_{\rm red} \sim 6-7$. The inclusion of a pure $K a$ term to \eq{Lambdabarlatt} gives a good fit\footnote{Unlike for the pole mass, it is not clear what is the operator of the OPE that would produce the NP correction  and the associated $u=1$ renormalon. Therefore, if for the pole mass we can be certain that the NP correction has the form $Ka$, without any anomalous dimensions nor any nontrivial $\ln(a)$ dependence, we can not exclude the possibility that this ${\cal O}(a)$ correction may have a non trivial anomalous dimension and/or $\ln(a)$ dependence.}. The statistical error is small and the $\chi^2_{\rm red}=1.17/1.06$ (for the smallest $|c|$ with positive/negative $c$ value) is good. Overall, we obtain (using the smallest $c$ positive, which means $N_P=7$ except for $\beta=5.7$ where we have $N_P=6$)
\be
\label{LPVlatt}
\bar \Lambda_{\rm PV}=1.42 \; r_0^{-1}({\rm stat.})^{-0.01}_{+0.04} (c)^{+0.05}_{-0.05}(Z_m) ^{+0.16}_{-0.16}
\,.
\ee
This number is not very different from the number obtained in \cite{Bali:2014sja} using a superasymptotic approximation truncated at the minimal term determined numerically (typically this always gives slightly better results than truncating at the minimal term predicted by theory). 

Let us now discuss the error budget in \eq{LPVlatt}. The first error is the statistical error of the fit. The remaining errors are different ways to estimate the error produced by the approximate knowledge of the hyperasymptotic expansion.  One possibility is to take the modulus of the difference with the evaluation using the $c$ negative with the smallest possible modulus. This is the second error we quote in \eq{LPVlatt}. The last error we include is due to the variation of $Z_m^{latt}(n_f=0)=17.9(1.0)$ \cite{Bali:2013qla} (correlated with the error of $c_n$). The error it produces in $\Omega_m$ is small. Comparatively, most of the error associated to $Z_m$ comes from the differences in $\sum_{N_P+1}^{N'=2N_P}\frac{1}{a}[c_n-c_n^{\rm (as)}]\alpha^{n+1}$ evaluated at different $Z_m$. Whereas $\sum_{N_P+1}^{N'=2N_P}\frac{1}{a}[c_n-c_n^{\rm (as)}]\alpha^{n+1}$ is quite small for the central value of $Z_m$, it significantly changes after variation of $Z_m$. This variation is only partially compensated by the variation of the coefficients $c_n$, which have smaller errors, producing a significant change in $\sum_{N_P+1}^{N'=2N_P}\frac{1}{a}[c_n-c_n^{\rm (as)}]\alpha^{n+1}$. We have also determined the central value in \eq{LPVlatt} not including the ${\cal O}(1/n^3)$ corrections in the asymptotic expressions for $c_n^{\rm (as)}$. The difference we obtain is -0.08. This is significant, showing that the $1/n$ corrections are sizable in the lattice scheme. On the other hand, the difference is well inside the error associated to $Z_m$. Actually, the difference with evaluations including the ${\cal O}(1/n^4)$ corrections in the asymptotic expressions for $c_n^{\rm (as)}$ is -0.03. This shows a convergent pattern, which we illustrate in Table \ref{Tab:LambdaMS}.  Overall, the largest source of uncertainty comes from the incomplete knowledge of $\sum_{N_P+1}^{N'=2N_P}\frac{1}{a}[c_n-c_n^{\rm (as)}]\alpha^{n+1}$, which is closely linked to the incomplete knowledge of $Z_m$. 
This discussion points to that more accurate determinations of $Z_m$ can be possible and, then, that the error of $\bar \Lambda_{\rm PV}$ associated to $Z_m$ could be made smaller. We believe these issues deserve further study that we leave for future work. 

\begin{table}[htb]
\begin{tabular}{|c||c|c|c||c|c|c|c|c|}
\hline
 {\rm latt}& ${\cal O}\left(\frac{1}{n^2}\right)$ & ${\cal O}(\frac{1}{n^3})$ & ${\cal O}(\frac{1}{n^4})$ & $\MS$ & $N_{tr}=7$& $N_{tr}=6$& $N_{tr}=5$& $N_{tr}=4$ \\
\hline
$\bar \Lambda_{\rm PV}$  &$1.33
 %-1.15766
$ & $
%-1.34875
1.42$ & $1.45$& $\bar \Lambda_{\rm PV}$ & $1.48$ & $1.52$ & $1.59$ & $1.68$ \\
%\hline
% Ref.\cite{Ayala:2014yxa} & $-1.18287\times 10^6$ & $-1.44889\times 10^8$ & $-1.72058\times 10^{10}$ & $-2.11111\times 10^{12}$ \\
\hline
\end{tabular}
\caption{Determinations of $\bar \Lambda_{\rm PV}$ in the lattice and $\MS$ scheme from fits of $\bar \Lambda_{\rm PV}-Ka$ to the right hand side of \eq{Lambdabarlatt}. The first three numbers show the impact in the fit of including the ${\cal O}(1/n^m)$ corrections for $m=2$, 3, 4 in the asymptotic expressions for $c_n^{(as)}$ in the lattice scheme (in the $\MS$ this effect is negligible). The other numbers are the fit of $\bar \Lambda_{\rm PV}$ in the $\MS$ scheme, using $\alpha_{\MS}=\alpha_{\rm latt}(1+\sum_{n=0}^{N_{tr}}d_n\alpha^n_{\rm latt})$ truncanted at $N_{tr}=4,5,6,7$. }
\label{Tab:LambdaMS}
\end{table}

One error that we do not include here is the error associated to $\alpha$. From the lattice point of view, we are talking of the relation between $\alpha(1/a)$ and $r_0$. We use the phenomenological formula deduced in \cite{Necco:2001xg}. The error of this formula is claimed to be around 0.5-1\% in the range $ \beta \in (5.7,6.92)$ (having a look to Fig. \ref{Fig:Lambda} a more conservative range could be (6,6.8)).

\subsubsection{Scheme dependence}
It is interesting to consider the scheme dependence of \eq{LPVlatt}. In \cite{Bali:2014sja} relative large differences were found for fits to $\bar \Lambda$ after (approximated) scheme conversion to the $\MS$ scheme. The real problem is not transforming the coefficients $c_n$ from the lattice to the $\MS$ scheme, but transforming $\al_{latt}$ to $\al_{\MS}$ with enough precision (in a way we need the relation between $\al_{latt}$ and $\al_{\MS}$ with NP, exponential, accuracy). This needs the coefficients of the $\beta$ function in the lattice scheme to high orders. We show estimates in Table \ref{Tab:alLatt}. The inclusion of these coefficients of the $\beta$ function makes that the determinations of $\bar \Lambda_{\rm PV}$ in the $\MS$ and lattice scheme approach each other as we include more terms in the perturbative expansion of the relation between $\al_{\MS}$ and $\al_{latt}$. We show the comparison in Table \ref{Tab:LambdaMS}.

\subsection{$\bar \Lambda^{\rm PV}_{pot}$ from lattice}
\begin{center}
\begin{figure}
\includegraphics[width=0.75\textwidth]{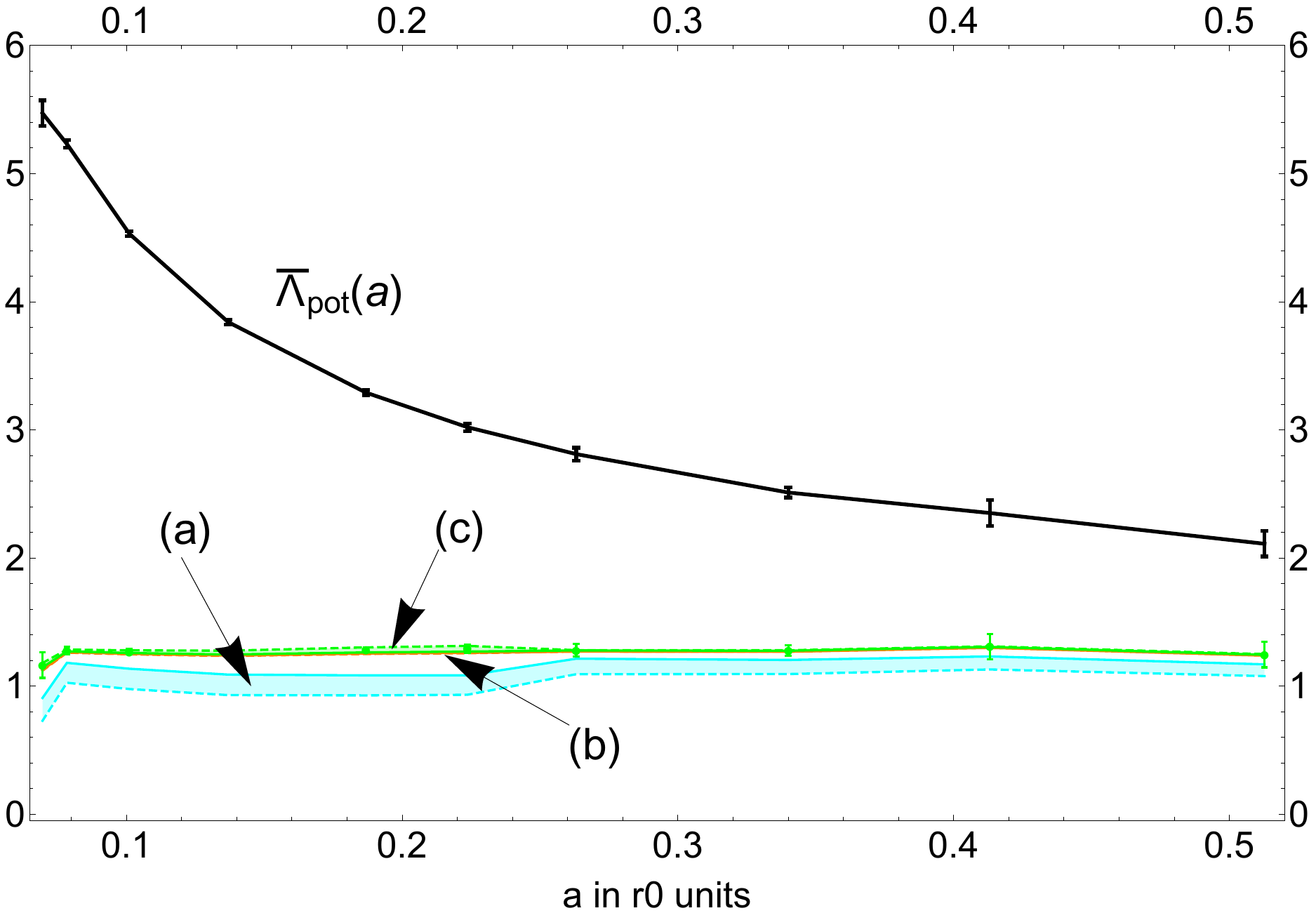}
\includegraphics[width=0.79\textwidth]{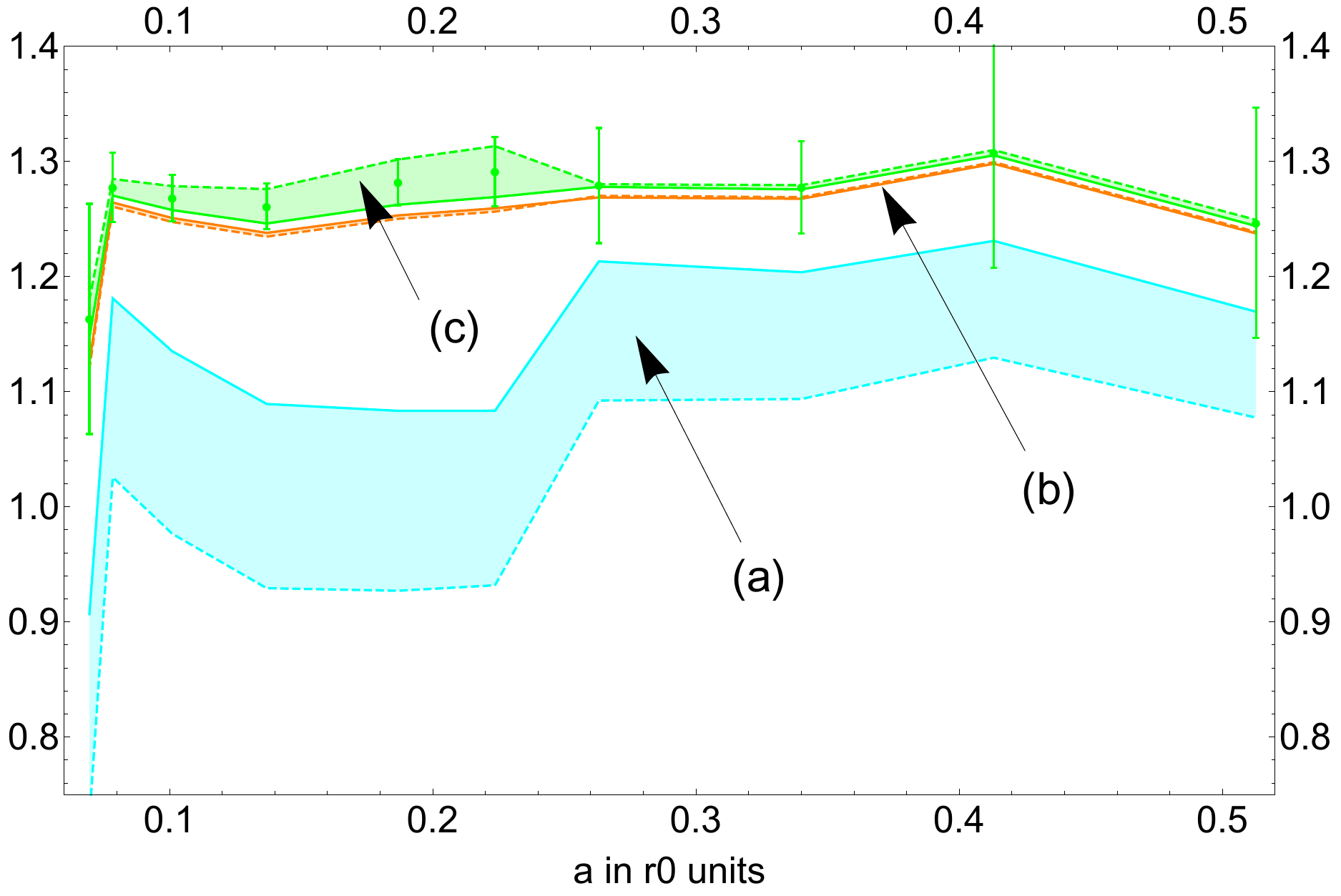}
\caption{{\bf Upper panel}: $\bar \Lambda_{pot}(a)$ is the Montecarlo lattice data \cite{statpot,Bali:1997am}, as analyzed in \cite{Bali:2003jq}. The continuous lines are drawn to guide the eye. The other lines correspond to \eq{LambdabarPot} truncated at different orders in the hyperasymptotic expansion: (a) $\bar \Lambda_{pot}(a)-\delta m_{P}(1/a)$, (b) $\bar \Lambda_{pot}(a)-\delta m_{P}(1/a)-\frac{1}{a}\Omega_m$, (c) $\bar \Lambda_{pot}(a)-\delta m_{P}(1/a)-\frac{1}{a}\Omega_m-\sum_{N_P+1}^{N'=2N_P}\frac{1}{a}[c_n-c_n^{\rm (as)}]\alpha^{n+1}$ (in this last case we include the error of the lattice points in the middle of the band). For each difference the bands are generated by the difference of the prediction produced by the smallest positive or negative possible values of $c$ that yield integer values for $N_P$. {\bf Lower panel}: As in the upper panel but in a smaller range. $r_0^{-1} \approx 400$ MeV.}
\label{Fig:barLambdaLatt2003}
\end{figure}
\end{center}

As an extra check of the method, we now consider the ground state energy of two static sources in the fundamental representation at a fixed distance $r_0$ computed in the lattice: $E_{\Sigma_g^+}(r_0;a)$. This object has the same renormalon as twice the pole mass. Following \cite{Bali:2003jq} we define the quantity
\be
\bar \Lambda_{pot}(a) \equiv \frac{E_{\Sigma_g^+}(r_0;a)}{2}+\Delta
\,,
\ee
where $\Delta$ is just a constant to fix the normalization at $r=r_0$. For $\bar \Lambda_{pot}(a)$ we perform an OPE assuming 
$r_0 \gg a$, and compute it using the PV prescription. We then have
\be
\bar \Lambda^{\rm PV}_{pot} \equiv \frac{E_{\Sigma_g^+}(r_0;a)}{2}+\Delta-\delta m^P_{\rm latt}-\frac{1}{a}\Omega_m-
\sum_{n=N_P+1}^{N'=2N_P}\frac{1}{a}[c_n-c_n^{\rm (as)}]\alpha^{n+1}+{\cal O}(a\lQ^2)
\,.
\label{LambdabarPot}
\ee 
We show the results in Fig. \ref{Fig:barLambdaLatt2003}. The lattice data is taken from \cite{statpot,Bali:1997am}, as analyzed in \cite{Bali:2003jq}. A nicely flat curve appears. 
This object does not show ${\cal O}(a\lQ^2)$ artifacts. This is consistent with the discussion in \cite{Necco:2001xg}, though there the discussion was only made for energy differences. This has the potentially important consequence that potentials computed with different $\beta$'s can be related with perturbation theory with good accuracy. There is no need to subtract independent constants for each $\beta \equiv 6/g^2$. 

\subsection{$\bar \Lambda_{\rm PV}$ from $B$ meson mass}
\label{Sec:Mesons}
We now move to the physical case with $n_f=3$ light quarks. Using HQET we approximate the $B/D$ meson mass by (we use spin averaged masses)
\be
\label{mB}
m_{B(D)}=m_{\rm PV}+\bar \Lambda_{\rm PV}+{\cal O}\left({1 \over m_{\rm PV}}\right)
\,.
\ee
It is not the aim of this paper to determine $\m_b$ (nor $\m_c$). We are rather interested to know the error associated to determinations of $m_{\rm PV}$ if $\m$ is known, and vice versa. We will then later use this analysis for the top quark mass determination. For this purpose we use its hyperasymptotic approximation 
\be
\label{eq:mbPV}
m_{\rm PV}(\m_{b/c})= m_{\rm P}(\m_{b/c})+\m_{b/c}\Omega_m+
\sum_{n=N_P+1}^{N'=2N_P}[r_n-r_n^{\rm (as)}]\alpha^{n+1}+\cdots\,.
\ee
 To make the error analysis we use the bottom case, and take $\m_b=4.186$ GeV from \cite{Peset:2018ria}. We obtain (we have added a -2 MeV to the relation between the $\MS$ bottom mass and the pole mass due to the charm quark \cite{Ayala:2014yxa})
\be
\label{mbPV}
m_{b,\rm PV}=4836(\mu)^{+8}_{-17}(Z_m)^{-11}_{+12}(\al)^{+8}_{-9}
\;{\rm MeV}
\,.
\ee
For the variation of $\mu$ we take the range $\mu \in (\m_b/2,2\m_b)$. For $Z_m$ we take $Z_m^{\MS}(n_f=3)=0.5626(260)$ from \cite{Ayala:2014yxa}. For the variation of $\alpha$ we take $\Lambda^{(n_f=3)}_{\MS}=332\pm17$ MeV from \cite{Tanabashi:2018oca}. The central value has been obtained with $N_P=3$ ($c=0.3611$). 
Therefore, the last term of \eq{eq:mbPV} is set to zero, as we do not have more terms of the perturbative expansion. In the counting of of \eq{SPVDN} the precision is then (1,0).  Within the hyperasymptotic counting the $\sum_{n=N_P+1}^{N'=2N_P}[r_n-r_n^{\rm (as)}]\alpha^{n+1}$ term should roughly scale as (assuming the next renormalon is located at $|u|=1$)
\be
\sim e^{ -\frac{2\pi}{\beta_0\alpha_X(\mu)}(1+\ln(2))}
\,.
\ee
This is the expected scaling if $\mu \sim \m$. Nevertheless, the dependence on $\mu$ will be quite different depending on whether the next renormalon is ultraviolet ($\sim \mu^{-2}$) or infrared ($\sim \mu^2$). Actually, the magnitude is also expected to be different, being more important for an eventual infrared renormalon. As the situation is somewhat uncertain we do not dwell further in this issue. To roughly estimate the size of subleading terms we could compute with  $N_P=2$ ($c=1.7935$). In the counting of of \eq{SPVDN} the precision is then $(1,N_{max}-N_P)$. The difference is below 1 MeV (after including $[r_3-r_3^{\rm (as)}]\alpha^{4}$, otherwise the difference is 7.5 MeV). Even computing with $N_P=1$, which formally allows us to reach the next renormalon located at $2N_P=2$ (i.e.  (1,$N_P$) precision in the counting of \eq{SPVDN}), the difference is $\sim 7$ MeV. 
Alternatively, the remaining $\mu$ scale dependence of $m_{\rm P}(\m_{b/c})-\m_{b/c}\Omega_m$ also gives a measure of  the uncomputed $\sum_{N_P+1}^{N'=2N_P}[r_n-r_n^{\rm (as)}]\alpha^{n+1}$ term, as such scale dependence should cancel in the total sum. We will then take it as the associated error. This is the first error quoted in \eq{LPVnumber}. Actually, the error associated to $Z_m$ is also a measure of the lack of knowledge of higher order terms in perturbation theory. Therefore, there is some degree of double counting by considering these two errors separately. 

It is interesting to analyze the error of the superasymptotic approximation of $m_{\rm PV}$ (only computing $m_P$). If we vary $\mu$ in the range $\mu \in (\m_b/2,2\m_b)$, we obtain ($N_P=3$, $c=0.3611$)
\be
m_{b,\rm P}=5077(\mu)^{+134}_{-242}\;{\rm MeV}
\,.
\ee
Nicely enough, it agrees with \eq{mbPV} within one sigma. This is also so if we take $N_P=2$ ($c=1.7935$): 
$m_{b,\rm P}=4922^{+107}_{-167}\;{\rm MeV}$. We find that the scale dependence of the superasymptotic approximation is large. Therefore the inclusion of $\m\Omega_m$ is crucial to make the result much more scale independent. On the other hand note that there is no error associated to $Z_m$ at this order (which is, in any case, comparatively small). 

Using \eq{mbPV} and \eq{mB} we can determine $\bar \Lambda_{\rm PV}$. We work at leading order in $1/m$. We obtain 
\be
\label{LPVnumber}
\bar \Lambda_{\rm PV}=477 (\mu)^{-8}_{+17}(Z_m)^{+11}_{-12}(\al)^{-8}_{+9}({\cal O}(1/m))^{+46}_{-46}
\;{\rm MeV}
\,,
\ee
where we have included an extra error source compared with \eq{mbPV}. This extra error is associated to the ${\cal O}(1/m)$ corrections. The existence or not of genuine NP $1/m$ corrections may introduce a significant error. In case they exist, if we take the hyperfine energy splitting as a measure of $1/m$ corrections, we find shifts from the central values of order $\sim 46$ MeV and $\sim 140$ MeV for $B$ and $D$ mesons respectively. As \eq{LPVnumber} has been obtained from the $B$ meson spin-average mass we conservatively estimate the error associated to genuine NP $1/m$ corrections to be of order $\sim 46$ MeV, as it is the most we can do from phenomenology and perturbation theory. Let us recall however that recent lattice simulations point to much smaller genuine NP $1/m$ corrections for the spin-independent average \cite{Bazavov:2018omf}. 

Earlier direct determinations of $\bar \Lambda_{\rm PV}$ or $m_{\rm PV}$ can be found in \cite{Lee:2003hh,Lee:2005hf}. The formulas are equivalent to those used here to one order less (using $N_P=N_{max}=2$). They also include less terms in the sum in \eq{rnas}. More recently, a determination of $\bar \Lambda$ has been obtained in \cite{Bazavov:2018omf} using lattice data. In this case the formulas are equivalent to those used here since $N_P=3$ (see Eq. (57) of Ref. \cite{HyperI}) except for the fact that the scale $\mu$ was always fixed equal to the heavy quark mass and that the mass was obtained in the MRS scheme \cite{Brambilla:2017hcq}. In this reference is also given the relation between the PV and the MRS mass. Using it we obtain (where we combine quadratically the error of $Z_m^{\MS}$ and $\Lambda_{\MS}$)
\be
\label{difference}
\bar \Lambda_{\rm PV}-\bar \Lambda_{\rm MRS}=\cos(\pi b)\frac{4\pi \Gamma(-b)}{2^{1+b}\beta_0}Z_m^X\Lambda_X\Bigg|_{n_f=3}=
%-4\pi*9.55631 \sim 
-120(8)\; {\rm MeV}
\,.
\ee
The prediction of \cite{Bazavov:2018omf} translates then to $\bar \Lambda_{\rm PV}=435(31)$, where we only include the error quoted in \cite{Bazavov:2018omf}. In particular, we do not include the error in \eq{difference}. Note that \eq{difference} scales like ${\cal O}(\lQ)$, whereas $\m\Omega_m$ scales like ${\cal O}(\sqrt{\alpha}\lQ)$. There is a  40 MeV difference with the number given in \eq{LPVnumber}. 10 MeV can be understood because the value of $\m_b$ used in \cite{Bazavov:2018omf} is around 10 MeV bigger. Another 10 MeV can be understood by the inclusion of $1/m$ nonperturbative effects. The remaining 20 MeV difference are more difficult to identify, though they are well inside uncertainties. Leaving aside the different $\alpha$'s used, another source of difference is the value of $Z_m$. The value used in \cite{Bazavov:2018omf} comes from \cite{Komijani:2017vep} (where the effect of scale variation was not included in the error analysis). This determination used a sum rule that is free of the leading pole mass renormalon. The possibility of using sum rules to determine the normalization of renormalons was first considered in \cite{Lee:1996yk}. For the determination of $Z_m$, sum rules were first used in \cite{Pineda:2001zq}. Later sum rule analyses can be found in \cite{Hoang:2017suc}.
 Alternatively one can use the ratio of the exact and asymptotic expression of the coefficients $r_n$ to determine $Z_m$ as in \cite{Bauer:2011ws,Bali:2013pla,Bali:2013qla,Ayala:2014yxa,Beneke:2016cbu}.
For an extra discussion on this issue see \cite{Pineda:2017uby}. Finally, it is worth mentioning that $Z_m$ can be determined either from the static potential or from the pole mass (and its relatives). The only value of $Z_m$ that uses the static potential is from \cite{Ayala:2014yxa}. A preference for determinations of $Z_m$ from the static potential can be theoretically motivated, as they are less affected by subleading renormalons. There are no ultraviolet renormalons and the next infrared renormalon is located at $u=3/2$. On the other hand, the pole mass is expected to have renormalons at $|u|=1$. Only in the event that there is no $u=1$ renormalon and the effect of the $u=-1$ renormalon is subleading both determinations would be on equal footing on theoretical grounds. In any case, irrespectively of this discussion, consistent numbers are obtained between different analyses. 

\subsection{$(N,\mu) \rightarrow \infty$. \eq{eq:muinfty}. Case 2B)}
All previous determinations of $\bar \Lambda_{\rm PV}$ have been obtained using limit 1). 
For completeness we have also explored how limit 2B) performs for $\bar \Lambda_{\rm PV}$, even though it is, in principle, less precise. We have considered the different methods to take the limit 2B) discussed at the end of \Sec{Sec:mPV}, and compared with the numbers obtained above. We first consider the evaluation of $m_{\rm PV}$ using the right hand side of \eq{eq:limit} with \eq{eq:Borelapprox}. The central value is determined using $c'_{min}=1.076
%1.076364808205274
$, which is the value that makes $K^{(A)}_{\MS}=0$, and $\mu_0=\m_b=4.186$ MeV. We obtain $\bar \Lambda_{\rm PV}=453$ MeV. The difference with \eq{LPVnumber} is 24 MeV, which is quite reasonable. We can also explore the $\mu$ dependence. Taking the variation $\mu_0 \in (\m_b/2,2\m_b)$, we obtain $\bar \Lambda_{\rm PV}=453^{-36}_{+55}(\mu)$ MeV. Comparatively with \eq{LPVnumber} the $\mu$ scale dependence is much larger. We next consider the limit as taken in \eq{1loop}. This requires the knowledge of the coefficients $r_n$ to all orders. For $n>3$ we take the asymptotic expression. On the other hand the running of the beta function is only needed to one loop. This allows us to go to orders as high as $N_A=3000$ (though it already converges at smaller values of $N_A$). Remarkably enough, we obtain the same result than before: 453 MeV. There is a residual dependence on $c'$. For illustration, if we take instead $c'=2$, we obtain $\bar \Lambda_{\rm PV}=438$ MeV  (the result using the right-hand side of \eq{eq:limit} with \eq{eq:Borelapprox} yields the same number), 
and the scale dependence is larger: $\bar \Lambda_{\rm PV}=438^{-55}_{+99}(\mu)$. The value of $c'$  we have used to make the analysis can be an issue. As discussed in \cite{VanAcoleyen:2003gc,HyperI}, taking $\chi-2$ very small deteriorates the convergence and larger values of $N_A$ are needed. This problem aminorates by taking larger values of $c'$. Since for the limit as taken in \eq{1loop} we can go to very large $N_A$ this is not a problem. We have also performed a similar analysis with $n_f=0$ and $r_0$ units, relevant for the analyses performed in \Sec{Sec:LambdaLatt}. The discussion follows parallel to the one we just had with the difference that we now know 20 coefficients of the perturbative expansion. The value we obtain: $\bar \Lambda_{\rm PV} = 1.37\, r_0^{-1}$ (using a quadratic fit) is indeed quite close to the value obtained in \Sec{Sec:LambdaLatt}, though less precise.

We have more problems with the other ways to take the $\mu \rightarrow \infty$ limit discussed at the end of \Sec{Sec:mPV}. The direct use of $N_A$ in \eq{eq:muinfty} or of $N_A$ in Eq. (67) in \cite{HyperI} requires, besides the coefficients $r_n$ to all orders, the $\beta$-function coefficients to all orders as well. We do not have them. Instead we use truncated version of the $\beta$ function. This makes the numerical calculation much more challenging, since the running in $\mu$ is more complicated. Therefore, we had problems to go to very large $N_A$. For $N_A \geq 200$ we find instabilities is some cases. As mentioned before, the value of $c'$  we use to make the analysis can be an issue. 
Taking $\chi-2$ very small deteriorates the convergence. This problem aminorates by taking larger values of $c'$. In the lattice scheme determination of quenched $\bar \Lambda_{\rm PV}$ we indeed observe convergence to the value obtained before using $c'=2$. Using $c'_{min}=1.076$ the convergence is less good. Determinations in the $\MS$ scheme do not show convergence if we stop at $N_A \leq 200$, though with an slight better behavior using $c'=2$ rather than $c'_{min}$. Overall, as the precision we get with method 2B) is worse than with method 1), we will not study this limit in more detail. 

\section{Top mass}
\label{Sec:Top}

\subsection{About the pole mass ambiguity}

The top quark mass is one of the key parameters of the standard model. A lot of experimental work has been devoted to its determination (see for instance \cite{ATLAS:2014wva,Khachatryan:2015hba,Aaboud:2016igd}). Whereas this is a matter of debate, it is typically assumed that the masses obtained from experiment correspond to the pole mass. Thus, there has been an ongoing discussion on the intrinsic uncertainty of these determinations (see for instance \cite{Beneke:2016cbu,Hoang:2017btd}, and \cite{Corcella:2019tgt} for a more recent discussion). We believe that, without further qualifications, the question is ill posed, or may lead to confusion. It is well known that the pole mass is well defined (infrared finite and gauge independent) at finite (albeit arbitrary) order in perturbation theory \cite{Kronfeld:1998di}. It is also well known that such series is divergent\footnote{Actually this is only proven in the large $\beta_0$ approximation \cite{Ball:1995ni,Neubert:1994vb}, and it is also supported by numerical analyses \cite{Bauer:2011ws,Bali:2013pla,Bali:2013qla}, but there is no analytic proof.}. Therefore, no numerical value can be assigned to the infinite sum of the perturbative series of the pole mass. Truncated sums are well defined but depend on the order of truncation (a detailed discussion relevant for the analysis made in the present paper can be found in \cite{HyperI}). These truncated sums can be related with observables or with intermediate definitions of the heavy quark mass, like the PV mass (which regulates via Borel resummation the infinite sum), in a well-defined way. 

In this context, the shortest answer to the above posed question is that the ambiguity (of a well-defined mass) is {\it zero}. As a matter of principle, $m_{\rm PV}$ (or $m_P$) can be {\it defined} with arbitrary accuracy (this also applies to any threshold mass), if one computes high enough orders of the perturbative series, and if $\m$ is given. One can discuss (actually one can compute) the scheme/scale dependence (if they have) of them. In this respect, there is no much conceptual difference with respect to asking about the scheme/scale dependence of minimal subtraction schemes for the heavy quark masses. 

A quite a different question is to determine the typical difference (that not ambiguity) between (reasonable) different definitions of the pole mass. The short answer to this question is that the differences are (at most) of order $\lQ$ for (reasonable) different definitions of the pole mass. We emphasize that one can not be more precise unless stating the specific definition used for the pole mass. For instance, the difference between $m_{\rm PV}$ and $m_P$ is of ${\cal O}(\sqrt{\al}\lQ)$ with a known prefactor. Truncating the perturbative series at order $N$ near $N^*$ are also legitimate definitions of the pole mass. The typical difference between truncating at different $N$ is of order $\lQ$: see for instance Eq. (62) of \cite{Bali:2013pla}. One could even use $M_B$ as a definition for the pole mass. Its difference with $m_{\rm PV}$ is of order $\lQ$. If one defines an imaginary mass by doing the Borel integral just above the positive real axis, the
 difference with $m_{\rm PV}$ is of ${\cal O}(i\lQ)$. The authors of \cite{Beneke:2016cbu} choose to divide this number by $\pi$ and take the modulus as their definition of the ambiguity. These examples illustrate that, even if the ambiguity is of ${\cal O}(\lQ)$, the coefficient multiplying $\lQ$ is arbitrary. Overall, it should be clear that no much more can be said, and we are indeed against of dwelling too much on this issue. Instead, we strongly advocate to avoid generic discussions about the pole mass, which is not well defined beyond perturbation theory, and restrict the discussion to the precision and errors of specific, NP well-defined, heavy quark masses the perturbative expansion of which can be related with the perturbative expansion of the pole mass. 

Once working with NP well-defined heavy quark masses like $m_{t,\rm PV}$ or $m_{t,P}$, we can address the more relevant question of  determining the precision with which $\m_{t}$ can be determined if $m_{t,\rm PV}$ or $m_{t, P}$ is known (and vice versa, if $\m_t$ is known what is the uncertainty of $m_{\rm PV}$) with nowadays knowledge of the perturbative expansion. In other words, with which precision the theoretical expression is known. For reference we will take the value $\m_t=163$ GeV in the following. We will see in the next section that indeed the precision is quite good and that the error is significantly smaller than typical numbers assigned for the ambiguity of the pole mass. We will not dwell in this paper on the precision with which $m_{t, \rm PV}$ or $m_{t,P}$ can be {\it determined} from experiment as such discussion is observable dependent. 

\subsection{Decoupling and running}

We now turn to an issue specific to the top quark (as compared with the bottom and charm quark). The top quark mass is much larger than $\lQ$. The latter is the scale that characterizes renormalon associated effects and it is the precision we want to achieve. This obviously generates ratios of quite disparate scales. In the context of threshold masses with an explicit infrared cutoff $\nu_f$, this calls for resummation of the large logarithms: $\ln \nu_f/m_t$. This is possible, and first done in \cite{Bali:2003jq} in the RS scheme (see also \cite{Pineda:2017uby} for an extra discussion on this issue). Here, we approach the problem in a different way. We want to work with expansions for the perturbative series of the pole mass truncated at the minimal term: $m_P$, and to improve upon it using hyperasymptotic expansions. Nevertheless, at the scale of the top mass, we do not have enough terms to reach the asymptotic behavior of the perturbative expansion. We use instead that the top quark pole mass and the pole mass of a fictitious top quark with mass $m_t'$ share the same leading infrared renormalon. Therefore, the leading infrared renormalon cancels in the difference. We can then decrease the top mass in a renormalon free way until we reach a top mass low enough that we can use the hyperasymptotic expansion. Such renormalon free running is determined by the following function (not compulsory to take $\mu=\m$ but it simplifies the computation) 
\be
\label{eq:calF}
{\cal F}(\m,n_f) \equiv \frac{d}{d \m}(m_{\rm PV}(\m)-\m)\simeq
\frac{d}{d \m}\sum_{n=0}r^{(n_f)}_n(\m;\mu=\m)\al_{(n_f)}^{n+1}(\m)
\equiv
\sum_{n=1}^{N+1}
f_n(\m)
\left(\frac{\al_{(n_f)}(\m)}{\pi}\right)^n
\,.
\ee
This formula is correct up to $N \sim 2N_P$, since $\m\Omega_m$ and $r_n^{(as)}$ are independent of $\m$ (see \eq{mPV}), so that their derivative with respect to $\m$ vanishes. The coefficients $r_n^{(n_f)}$ are evaluated for $n_f$ massless particles. In the context of the MSR threshold mass the running is implemented in a similar way (see, for instance, \cite{Hoang:2017suc}).  \eq{eq:calF} makes explicit that such running is just a natural consequence of the relation between observables and their OPEs (for illustration, it follows from the fact that $M_B-M_D$, the $B$ minus $D$ meson mass difference is free of the leading infrared renormalon), and not linked to an specific threshold mass definition.

There is still another issue specific to the top quark: there are two heavy quarks (the bottom and charm), with masses much larger than $\lQ$, that generate extra corrections to the pole-$\MS$ mass relation due to the finite mass of the bottom and charm quark. Therefore, we have for $\m \sim \m_t$ 
\be
m_{\rm PV}(\m)=\m+\sum_{n=0}^{N_{max}}r^{(n_f)}_n(\m;\mu=\m)\al_{(n_f)}^{n+1}(\m)
+\delta m_b^{(n_f)}(\m)+\delta m_c^{(n_f)}(\m)+\delta m_{bc}^{(n_f)}(\m)
\,,
\ee
where it is implicit that $N_{max}$ (the number of known terms of the perturbative expansion) is not large enough to see the decoupling of the bottom nor charm and certainly $N_{max}<N_P$. 
$n_f$ stands for the number of active flavours. At the top mass scale we take $n_f=5$. The ${\cal O}(\al^2)$ term of 
$\delta m_Q^{(n_f)}$ was computed in \cite{Gray:1990yh} and the ${\cal O}(\al^3)$ term in \cite{Bekavac:2007tk}. Note as well that at ${\cal O}(\al^3)$ there is a new contribution including a vacuum polarization of the bottom and charm at the same time. We name it $\delta m^{(n_f)}_{bc}$ and it has been computed in \cite{Hoang:2017btd}. 

As we decrease the value of $\m_t$ the bottom and charm quark will decouple. This decoupling will be absorbed in $\delta m^{(n_f)}_{b/c/bc}$, which are polynomials in powers of $\al^{(n_f)}$. In general this is not just changing $n_f$ in the original expressions  from $n_f=5$ to $n_f=4$ or 3. The explicit expressions can be found in the Appendix \ref{Sec:light}. 
 
The renormalon is associated to scales smaller than the bottom and charm quark masses. Therefore, such scales should be decoupled before we talk about the hyperasymptotic expansion. As we have mentioned above we do such decoupling by varying the mass of the top till reaching a fictitious top with a mass small enough such that, first the bottom, and later the charm, decouple. Overall, our final formula is the following:
\bea
m_{\rm PV}(\m_t)=\m_{t}&+& 
\int^{\m_t}_{\mu_b} d \m
\left({\cal F}(\m,5)+\frac{d}{d \m}(\delta m_b^{(5)}(\m)+\delta m_c^{(5)}(\m)+\delta m_{(bc)}^{(5)}(\m))\right)
\nn
\\
&+&
\int^{\mu_b}_{\mu_c} d \m 
\left({\cal F}(\m,4)+\frac{d}{d \m}(\delta m_b^{(4)}(\m)+\delta m_c^{(4)}(\m)+\delta m_{(bc)}^{(4)}(\m))\right)
\nn
\\
&&
\qquad
\qquad
+m_{\rm PV}(\mu_c)-\mu_c
\,.
\label{mPVt}
\eea
We emphasize that ${\cal F}(\m,n_f)$  is expanded in powers of $\al$ before integration. 
We take $\mu_b$ small enough such that the bottom decouples and $\mu_c$ small enough such that the bottom and charm decouple, and also such that we reach the asymptotic limit of the pole-$\MS$ mass perturbative expansion with the existing known coefficients. Therefore, 
\be
\label{mPVtp}
m_{\rm PV}(\mu_c)=m_{P}(\mu_c) +\mu_c \Omega_m
+\delta m^{(3)}_b(\mu_c)+\delta m^{(3)}_c(\mu_c)+\delta m^{(3)}_{(bc)}(\mu_c)
+{\cal O}(\mu_c e^{ -\frac{2\pi}{\beta_0\alpha(\mu_c)}(1+\ln(2))})
\,.
\ee
The ${\cal O}(\mu_c e^{ -\frac{2\pi}{\beta_0\alpha(\mu_c)}(1+\ln(2))})$ term stands for subleading corrections in the hyperasymptotic expansions, which are not known. 

\begin{center}
\begin{figure}
\includegraphics[width=0.75\textwidth]{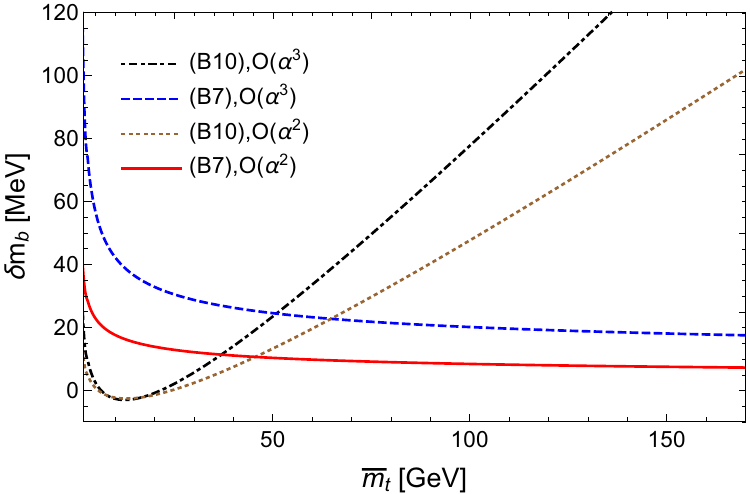}
\includegraphics[width=0.75\textwidth]{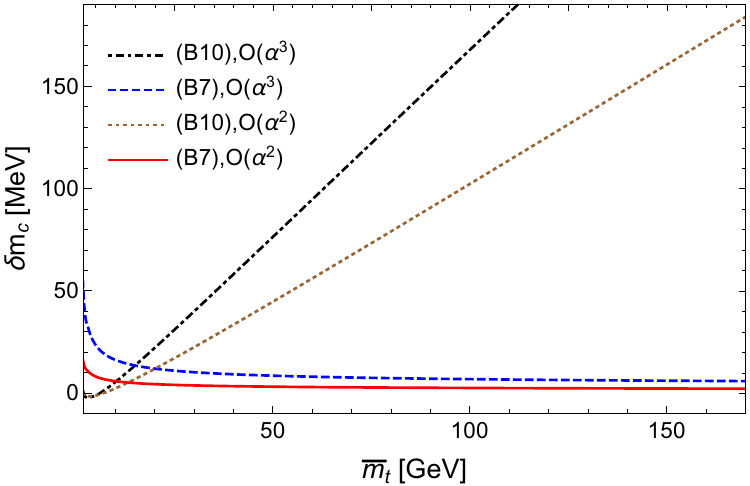}
\caption{{\bf Upper panel}: Plot of the correction to the PV mass of a top mass with varying $\m_t$ mass due to a heavy quark with $\MS$ mass equal to 4.185 GeV (bottom)  with and without decoupling (assuming a single heavy quark). {\bf Lower panel}: As in the upper panel with a heavy quark with $\MS$ equal to 1.223 GeV (charm).  We use \eqs{A7}{A10}.}
\label{Fig:deltamQ}
\end{figure}
\end{center}
Let us now discuss in more detail the dependence on the bottom and charm quark, in particular the effects associated to the fact that they have masses much bigger than $\lQ$ (for the analysis we take $\m_b=4.186$ GeV and $\m_c=1.223$ GeV \cite{Peset:2018ria} but the sensitivity to the specific values we use is very tiny). As already discussed in \cite{Ball:1995ni}, the natural scale of a $n$-loop integral is not $\m_t$ but $\m_te^{-n}$. For the case of the bottom versus charm quark it was observed in \cite{Ayala:2014yxa}\footnote{In that reference MeV should read GeV instead from Eq. (8) to Eq. (12).}  that the charm quark effectively decouples at order $\alpha^{2}/\alpha^3$ for the case of the charm quark effects in the bottom pole mass-$\MS$ mass relation. If we lower the mass of the top we can also observe at which scales it is more convenient to decouple the bottom and charm quark in the top pole mass-$\MS$ mass relation. This can be illustrated in Fig. \ref{Fig:deltamQ}, where we plot the corrections associated to the bottom and charm with and without decoupling in terms of the fictitious top mass (assuming a single heavy quark). Obviously for very large top masses it is not convenient to do the decoupling. Nevertheless, as we decrease the mass of the top it becomes much more effective to decouple, first the bottom, and afterwards the charm quark. Once this is done the corrections due to the bottom and charm masses to \eq{mPVtp} are very small. Comparatively to other errors, the uncertainty associated to the ${\cal O}(\alpha^4)$ corrections is negligible. Also the correction associated to the bottom and charm quark masses to \eq{mPVt} is, comparatively to the total running, very small.  From this analysis we will take as central values $\mu_b=20$ GeV and $\mu_c=5$ GeV. For these values we obtain
\bea 
&&
\int^{\m_t}_{\mu_b} d \m
\frac{d}{d \m}(\delta m_b^{(5)}(\m)+\delta m_c^{(5)}(\m)+\delta m_{(bc)}^{(5)}(\m))
\nn
\\
&+&
\int^{\mu_b}_{\mu_c} d \m 
\frac{d}{d \m}(\delta m_b^{(4)}(\m)+\delta m_c^{(4)}(\m)+\delta m_{(bc)}^{(4)}(\m))
\nn
\\
&+&
\delta m^{(3)}_b(\mu_c)+\delta m^{(3)}_c(\mu_c)+\delta m^{(3)}_{(bc)}(\mu_c)
=
-2.5\Bigg|_{{\cal O}(\alpha^2)}+0.8\Bigg|_{{\cal O}(\alpha^3)}=-1.7 \; {\rm MeV}
\,.
\label{eq:deltamTot}
\eea
The specific value depends on $\mu_b$ and $\mu_c$ but the good convergence and smallness of this correction holds true for other values of $\mu_b$ and $\mu_c$. The implementation of the decoupling of the bottom and charm in \cite{Hoang:2017btd} produces a much larger correction. An even larger effect is observed in the implementation performed in \cite{Beneke:2016cbu}, where the perturbative expansion is always performed at the scale of the top mass (using renormalon based estimates for the higher order coefficients), decoupling the bottom, and later the charm, depending on the order of perturbation theory. Therefore, we take our numbers as optima, and the error negligible compared with other uncertainties. 

\begin{center}
\begin{figure}
\includegraphics[width=0.75\textwidth]{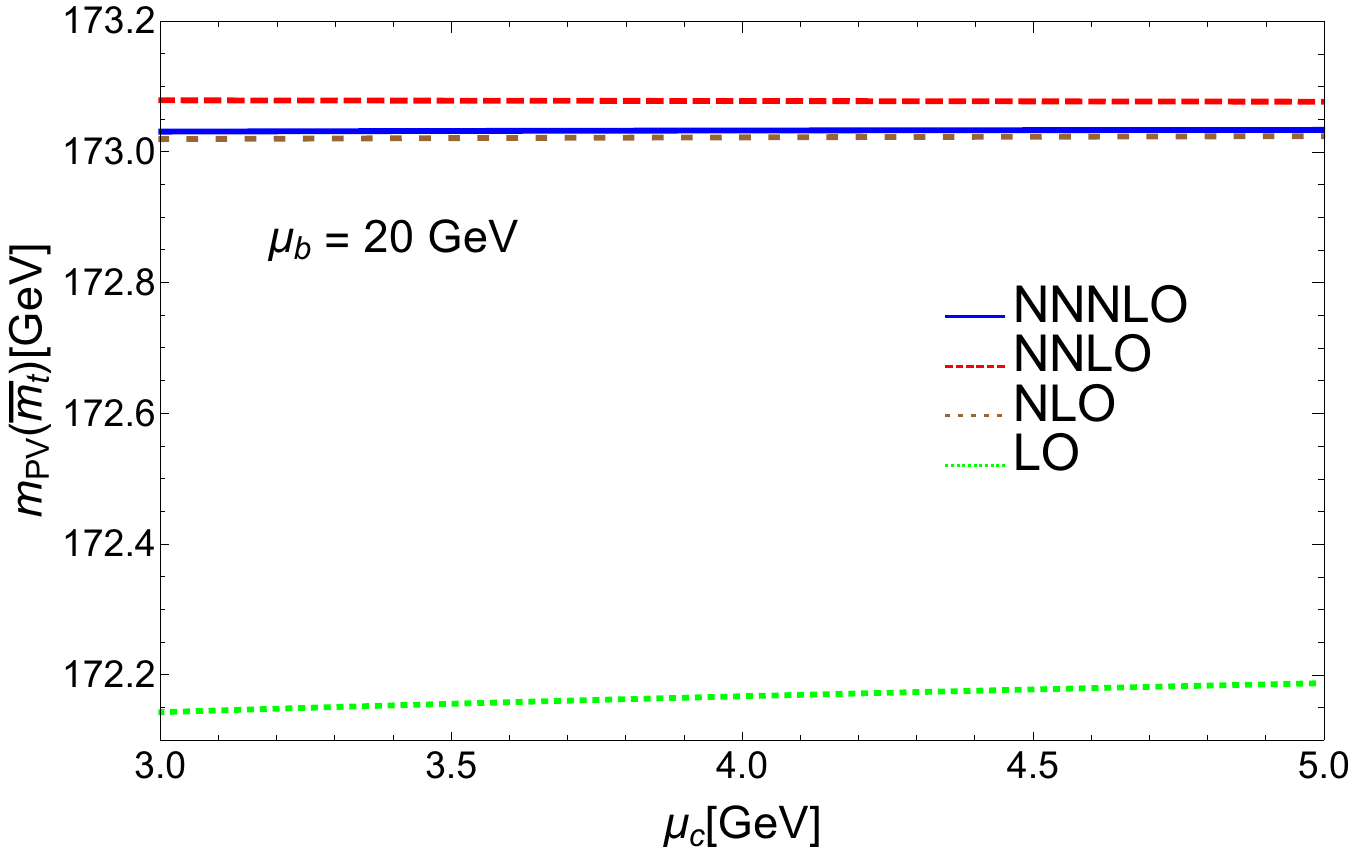}
\includegraphics[width=0.75\textwidth]{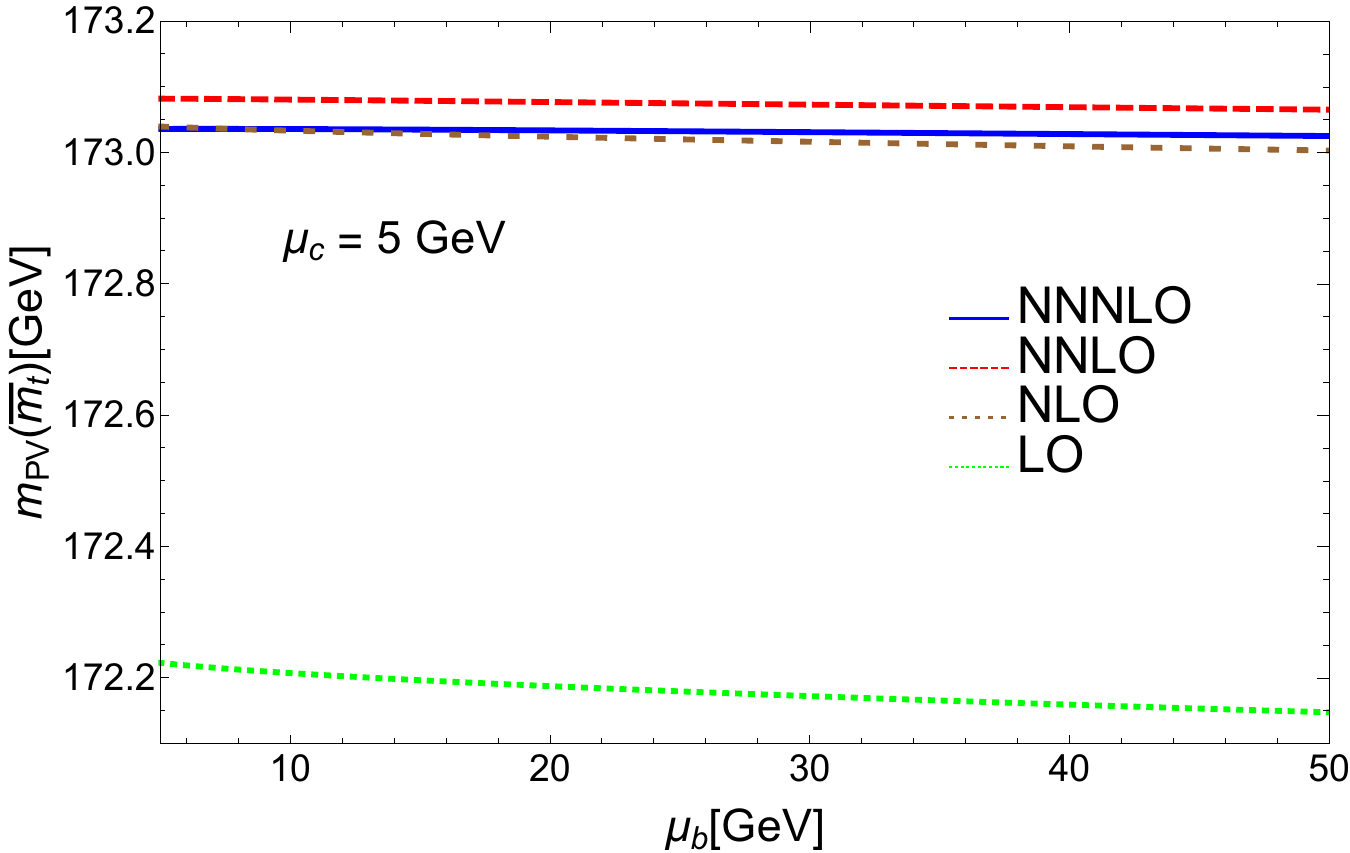}
\caption{Plots of \eq{mPVt} in terms of $\mu_b$ (upper panel) and $\mu_c$ (lower panel) truncating the perturbative expansion of ${\cal F}(\m,n_f)$ at different orders in $\al$ in \eq{calFnum}. In the upper figure we set $\mu_c=5$ GeV. In the lower figure we set $\mu_b=20$ GeV.}
\label{Fig:mPVt}
\end{figure}
\end{center}

We next explore the convergence pattern of the perturbative expansion. We first consider the perturbative expansion associated to ${\cal F}$. We find
\be
\label{calFnum}
\int^{\m_t}_{\mu_b} d \m {\cal F}(\m,5)
+
\int^{\mu_b}_{\mu_c} d \m 
{\cal F}(\m,4)=8445+837+53-43=9291(22) \; {\rm MeV}
\,.
\ee
We observe a convergent pattern. For the last two terms the convergence deteriorates. On the other hand the perturbative expansion becomes sign alternating. This may indicate sensitivity to the $u=-1$ renormalon.  We discuss this further in the next section. For sign-alternating asymptotic perturbative expansion the left-over is $\sim -1/2 \times$(the last computed term) (see \cite{Dingle}).\footnote{We emphasize that these arguments do not apply to IR renormalons (and in particular to the $u=1/2$ renormalon).} Therefore, we take it as the error of the truncation of the perturbative expansion, which is the error we quote in \eq{calFnum}. We also explore the dependence of \eq{mPVt} on $\mu_b$ and $\mu_c$. The dependence is very small, as we can see in Fig. \ref{Fig:mPVt}. For $\mu_c$ the variation is negligible, and for $\mu_b$ one gets variations of $\sim 5$ MeV for a central value of $\mu_b$ or around 20 GeV. Therefore, we will neglect it for the total error budget. 

The other source of error is associated to the approximate determination of \eq{mPVtp} (except for the $\delta m_q$ terms, which have already been taken into account in \eq{eq:deltamTot}). The error analysis is equal to the one in \eq{LPVnumber} adapted by changing $\m_b=4.186\; {\rm GeV} \rightarrow \mu_c=5$ GeV (the error associated to $\alpha$ is only computed for the full \eq{mPVt})
\be
\label{mPVtopnumber}
\left(m_P(\mu_c)+\mu_c\Omega_m\right)\Bigg|_{\mu_c=5\; {\rm GeV}}=5744 (\mu)^{+7}_{-15}(Z_m)^{+9}_{-9}
\; {\rm MeV}
\,.
\ee
 
Finally, we also include the error associated to $\alpha$. Combining all errors we obtain 
\be
\label{mtPVfinalnumber}
m_{t,\rm PV}(163{\rm MeV})=173033 ({\rm h.o.})^{+22}_{-22} (\mu)^{+7}_{-15}(Z_m)^{+9}_{-9}(\alpha)^{+119}_{-123}\;{\rm MeV}
\,.
\ee
By far the largest uncertainty is associated to $\al$. For the purely theoretical error budget, the error is associated to higher order corrections in perturbation theory. They show up in different ways. One is the approximate knowledge of $Z_m$, which shows up in $\Omega_m$. The other is the error in $\mu$, which is a measure of the ${\cal O}(e^{ -\frac{2\pi}{\beta_0\alpha_X(\mu)}(1+\ln(2))})$ corrections to \eq{mPVtp}.  h.o. stands for the error associated to higher order terms in perturbation theory of \eq{calFnum}.  All these errors would profit from higher order perturbative computations. 
We have also explored other sources of uncertainty, and find them to be comparatively very small: the error (and the effect) associated to the finite mass of the bottom and charm quark is found to be very small, and similarly for variations in the values of $\mu_b$ and $\mu_c$.

It is also useful to make the error estimate of the ratio of the PV and $\MS$ top mass. We obtain ($\m_t=163$ GeV)
\be
\label{eq:mtnum}
[\frac{m_{t,\rm PV}}{\m_t}-1]\times 10^5= 6155 ({\rm h.o.})^{+13}_{-13} (\mu)^{+4}_{-9}(Z_m)^{+6}_{-6}(\alpha)^{+73}_{-75}
\,.
\ee
Note that there is no ambiguity error associated to this number. Except for $\alpha$ all errors are associated to the lack of knowledge of higher order terms of the perturbative expansion. In comparison with \cite{Beneke:2016cbu} we find that our result is less sensitive to $Z_m$ and to its associated error. 

\subsection{$|u|=1$ renormalons}

The perturbative expansion of ${\cal F}(\m,n_f)$ is free of the $u=1/2$ renormalon. Therefore, it is the ideal object on which to study the subleading renormalons of the pole mass. In principle, these are located at $u=1$ and $u=-1$. The existence of an infrared renormalon at $u=1$ has been a matter of debate \cite{Neubert:1996zy}. The existence of an ultraviolet renormalon at $u=-1$ can be established in the large $\beta_0$ approximation \cite{Ball:1995ni,Neubert:1994vb} but not beyond. With respect to this discussion some interesting observations can be drawn out of our analysis. The coefficients $f_n$ show an interesting dependence in $n_f$ (with changes of sign of different powers of $n_f$). 
In Table \ref{Tab:alCalF} we give the numbers of $f_n$ for different values of $n_f$ and also in the large $\beta_0$ approximation. We observe that for $n_f=3$ the ${\cal O}(\al^4)$ flips sign. For $n_f=6$, the ${\cal O}(\al^3)$ and ${\cal O}(\al^4)$ flip sign.  The situation is somewhat puzzling. Let us first note that the sign of the coefficients would be interchanged compared with the large $\beta_0$ predictions (for $n_f=3$). This could still be understood from a $u=-1$ renormalon if $Z_{-2}^X$ flips sign from the large $\beta_0$ prediction to real QCD. This would indicate a large dependence of $Z_{-2}^X$ on $n_f$ compared with what has been seen for $Z_m^X$, where the large $\beta_0$ approximation gave the right sign and order of magnitude. For $n_f \rightarrow \infty$, the results agree with QED expectations ($\beta_0$ becomes negative and the perturbative series is non sign-alternating). For $n_f=6$ we observe that the last two terms are negative. One may then wonder if what we are seeing for $n_f=6$ (and maybe also for $n_f=3$) is that the $u=-1$ renormalon becomes effectively infrared. Obviously, we need higher order coefficients $f_n$ to clarify this issue.\footnote{The coefficients of the perturbative expansion of the pole mass itself are also a polynomial in powers of $n_f$. The sign dependence of the different powers of $n_f$ has been studied in \cite{Kataev:2018gle,Kataev:2018mob}.}  

 \begin{table}
\begin{tabular}{|c|c|c|c|c|c|}
\hline
${\cal F}(\m,n_f)$
&  $f_1$ & $f_2$ & $f_3$ & $f_4$ & $f_5$\\
\hline
$n_f=0$ &$4/3$ & 
$
6.11$ & $25.52$ & 
$18.46$ &\\
\hline
$n_f=3$ &$4/3$ & 
$4.32$ & $12.76$ & 
$-63.37$ &\\
\hline
$n_f=6$ &$4/3$ & 
$2.53$ & $-0.74$ & 
$-105.70$ &\\
\hline
(Large $\beta_0$/exact) $n_f=10^{20}$  & $4/3$ & $-5.97\times 10^{19}$ & $-4.15\times 10^{38}$ & $-2.54\times 10^{58}$ & $-5.09 \times 10^{77}$\\
\hline
(Large $\beta_0$) $n_f=0$ &$4/3$ & 
$9.85$ & $-11.31$ & 
$114.33$ &-377.22\\
\hline
(Large $\beta_0$) $n_f=3$ &$4/3$ & 
$8.06$ & $-7.57$ & 
$62.62$ &-169.04\\
\hline
(Large $\beta_0$) $n_f=6$ &$4/3$ & 
$6.27$ & $-4.58$ & 
$29.46$ &-61.86\\
\hline
\end{tabular}
\caption{The coefficients $f_n$ of ${\cal F}(\m,n_f)$. Note that $f_4(n_f=0)$ has a $9\%$ error from the determination in \cite{Marquard:2015qpa}. The $n_f=10^{20}$ case is used as a test for comparison with the large $\beta_0$. The last three (four) rows are the coefficients $f_n$ in the large $\beta_0$ approximation.}
\label{Tab:alCalF}
\end{table}

It is usual lore that infrared renormalons dominate over ultraviolet ones (this is somewhat based on large $\beta_0$ analyses where ultraviolet renormalons are typically suppressed by the factor $\sim e^{d \frac{c_X}{2}}$ whereas infrared renormalons are enhanced by the factor $\sim e^{-d \frac{c_X}{2}}$). If we take this seriously, and also the numbers we obtain for $f_n$ as an indication of the existence of the $u=-1$ renormalon, this may indicate that the $u=1$ renormalon is indeed zero. In this respect, it is worth mentioning the analysis of \cite{Bazavov:2018omf} where the NP correction associated to the $u=1$ renormalon was found to be zero within errors. This is consistent with this discussion.

On the theoretical side it is also interesting to see where the $u=1$ renormalon would show up in a perturbative computation of the heavy quarkonium mass. For the purposes of this discussion, the heavy quarkonium mass would read 
\be
M_{nl}=2m_Q+\langle \frac{{\bf p}^2}{m_Q} \rangle_{nl}+\langle V_0 \rangle_{nl}+\langle \frac{V_1}{m_Q}\rangle_{nl}+{\cal O}\left(\frac{1}{m_Q^2}\right)
\,,
\ee 
where $V_0$ is the static potential, and $V_1$ is the $1/m_Q$ potential. OPE analyses in the static limit show that $V_0$ does not have renormalon at $u=1$. The virial theorem: $\langle \frac{{\bf p}^2}{m_Q} \rangle_{nl}=\langle r V'_0 \rangle_{nl}$, also guaranties that the kinetic term does not have such $u=1$ renormalon. Therefore, any possible $u=1$ infrared renormalon of the pole mass should cancel with the analogous infrared renormalon of the $V_1/m_Q$ potential. The fact that the latter can be written in a closed way in terms of Wilson loops \cite{Brambilla:2000gk} may open a venue on which to study this issue in further detail. This is postponed to future work.  
 
\section{Conclusions}

In this paper we have constructed hyperasymptotic expansions for the heavy quark pole mass (and for associated quantities) regulated using the PV prescription along the lines of \cite{HyperI}. We generalize the discussion of that reference by including possible ultraviolet renormalons. Such organization of the computation allows us to have a parametric control of the error committed when truncating the hyperasymptotic expansion. 

In Sec. \ref{Sec:mOS} the hyperasymptotic expansion of the pole mass of a heavy quark in the large $\beta_0$ is computed. We use it as a toy-model observable to test our methods. It works as expected. We can see the $u=1/2$ infrared renormalon and the $u=-1$ ultraviolet renormalon. The next infrared renormalon is located at $u=3/2$. Compared with the static potential case studied in \cite{HyperI} in the large $\beta_0$ approximation, infrared renormalons are located at the same points in the Borel plane. On the other hand, the pole mass has ultraviolet renormalons, whereas the static potential does not. In practice the main difference comes from the relevance of the $u=-1$ renormalon. In general,   because of the $u=-1$ renormalon,
it is necessary to stop the second perturbative expansion (see \eq{mPV})  at $N \sim 2 \times \frac{2\pi}{\beta_0\alpha}$, otherwise the perturbative series would start to diverge, as we can observe in Fig. \ref{Fig:Boyd} in the $\MS$ scheme. Nevertheless, the importance of this renormalon heavily depends on the factorization scale $\mu$ one uses. If one takes $\mu$ high enough, one could indeed do perturbation theory until $N \sim 3 \times \frac{2\pi}{\beta_0\alpha}$, where the $u=3/2$ renormalon shows up. We can see the irrelevance of the $u=-1$ renormalon in the lattice scheme, which is equivalent to the $\MS$ scheme with a much larger $\mu$, in Fig. \ref{Fig:Boyd}. One should keep in mind, though, that one needs perturbation theory to a much higher order in the lattice scheme to reach the same precision than in the $\MS$ scheme. We expect this qualitative behavior of ultraviolet renormalons to also hold true beyond the large $\beta_0$ approximation. 

We next move to real QCD. We have performed determinations of $\bar \Lambda_{\rm PV}$ using quenched lattice QCD. For these observables perturbative expansions to high orders are available \cite{Bauer:2011ws,Bali:2013pla,Bali:2013qla}. 
This allows us to test the method and go beyond the superasymptotic and the leading term in the hyperasymptotic approximation. We observe ${\cal O}(a\lQ^2)$ corrections for the $B$ meson mass in the static approximation, but not for an analogous observable from the static potential. Nevertheless, we do not have enough precision to quantitatively study these effects. The limiting factor is the error of the normalization of the leading renormalon, and, related, the lack of knowledge of the higher order beta function coefficients. The latter affects the ${\cal O}(1/n)$ corrections to the asymptotic formula of the perturbative series coefficients. These effects are sizable in the lattice scheme. On the other hand they are quite small in the $\MS$ scheme. On top of that the higher order coefficients of the perturbative expansion of $\delta m_{\rm latt}$ are not known with enough precision to disentangle the subleading renormalon (their error is strongly correlated with the error of $Z_m$). All these considerations forbid quantitative analyses beyond the leading term in the hyperasymptotic approximation. Further investigations are needed to improve on these issues, particularly on the error of $Z_m$, which also affects the discussion below. 

We also determine $\bar \Lambda_{\rm PV}$ from the physical $B$ meson mass assuming that the $\MS$ heavy quark mass is known. The result can be found in \eq{LPVnumber}. In this analysis, we determine the error associated to the incomplete knowledge of the perturbative expansion in determinations of the heavy quark mass. We translate this result to the case of the top mass, which we study in detail in Sec. \ref{Sec:Top}. 
In this section the issue of the uncertainty of the (top) pole mass is critically reexamined. In particular, the bottom and charm quark finite mass effects are carefully incorporated. In our implementation we find these to be very small. We find the present uncertainty in the relation between $\m_t$ and $m_{\rm PV}$  to be (for $\m_t=163$ GeV)
\be
m_{t,\rm PV}(163{\rm MeV})=173033 ({\rm th})^{+25}_{-28}(\alpha)^{+119}_{-123}\;{\rm MeV}
\,,
\ee
\be
[\frac{m_{t,\rm PV}}{\m_t}-1]\times 10^5= 6155\, ({\rm th})^{+15}_{-17}\,(\alpha)^{+73}_{-75}
\,,
\ee
where we have combined the theoretical errors quoted in \eqs{mtPVfinalnumber}{eq:mtnum} in quadrature. There is no ambiguity associated to the renormalon in this number. The precision is systematically improvable the more terms of the perturbative expansion get to be known in the future. 
Interestingly enough, it seems we have found some evidence for the existence of the next renormalon at $u=-1$ but not of a possible renormalon at $u=1$. We believe this makes very timely a quantitative determination of the renormalization group structure of the $u=-1$ renormalon, which to our knowledge is lacking. We leave this for future work. 

\medskip

{\bf Acknowledgments}\\
\noindent
We thank M. Steinhauser for comments on the manuscript. 
C.A. thanks the IFAE group at Universitat Aut\`onoma de Barcelona for warm hospitality
during part of this work. This work was supported in part by the Spanish FPA2017-86989-P and SEV-2016-0588 grants from the ministerio de Ciencia, Innovaci\'on y Universidades, and the 2017SGR1069 grant from the Generalitat de Catalunya; and by the Chilean FONDECYT Postdoctoral Grant No. 3170116, and by FONDECYT Regular Grant No. 1180344.

\begin{appendix}
\appendix

\section{Evaluation of $\Omega_d$}
\label{Sec:Integral}
We here briefly sketch how we compute the integrals that appear in $\Omega_d$. For $d<0$ we use the recursion formulas developed in \cite{Dingle} (see for instance Eq. (46) of Chapter XXI). For $d>0$, we can use also such formulas (see for instance Eq. (47) of Chapter XXI). In this case, we can also alternatively perform the integration in the following way. For simplicity, we take the case $b=0$ and $d=1$, as the method is similar for the more general case. 
\be
I=\int_{0,\rm PV}^{\infty}due^{-\frac{4\pi}{\beta_0\al}u}\frac{(2u)^{N+1}}{1-2u}
=
-\frac{1}{2}e^{-\frac{2\pi}{\beta_0\al}}
\int_{-\frac{1}{2},\rm PV}^{\infty}\frac{dy}{y}e^{-\frac{4\pi}{\beta_0\al}y}
e^{\left(\frac{2\pi}{\beta_0\al}(1-c\al)+1\right)\ln(1+2y)}
\,,
\ee
where in the second equality we set $N=N_P$ according to \eq{eq:NP}. We also do the change of variables (where $K=c-\frac{\beta_0}{2\pi}$)
\be
y=\frac{x}{2}\frac{1}{\sqrt{1-K\al}}\sqrt{\frac{\beta_0\al}{\pi}}
\,.
\ee
We can then expand the exponent in powers of $\al$ and $x$:
\be
I=
-\frac{1}{2}e^{-\frac{2\pi}{\beta_0\al}}
\int_{-\sqrt{\frac{\pi}{\beta_0\al}(1-K\al)},\rm PV}^{\infty}
\frac{dx}{x}e^{-x^2}
e^{-\frac{4\pi}{\beta_0\al}K\frac{x}{2}\sqrt{\frac{\beta_0\al}{\pi}}\frac{1}{\sqrt{1-K\al}}+\frac{2}{3}x^3\sqrt{\frac{\beta_0\al}{\pi}}\frac{1}{\sqrt{1-K\al}}+\cdots}
\,.
\ee
The inferior  limit of the integral is then extended to $-\infty$. We then have
\bea
I&\simeq&
-\frac{1}{2}e^{-\frac{2\pi}{\beta_0\al}}
\int_{-\infty,\rm PV}^{\infty}
\frac{dx}{x}e^{-x^2}
e^{-\frac{4\pi}{\beta_0\al}K\frac{x}{2}\sqrt{\frac{\beta_0\al}{\pi}}\frac{1}{\sqrt{1-K\al}}+\frac{2}{3}x^3\sqrt{\frac{\beta_0\al}{\pi}}\frac{1}{\sqrt{1-K\al}}+\cdots}
\nn
\\
&\simeq&
e^{-\frac{2\pi}{\beta_0\al}}\frac{1}{6}\sqrt{\beta_0\al}\left(-1+\frac{6\pi}{\beta_0}K\right)+\cdots \,.
\eea

Irrespectively of considering $d>0$ or $d>0$, it is not clear to us what is the asymptotic structure of this expansion. This is something that we are investigating. In any case, at present, we have not seen evidence of asymptotic behavior of this expansion for all cases we have considered. This does not preclude however that if we go to higher orders we will find an asymptotic behavior for this perturbative expansion. 

\section{bottom and charm finite mass contributions to $m_{t,\rm PV}$}
\label{Sec:light}
We define
\begin{eqnarray}
\delta m_q^{(1)} &\equiv& 
\frac{{\m}}{3} \left(\left(1-\frac{{\m}_q}{{\m}}\right) \left(1-\frac{{\m}_q^3}{{\m}^3}\right) \left(\text{Li}_2\left(\frac{{\m}_q}{{\m}}\right)-\frac{1}{2} {\rm ln} ^2\left(\frac{{\m}_q}{{\m}}\right)+{\rm ln}\left(1-\frac{{\m}_q}{{\m}}\right) \right. \right.
\nonumber\\ 
&&
\left.\left.
\times  {\rm ln}\left(\frac{{\m}_q}{{\m}}\right)-\frac{\pi ^2}{3}\right)+\left(1+\frac{{\m}_q}{{\m}}\right) \left(1+\frac{{\m}_q^3}{{\m}^3}\right) \left(\text{Li}_2\left(-\frac{{\m}_q}{{\m}}\right)-\frac{1}{2} {\rm ln}^2\left(\frac{{\m}_q}{{\m}}\right)\right.\right.
\nonumber\\ 
&&
\left.\left.
+{\rm ln} \left(1+\frac{{\m}_q}{{\m}}\right) {\rm ln} \left(\frac{{\m}_q}{{\m}}\right)+\frac{\pi ^2}{6}\right)-\frac{{\m}_q^2}{{\m}^2}\left({\rm ln} \left(\frac{{\m}_q}{{\m}}\right)+\frac{3}{2}\right)+{\rm ln}^2\left(\frac{{\m}_q}{{\m}}\right) \right.
\nonumber\\ 
&& \left.
+\frac{\pi ^2}{6}\right)
\end{eqnarray}
(note that this coefficient is $n_f$-independent), 
\begin{eqnarray}
\delta m_q^{(2,n_f)} &=& \frac{{\m}}{64} \left[h\left(\frac{{\m}_q}{{\m}}\right) +w\left(1,\frac{{\m}_q}{{\m}}\right) +n_f\ p\left(\frac{{\m}_q}{{\m}}\right) \right]
\,,
\end{eqnarray}
where $n_f=5$ for $q=b$ and $n_f=4$ for $q=c$, and we use the representation for the functions $h(x)$, $w(x,y)$ and $p(x)$ given in Ref.~\cite{Hoang:2017btd}, and
\be
\delta m_{bc}^{(2)} = \frac{\m}{64}\ w\left(\frac{\m_{b}}{\m},\frac{\m_{c}}{\m}\right) \,.
\ee
%\begin{eqnarray}
%\delta m_{bc}^{(2,n_f)} &=& \frac{\m}{64}  \left[w\left(1,\frac{\m_{b}}{\m}\right)+w\left(1,\frac{\m_{c}}{\m}\right)+w\left(\frac{\m_{b}}{\m},\frac{\m_{c}}{\m}\right)+h\left(\frac{\m_{b}}{\m}\right)+h\left(\frac{\m_{c}}{\m}\right) \right.
%\nonumber\\ 
%&& \left.
%+n_f \left(p\left(\frac{\m_{b}}{\m}\right)+p\left(\frac{\m_{c}}{\m}\right)\right)-p\left(\frac{\m_{c}}{\m}\right)
%\right]
% \,.
%\end{eqnarray}
%in our case $q_1=b$, $q_2=c$, and $Q=t$. 

We then have 
\bea
\delta m_{b/c}^{(5)}&=&\delta m_{b/c}^{(1)}\frac{\al_{(5)}^{2}(\m)}{\pi^2}+\delta m^{(2,5/4)}_{b/c}\frac{\al_{(5)}^{3}(\m)}{\pi^3}
\\
\delta m_{bc}^{(5)}&=&\delta m^{(2)}_{bc}\frac{\al_{(5)}^{3}(\m)}{\pi^3}
\\
\delta m_{b}^{(4)}&=&\left[\delta m_{b}^{(1)}+\delta m_{b,dec}^{(1)}\right]\frac{\al_{(4)}^{2}(\m)}{\pi^2}
+\left[\delta m^{(2,5)}_{b}+\delta m_{b,dec}^{(2)}\right]\frac{\al_{(4)}^{3}(\m)}{\pi^3}
\\
\label{A7}
\delta m_{c}^{(4)}&=&\delta m_{c}^{(1)}\frac{\al_{(4)}^{2}(\m)}{\pi^2}+\delta m^{(2,4)}_{c}\frac{\al_{(4)}^{3}(\m)}{\pi^3}
\\
\delta m_{bc}^{(4)}&=&\left[\delta m^{(2)}_{bc}+\delta m^{(2)}_{bc,dec}\right]\frac{\al_{(4)}^{3}(\m)}{\pi^3}
\\
\delta m_{b}^{(3)}&=&\left[\delta m_{b}^{(1)}+\delta m_{b,dec}^{(1)}\right]\frac{\al_{(3)}^{2}(\m)}{\pi^2}
+\left[\delta m^{(2,5)}_{b}+\delta m_{b,dec}^{(2)}\right]\frac{\al_{(3)}^{3}(\m)}{\pi^3}
\\
\label{A10}
\delta m_{c}^{(3)}&=&\left[\delta m_{c}^{(1)}+\delta m_{c,dec}^{(1)}\right]\frac{\al_{(3)}^{2}(\m)}{\pi^2}
+\left[\delta m^{(2,4)}_{c}+\delta m_{c,dec}^{(2)}\right]\frac{\al_{(3)}^{3}(\m)}{\pi^3}
\\
\delta m_{bc}^{(3)}&=&\left[\delta m^{(2)}_{bc}+\delta m^{(2)}_{bc,dec}
+\delta m^{(2)}_{cb,dec}\right]\frac{\al_{(3)}^{3}(\m)}{\pi^3}
\,,
\eea
where $\delta m_{(q,\rm dec)}^{(i)}$ are generated by the decoupling and read
\bea
 \delta  m_{(q,\rm dec)}^{(1)}=
-\frac{2}{9}{\m}\left(\frac{71}{32}+{\rm ln}\left(\frac{\m_{q}^2}{\m^2}\right)+\frac{\pi^2}{4} \right)
\,,
%&&
%\frac{1}{3} \left(\frac{2}{3} \log \left(
%  \m_b^2/\m_c^2\right)
%   \right.
%   \\
%   &&
%    \left.
%   +3 \left(-\frac{\zeta (3)}{6}+\frac{\pi ^2}{9}+\frac{2195}{288}+\frac{1}{9} \pi ^2 \log
%   (2)\right)\right)
%   \\
%   &&
% +\frac{\zeta (3)}{6}-\frac{\pi ^2}{6}-\frac{779}{96}-\frac{1}{9} \pi ^2 \log (2)
\eea
\begin{eqnarray}
\delta m_{(q,dec)}^{(2,n_f)} &=& {\m}\left\{\left[\frac{2353}{11664}+\frac{7}{27}\zeta(3)+\frac{13 \pi ^2}{162}-\left(\frac{\pi ^2}{54}+\frac{71}{432}\right){\rm ln}\left(\frac{{\m}^2}{{\m}_{q}^2} \right)\right] n_f +\frac{8 \text{Li}_4\left(\frac{1}{2}\right)}{27}\right.
\nonumber\\ 
&& \left.
-\frac{751 }{216}\zeta (3) +\frac{61 \pi ^4}{1944}-\frac{113 \pi ^2}{72}-\frac{29869}{2916}+\frac{{\rm ln}^4(2)}{81}+\frac{2}{81} \pi ^2 {\rm ln}^2(2)-\frac{11}{81} \pi ^2 {\rm ln}(2) \right.
\nonumber\\ 
&& \left.
+\left(\frac{1225}{288}-\frac{1}{18}\zeta (3)+\frac{\pi^2}{9}+\frac{1}{27} \pi ^2 {\rm ln}(2)\right){\rm ln}\left(\frac{{\m}^2}{{\m}_{q}^2} \right)+\frac{1}{27}{\rm ln}^2\left(\frac{{\m}^2}{{\m}_{q}^2} \right)
 \right\}
\nonumber\\ 
&&
+\frac{1}{3}{\rm ln}\left(\frac{{\m}^2}{{\m}_{q}^2} \right)\delta m_{q}^{(1)}
\,.
\end{eqnarray}
Note that $\delta m_{(b,dec)}^{(2)}=\delta m_{(b,dec)}^{(2,5)}$ and $\delta m_{(c,dec)}^{(2)}=\delta m_{(b,dec)}^{(2,4)}$. This last expression indeed corresponds to Eq. (17)  of \cite{Ayala:2014yxa} changing $\m_b$ by $\m$. 

Finally, we also have
\be
\delta m^{(2)}_{bc,dec}=\frac{1}{3}\ln \left(\frac{\m^2}{\m_b^2}\right)\delta m_c^{(1)}
\,,
\ee
\be
\delta m^{(2)}_{cb,dec} =\frac{1}{3}\ln \left(\frac{\m^2}{\m_c^2}\right)[\delta m_b^{(1)}+\delta m_{b,dec}^{(1)}]
\,.
\ee

\end{appendix}


\begin{thebibliography}{13}

\bibitem{HyperI}
   C.~Ayala, X.~Lobregat and A.~Pineda,
  %``Superasymptotic and hyperasymptotic approximation to the operator product expansion,''
  Phys.\ Rev.\ D {\bf 99}, no. 7, 074019 (2019)
 % doi:10.1103/PhysRevD.99.074019
  [arXiv:1902.07736 [hep-th]].

\bibitem{BerryandHowls}
M. V. Berry and C. J. Howls, Hyperasymptotics, Proc. Roy. Soc. London
A, 430 (1990), pp. 653-668.

\bibitem{Boyd99}
J. P. Boyd, {\it The Devil's Invention: Asymptotic, Superasymptotic and Hyperasymptotic Series},
Acta Applicandae Mathematica, Vol. {\bf 56}, 1 (1999).

%\cite{Ayala:2019lak}
\bibitem{Ayala:2019lak} 
  C.~Ayala, X.~Lobregat and A.~Pineda,
  %``Hyperasymptotic approximation to the operator product expansion,''
  arXiv:1910.04090 [hep-ph].
  %%CITATION = ARXIV:1910.04090;%%

\bibitem{Dingle} R.B. Dingle, {\it Asymptotic Expansions: Their
Derivation and Interpretation} (Academic Press, London, 1973).

%\cite{Tarrach:1980up}
\bibitem{Tarrach:1980up} 
  R.~Tarrach,
  %``The Pole Mass in Perturbative QCD,''
  Nucl.\ Phys.\ B {\bf 183}, 384 (1981).
  %doi:10.1016/0550-3213(81)90140-1
  %%CITATION = doi:10.1016/0550-3213(81)90140-1;%%

%\cite{Chetyrkin:1999ys}
\bibitem{Chetyrkin:1999ys} 
  K.~G.~Chetyrkin and M.~Steinhauser,
  %``Short distance mass of a heavy quark at order $\alpha_s^3$,''
  Phys.\ Rev.\ Lett.\  {\bf 83}, 4001 (1999)
  %doi:10.1103/PhysRevLett.83.4001
  [hep-ph/9907509].
  %%CITATION = doi:10.1103/PhysRevLett.83.4001;%%

%\cite{Melnikov:2000qh}
\bibitem{Melnikov:2000qh} 
  K.~Melnikov and T.~v.~Ritbergen,
  %``The Three loop relation between the MS-bar and the pole quark masses,''
  Phys.\ Lett.\ B {\bf 482}, 99 (2000)
 % doi:10.1016/S0370-2693(00)00507-4
  [hep-ph/9912391].
  %%CITATION = doi:10.1016/S0370-2693(00)00507-4;%%
  
%\cite{Marquard:2015qpa}
\bibitem{Marquard:2015qpa} 
  P.~Marquard, A.~V.~Smirnov, V.~A.~Smirnov and M.~Steinhauser,
  %``Quark Mass Relations to Four-Loop Order in Perturbative QCD,''
  Phys.\ Rev.\ Lett.\  {\bf 114}, no. 14, 142002 (2015)
  %doi:10.1103/PhysRevLett.114.142002
  [arXiv:1502.01030 [hep-ph]].
  %%CITATION = doi:10.1103/PhysRevLett.114.142002;%%

%\cite{Beneke:1994rs}
\bibitem{Beneke:1994rs} 
  M.~Beneke,
  %``More on ambiguities in the pole mass,''
  Phys.\ Lett.\ B {\bf 344}, 341 (1995)
  %doi:10.1016/0370-2693(94)01505-7
  [hep-ph/9408380].
  %%CITATION = doi:10.1016/0370-2693(94)01505-7;%%

%\cite{Beneke:1998ui}
\bibitem{Beneke:1998ui} 
  M.~Beneke,
  %``Renormalons,''
  Phys.\ Rept.\  {\bf 317}, 1 (1999)
  %doi:10.1016/S0370-1573(98)00130-6
  [hep-ph/9807443].
  %%CITATION = doi:10.1016/S0370-1573(98)00130-6;%%

%\cite{Pineda:2001zq}
\bibitem{Pineda:2001zq} 
  A.~Pineda,
  %``Determination of the bottom quark mass from the Upsilon(1S) system,''
  JHEP {\bf 0106}, 022 (2001)
%  doi:10.1088/1126-6708/2001/06/022
  [hep-ph/0105008].
  %%CITATION = doi:10.1088/1126-6708/2001/06/022;%%

%\cite{Ayala:2014yxa}
\bibitem{Ayala:2014yxa} 
  C.~Ayala, G.~Cvetic and A.~Pineda,
  %``The bottom quark mass from the $ \boldsymbol{\Upsilon} (1S) $ system at NNNLO,''
  JHEP {\bf 1409}, 045 (2014)
  %doi:10.1007/JHEP09(2014)045
  [arXiv:1407.2128 [hep-ph]].
  %%CITATION = doi:10.1007/JHEP09(2014)045;%%

 %\cite{Luke:1992cs}
\bibitem{Luke:1992cs} 
  M.~E.~Luke and A.~V.~Manohar,
  %``Reparametrization invariance constraints on heavy particle effective field theories,''
  Phys.\ Lett.\ B {\bf 286}, 348 (1992)
  %doi:10.1016/0370-2693(92)91786-9
  [hep-ph/9205228].
  %%CITATION = doi:10.1016/0370-2693(92)91786-9;%%

%\cite{Neubert:1996zy}
\bibitem{Neubert:1996zy} 
  M.~Neubert,
  %``Exploring the invisible renormalon: Renormalization of the heavy quark kinetic energy,''
  Phys.\ Lett.\ B {\bf 393}, 110 (1997)
  %doi:10.1016/S0370-2693(96)01600-0
  [hep-ph/9610471].
  %%CITATION = doi:10.1016/S0370-2693(96)01600-0;%%

%\cite{Sumino:2005cq}
\bibitem{Sumino:2005cq} 
  Y.~Sumino,
  %``Static QCD potential at r < Lambda(QCD)-1: Perturbative expansion and operator-product expansion,''
  Phys.\ Rev.\ D {\bf 76}, 114009 (2007)
 % doi:10.1103/PhysRevD.76.114009
  [hep-ph/0505034].
  %%CITATION = doi:10.1103/PhysRevD.76.114009;%%

%\cite{VanAcoleyen:2003gc}
\bibitem{VanAcoleyen:2003gc} 
  K.~Van Acoleyen and H.~Verschelde,
  %``QCD perturbation theory at large orders with large renormalization scales in the large beta(0) limit,''
  Phys.\ Rev.\ D {\bf 69}, 125006 (2004)
  %doi:10.1103/PhysRevD.69.125006
  [hep-ph/0307070].
  %%CITATION = doi:10.1103/PhysRevD.69.125006;%%

%\cite{Beneke:1994sw}
\bibitem{Beneke:1994sw} 
  M.~Beneke and V.~M.~Braun,
  %``Heavy quark effective theory beyond perturbation theory: Renormalons, the pole mass and the residual mass term,''
  Nucl.\ Phys.\ B {\bf 426}, 301 (1994)
  %doi:10.1016/0550-3213(94)90314-X
  [hep-ph/9402364].
  %%CITATION = doi:10.1016/0550-3213(94)90314-X;%%

  %%%\cite{Ball:1995ni}
\bibitem{Ball:1995ni} 
  P.~Ball, M.~Beneke and V.~M.~Braun,
  %``Resummation of (beta0 alpha-s)**n corrections in QCD: Techniques and applications to the tau hadronic width and the heavy quark pole mass,''
  Nucl.\ Phys.\ B {\bf 452}, 563 (1995)
  %doi:10.1016/0550-3213(95)00392-6
  [hep-ph/9502300].
  %%CITATION = doi:10.1016/0550-3213(95)00392-6;%%
  %%
 
  
 %\cite{Neubert:1994vb}
\bibitem{Neubert:1994vb} 
  M.~Neubert,
  %``Scale setting in QCD and the momentum flow in Feynman diagrams,''
  Phys.\ Rev.\ D {\bf 51}, 5924 (1995)
  %doi:10.1103/PhysRevD.51.5924
  [hep-ph/9412265].
  %%CITATION = doi:10.1103/PhysRevD.51.5924;%%

%\cite{Bauer:2011ws}
\bibitem{Bauer:2011ws} 
  C.~Bauer, G.~S.~Bali and A.~Pineda,
  %``Compelling Evidence of Renormalons in QCD from High Order Perturbative Expansions,''
  Phys.\ Rev.\ Lett.\  {\bf 108}, 242002 (2012)
  %doi:10.1103/PhysRevLett.108.242002
  [arXiv:1111.3946 [hep-ph]].
  %%CITATION = doi:10.1103/PhysRevLett.108.242002;%%

%\cite{Bali:2013pla}
\bibitem{Bali:2013pla} 
  G.~S.~Bali, C.~Bauer, A.~Pineda and C.~Torrero,
  %``Perturbative expansion of the energy of static sources at large orders in four-dimensional SU(3) gauge theory,''
  Phys.\ Rev.\ D {\bf 87}, 094517 (2013)
  %doi:10.1103/PhysRevD.87.094517
  [arXiv:1303.3279 [hep-lat]].
  %%CITATION = doi:10.1103/PhysRevD.87.094517;%%

%\cite{Bali:2013qla}
\bibitem{Bali:2013qla} 
  G.~S.~Bali, C.~Bauer and A.~Pineda,
  %``The static quark self-energy at $O(\alpha^20)$ in perturbation theory,''
  PoS LATTICE {\bf 2013}, 371 (2014)
  [arXiv:1311.0114 [hep-lat]].
  %%CITATION = ARXIV:1311.0114;%%

%\cite{Hayashi:2019mlb}
\bibitem{Hayashi:2019mlb} 
  Y.~Hayashi and Y.~Sumino,
  %``UV contributions to energy of a static quark-antiquark pair in large $\beta_0$ approximation,''
  Phys.\ Lett.\ B {\bf 795}, 107 (2019)
  %doi:10.1016/j.physletb.2019.05.049
  [arXiv:1904.02563 [hep-ph]].

%\cite{Hasenfratz:1980kn}
\bibitem{Hasenfratz:1980kn} 
  A.~Hasenfratz and P.~Hasenfratz,
  %``The Connection Between the Lambda Parameters of Lattice and Continuum QCD,''
  Phys.\ Lett.\  {\bf 93B}, 165 (1980).
  %doi:10.1016/0370-2693(80)90118-5
  %%CITATION = doi:10.1016/0370-2693(80)90118-5;%%
  
%\cite{DiRenzo:1994sy}
\bibitem{DiRenzo:1994sy} 
  F.~Di Renzo, E.~Onofri, G.~Marchesini and P.~Marenzoni,
  %``Four loop result in SU(3) lattice gauge theory by a stochastic method: Lattice correction to the condensate,''
  Nucl.\ Phys.\ B {\bf 426}, 675 (1994)
  %doi:10.1016/0550-3213(94)90026-4
  [hep-lat/9405019].
  %%CITATION = doi:10.1016/0550-3213(94)90026-4;%%

%\cite{DiRenzo:2004hhl}
\bibitem{DiRenzo:2004hhl} 
  F.~Di Renzo and L.~Scorzato,
  %``Numerical stochastic perturbation theory for full QCD,''
  JHEP {\bf 0410}, 073 (2004)
  %doi:10.1088/1126-6708/2004/10/073
  [hep-lat/0410010].
  %%CITATION = doi:10.1088/1126-6708/2004/10/073;%%

%\cite{Necco:2001xg}
\bibitem{Necco:2001xg} 
  S.~Necco and R.~Sommer,
  %``The N(f) = 0 heavy quark potential from short to intermediate distances,''
  Nucl.\ Phys.\ B {\bf 622}, 328 (2002)
  %doi:10.1016/S0550-3213(01)00582-X
  [hep-lat/0108008].
  %%CITATION = doi:10.1016/S0550-3213(01)00582-X;%%

%\cite{Bali:2014sja}
\bibitem{Bali:2014sja} 
  G.~S.~Bali, C.~Bauer and A.~Pineda,
  %``Model-independent determination of the gluon condensate in four-dimensional SU(3) gauge theory,''
  Phys.\ Rev.\ Lett.\  {\bf 113}, 092001 (2014)
  %doi:10.1103/PhysRevLett.113.092001
  [arXiv:1403.6477 [hep-ph]].
  %%CITATION = doi:10.1103/PhysRevLett.113.092001;%%

%\cite{Duncan:1994uq}
\bibitem{Duncan:1994uq} 
  A.~Duncan, E.~Eichten, J.~Flynn, B.~R.~Hill, G.~Hockney and H.~Thacker,
  %``Properties of B mesons in lattice QCD,''
  Phys.\ Rev.\ D {\bf 51}, 5101 (1995)
  %doi:10.1103/PhysRevD.51.5101
  [hep-lat/9407025].
  %%CITATION = doi:10.1103/PhysRevD.51.5101;%%

%\cite{Allton:1994tt}
\bibitem{Allton:1994tt} 
  C.~R.~Allton {\it et al.} [APE Collaboration],
  %``Results for the B meson decay constant from the APE Collaboration,''
  Nucl.\ Phys.\ Proc.\ Suppl.\  {\bf 42}, 385 (1995)
 % doi:10.1016/0920-5632(95)00257-A
  [hep-lat/9502013].
  %%CITATION = doi:10.1016/0920-5632(95)00257-A;%%

%\cite{Ewing:1995ih}
\bibitem{Ewing:1995ih} 
  A.~K.~Ewing {\it et al.} [UKQCD Collaboration],
  %``Heavy quark spectroscopy and matrix elements: a lattice study using the static approximation,''
  Phys.\ Rev.\ D {\bf 54}, 3526 (1996)
  %doi:10.1103/PhysRevD.54.3526
  [hep-lat/9508030].
  %%CITATION = doi:10.1103/PhysRevD.54.3526;%%

%\cite{Bali:2003jq}
\bibitem{Bali:2003jq} 
  G.~S.~Bali and A.~Pineda,
  %``QCD phenomenology of static sources and gluonic excitations at short distances,''
  Phys.\ Rev.\ D {\bf 69}, 094001 (2004)
  %doi:10.1103/PhysRevD.69.094001
  [hep-ph/0310130].
  %%CITATION = doi:10.1103/PhysRevD.69.094001;%%

\bibitem{statpot}
G.~S.~Bali and K.~Schilling,
%``Static Quark - Anti-Quark Potential: Scaling Behavior And Finite Size Effects In SU(3) Lattice Gauge Theory,''
Phys.\ Rev.\ D {\bf 46}, 2636 (1992);
%%CITATION = PHRVA,D46,2636;%%
%``Running coupling and the Lambda parameter from SU(3) lattice simulations,''
Phys.\ Rev.\ D {\bf 47}, 661 (1993)
[arXiv:hep-lat/9208028];
%%CITATION = HEP-LAT 9208028;%%
%``The Static quark - anti-quark potential: A 'Classical' experiment on the connection machine CM-2,''
Int.\ J.\ Mod.\ Phys.\ C {\bf 4}, 1167 (1993)
[arXiv:hep-lat/9308014].
%%CITATION = HEP-LAT 9308014;%%

\bibitem{Bali:1997am}
G.~S.~Bali, K.~Schilling and A.~Wachter,
%``Complete O(v**2) corrections to the static interquark potential from  SU(3) gauge theory,''
Phys.\ Rev.\ D {\bf 56}, 2566 (1997)
[arXiv:hep-lat/9703019].
%%CITATION = HEP-LAT 9703019;%%

%\cite{Peset:2018ria}
\bibitem{Peset:2018ria} 
  C.~Peset, A.~Pineda and J.~Segovia,
  %``The charm/bottom quark mass from heavy quarkonium at N$^{3}$LO,''
  JHEP {\bf 1809}, 167 (2018)
  %doi:10.1007/JHEP09(2018)167
  [arXiv:1806.05197 [hep-ph]].
  %%CITATION = doi:10.1007/JHEP09(2018)167;%%

%\cite{Tanabashi:2018oca}
\bibitem{Tanabashi:2018oca} 
  M.~Tanabashi {\it et al.} [Particle Data Group],
  %``Review of Particle Physics,''
  Phys.\ Rev.\ D {\bf 98}, no. 3, 030001 (2018).
  %doi:10.1103/PhysRevD.98.030001
  %%CITATION = doi:10.1103/PhysRevD.98.030001;%%

 %\cite{Bazavov:2018omf}
\bibitem{Bazavov:2018omf} 
  A.~Bazavov {\it et al.} [Fermilab Lattice and MILC and TUMQCD Collaborations],
  %``Up-, down-, strange-, charm-, and bottom-quark masses from four-flavor lattice QCD,''
  Phys.\ Rev.\ D {\bf 98}, no. 5, 054517 (2018)
  %doi:10.1103/PhysRevD.98.054517
  [arXiv:1802.04248 [hep-lat]].
  %%CITATION = doi:10.1103/PhysRevD.98.054517;%%
  
%\cite{Lee:2003hh}
\bibitem{Lee:2003hh} 
  T.~Lee,
  %``Heavy quark mass determination from the quarkonium ground state energy: A Pole mass approach,''
  JHEP {\bf 0310}, 044 (2003)
%  doi:10.1088/1126-6708/2003/10/044
  [hep-ph/0304185].
  %%CITATION = doi:10.1088/1126-6708/2003/10/044;%%

%\cite{Lee:2005hf}
\bibitem{Lee:2005hf} 
  T.~Lee,
  %``Constraining renormalon effects in lattice determination of heavy quark mass,''
  Phys.\ Rev.\ D {\bf 73}, 054505 (2006)
  %doi:10.1103/PhysRevD.73.054505
  [hep-ph/0511238].
  %%CITATION = doi:10.1103/PhysRevD.73.054505;%%

%\cite{Brambilla:2017hcq}
\bibitem{Brambilla:2017hcq} 
  N.~Brambilla {\it et al.} [TUMQCD Collaboration],
  %``Relations between Heavy-light Meson and Quark Masses,''
  Phys.\ Rev.\ D {\bf 97}, no. 3, 034503 (2018)
  %doi:10.1103/PhysRevD.97.034503
  [arXiv:1712.04983 [hep-ph]].
  %%CITATION = doi:10.1103/PhysRevD.97.034503;%%
  
%\cite{Komijani:2017vep}
\bibitem{Komijani:2017vep} 
  J.~Komijani,
  %``A discussion on leading renormalon in the pole mass,''
  JHEP {\bf 1708}, 062 (2017)
%  doi:10.1007/JHEP08(2017)062
  [arXiv:1701.00347 [hep-ph]].
  %%CITATION = doi:10.1007/JHEP08(2017)062;%%
  
  %\cite{Lee:1996yk}
\bibitem{Lee:1996yk} 
  T.~Lee,
  %``Renormalons beyond one loop,''
  Phys.\ Rev.\ D {\bf 56}, 1091 (1997)
  %doi:10.1103/PhysRevD.56.1091
  [hep-th/9611010].
  %%CITATION = doi:10.1103/PhysRevD.56.1091;%%

%\cite{Hoang:2017suc}
\bibitem{Hoang:2017suc} 
  A.~H.~Hoang, A.~Jain, C.~Lepenik, V.~Mateu, M.~Preisser, I.~Scimemi and I.~W.~Stewart,
  %``The MSR mass and the $ \mathcal{O}\left({\Lambda}_{\mathrm{QCD}}\right) $ renormalon sum rule,''
  JHEP {\bf 1804}, 003 (2018)
  %doi:10.1007/JHEP04(2018)003
  [arXiv:1704.01580 [hep-ph]].
  %%CITATION = doi:10.1007/JHEP04(2018)003;%%

%\cite{Beneke:2016cbu}
\bibitem{Beneke:2016cbu} 
  M.~Beneke, P.~Marquard, P.~Nason and M.~Steinhauser,
  %``On the ultimate uncertainty of the top quark pole mass,''
  Phys.\ Lett.\ B {\bf 775}, 63 (2017)
  %doi:10.1016/j.physletb.2017.10.054
  [arXiv:1605.03609 [hep-ph]].
  %%CITATION = doi:10.1016/j.physletb.2017.10.054;%%

%\cite{Pineda:2017uby}
\bibitem{Pineda:2017uby} 
  A.~Pineda,
  %``Comment on "The MSR Mass and the ${\cal O}(\Lambda_{\rm QCD})$ Renormalon Sum Rule",''
  arXiv:1704.05095 [hep-ph].
  %%CITATION = ARXIV:1704.05095;%%

%\cite{ATLAS:2014wva}
\bibitem{ATLAS:2014wva} 
  [ATLAS and CDF and CMS and D0 Collaborations],
  %``First combination of Tevatron and LHC measurements of the top-quark mass,''
  arXiv:1403.4427 [hep-ex].
  %%CITATION = ARXIV:1403.4427;%%

%\cite{Khachatryan:2015hba}
\bibitem{Khachatryan:2015hba} 
  V.~Khachatryan {\it et al.} [CMS Collaboration],
  %``Measurement of the top quark mass using proton-proton data at ${\sqrt{(s)}}$ = 7 and 8 TeV,''
  Phys.\ Rev.\ D {\bf 93}, no. 7, 072004 (2016)
  %doi:10.1103/PhysRevD.93.072004
  [arXiv:1509.04044 [hep-ex]].
  %%CITATION = doi:10.1103/PhysRevD.93.072004;%%

%\cite{Aaboud:2016igd}
\bibitem{Aaboud:2016igd} 
  M.~Aaboud {\it et al.} [ATLAS Collaboration],
  %``Measurement of the top quark mass in the $t\bar{t}\to$ dilepton channel from $\sqrt{s}=8$ TeV ATLAS data,''
  Phys.\ Lett.\ B {\bf 761}, 350 (2016)
  %doi:10.1016/j.physletb.2016.08.042
  [arXiv:1606.02179 [hep-ex]].
  %%CITATION = doi:10.1016/j.physletb.2016.08.042;%%

%\cite{Hoang:2017btd}
\bibitem{Hoang:2017btd} 
  A.~H.~Hoang, C.~Lepenik and M.~Preisser,
  %``On the Light Massive Flavor Dependence of the Large Order Asymptotic Behavior and the Ambiguity of the Pole Mass,''
  JHEP {\bf 1709}, 099 (2017)
 % doi:10.1007/JHEP09(2017)099
  [arXiv:1706.08526 [hep-ph]].
  %%CITATION = doi:10.1007/JHEP09(2017)099;%%

%\cite{Corcella:2019tgt}
\bibitem{Corcella:2019tgt} 
  G.~Corcella,
  %``The top-quark mass: challenges in definition and determination,''
  Front.\ in Phys.\  {\bf 7}, 54 (2019)
  %doi:10.3389/fphy.2019.00054
  [arXiv:1903.06574 [hep-ph]].
  %%CITATION = doi:10.3389/fphy.2019.00054;%%

%\cite{Kronfeld:1998di}
\bibitem{Kronfeld:1998di} 
  A.~S.~Kronfeld,
  %``The Perturbative pole mass in QCD,''
  Phys.\ Rev.\ D {\bf 58}, 051501 (1998)
  %doi:10.1103/PhysRevD.58.051501
  [hep-ph/9805215].
  %%CITATION = doi:10.1103/PhysRevD.58.051501;%%

%\cite{Gray:1990yh}
\bibitem{Gray:1990yh} 
  N.~Gray, D.~J.~Broadhurst, W.~Grafe and K.~Schilcher,
  %``Three Loop Relation of Quark (Modified) Ms and Pole Masses,''
  Z.\ Phys.\ C {\bf 48}, 673 (1990).
 % doi:10.1007/BF01614703
  %%CITATION = doi:10.1007/BF01614703;%

%\cite{Bekavac:2007tk}
\bibitem{Bekavac:2007tk} 
  S.~Bekavac, A.~Grozin, D.~Seidel and M.~Steinhauser,
  %``Light quark mass effects in the on-shell renormalization constants,''
  JHEP {\bf 0710}, 006 (2007)
 % doi:10.1088/1126-6708/2007/10/006
  [arXiv:0708.1729 [hep-ph]].
  %%CITATION = doi:10.1088/1126-6708/2007/10/006;%%

%\cite{Kataev:2018gle}
\bibitem{Kataev:2018gle} 
  A.~L.~Kataev and V.~S.~Molokoedov,
  %``Multiloop contributions to the $\overline{\rm{MS}}$-on-shell mass relation for heavy quarks in QCD and charged leptons in QED and the asymptotic structure of the perturbative QCD series,''
  arXiv:1807.05406 [hep-ph].
  %%CITATION = ARXIV:1807.05406;%%

%\cite{Kataev:2018mob}
\bibitem{Kataev:2018mob} 
  A.~L.~Kataev and V.~S.~Molokoedov,
  %``Dependence of Five- and Six-Loop Estimated QCD Corrections to the Relation between Pole and Running Masses of Heavy Quarks on the Number of Light Flavors,''
  JETP Lett.\  {\bf 108}, no. 12, 777 (2018)
  %doi:10.1134/S0021364018240050
  [arXiv:1811.02867 [hep-ph]].
  %%CITATION = doi:10.1134/S0021364018240050;%% 

 %\cite{Brambilla:2000gk}
\bibitem{Brambilla:2000gk} 
  N.~Brambilla, A.~Pineda, J.~Soto and A.~Vairo,
  %``The QCD potential at O(1/m),''
  Phys.\ Rev.\ D {\bf 63}, 014023 (2001)
  %doi:10.1103/PhysRevD.63.014023
  [hep-ph/0002250].
  %%CITATION = doi:10.1103/PhysRevD.63.014023;%% 

\end{thebibliography}
\end{document}